\documentclass{aa}
\usepackage{array}
\usepackage{tabularx}
\usepackage{txfonts}
\usepackage{natbib}
\usepackage{graphicx}
\usepackage{longtable}

\begin{document}
   \title{On the observability of resonant structures in planetesimal disks due to planetary migration}

   \author{R. Reche
          \and
          H. Beust
          \and
          J.-C. Augereau
          \and
          O. Absil
          }

   \offprints{R. Reche \email{Remy.Reche@obs.ujf-grenoble.fr}}

   \institute{Laboratoire d'Astrophysique de Grenoble, CNRS, Universite Joseph-Fourier,
UMR 5571, Grenoble, France}

\date{Received 24 May 2007; accepted 18 December 2007}

 
\abstract
    {The observed clumpy structures in debris disks are commonly  
interpreted as particles trapped in mean-motion resonances with an  
unseen exo-planet. Populating the resonances requires a migrating  
process of either the particles (spiraling inward due to drag  
forces) or  the planet (moving outward). Because the drag
time-scale  in resolved debris disks is generally long compared to  
the collisional time-scale, the planet migration scenario might be more  
likely, but this model has so far only been investigated for planets  
on circular orbits.}
    {We present a thorough study of the impact of a  
migrating planet on a planetesimal disk, by exploring a broad range  
of masses and eccentricities for the planet. We discuss the  
sensitivity of the structures generated in debris disks to the basic  
planet parameters.}
    {We perform many N-body numerical simulations, using the symplectic  
integrator SWIFT, taking into account the gravitational influence of  
the star and the planet on massless test particles. A constant
migration rate is assumed for the planet.}
    {The effect of planetary migration on the trapping of particles  
in mean motion resonances is found to be very sensitive to the initial
eccentricity of the planet and of the planetesimals. A planetary  
eccentricity as low as $0.05$ is enough to smear out all the resonant  
structures, except for the most massive planets. The planetesimals  
also initially have to be on orbits with a mean eccentricity of less than  
than $0.1$ in order to keep the resonant clumps visible.}
    {This numerical work extends previous analytical studies and  
provides a collection of disk images that may help in interpreting  
the observations of structures in debris disks. Overall, it shows  
that stringent conditions must be fulfilled to obtain observable  
resonant structures in debris disks. Theoretical models
  of the origin of planetary migration will
  therefore have to explain how planetary systems remain in a suitable
  configuration to reproduce the observed structures.}

  \keywords{Method: N-body simulations -- Celestial mechanics --
    (Stars:) planetary systems -- (Stars: individual:) \object{Vega}}

   \maketitle
%

\section{Introduction}
\begin{figure}
\resizebox{\hsize}{!}{\includegraphics[angle=-90]{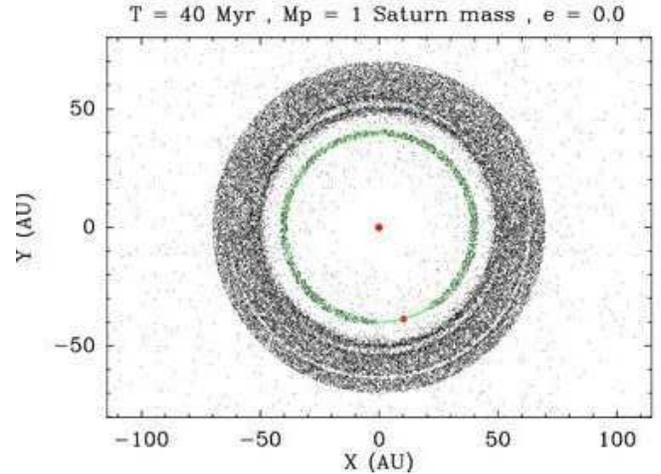}} 
\caption{\label{withoutMigration} Example of a planetesimal disk without
  outward planet migration, nor inward P-R drag migration of the test
  particles, according to our numerical simulations as explained in
  Section \ref{model}. The star and planet locations,
  projected onto the orbital plane of the planet, are represented by large
  red points, and the planet orbit by a thin green line. The initial
  planetesimal disk consists of $50\,000$ planetesimals distributed between
  40 and 75 AU, with the surface density distribution proportional to $r^{-1}$. Although
  some planetesimals are trapped in MMRs with the planet, they are not
  sufficiently numerous to
  generate spatial structures (besides the $1$:$1$ MMR). \thanks{See the electronic edition of the
    Journal for a color version of this figure.}} 
\end{figure}

Since the first direct imaging of a debris disk around \object{$\beta$ Pictoris}
 by \citet{1984Sci...226.1421S}, a dozen other optically thin dust disks have been
spatially resolved around nearby main-sequence stars showing an
infrared excess \citep[][and references there
in]{2007ApJ...661L..85K,2006ApJ...650..414S}.  The images often reveal
asymmetric structures and clumps, interpreted as the signature of gravitational perturbations. 
 A planet immersed in a debris disk usually produces structures such
 as a gap along  its orbit,
 by ejecting particles during close encounters, or density waves
 (e.g. a one-arm
spiral), by modifying the precession rate of the dust particles
\citep{2005A&A...440..937W}. However, such
structures cannot explain the observations of clumpy,
non-axisymmetric disks \citep{2004ASPC..321..305A,2007prpl.conf..573M}, and resonant
mechanisms with unseen planets have been proposed to
account for the observed asymmetries \citep{2000ApJ...537L.147O,2002ApJ...578L.149Q,2003ApJ...588.1110K,2003ApJ...598.1321W}. A
particle belongs to a mean motion resonance (MMR) when the particle to planet period ratio is
a rational number, $m:n$ with $m$ and $n$
integers. An MMR is located at a semi-major axis $a$ given by
$a/a_p=(m/n)^{2/3}$, where $a_p$ is the planet semi-major axis.
In the Solar System,  for example, about $15\%$ of the known Kuiper Belt
objects, including Pluto, are trapped in the $3$:$2$ resonance with
Neptune \citep{2007prpl.conf..895C}. The
interesting property of MMRs  for modeling asymmetric disks
is that, as explained for example in \citet{2000ssd..book.....M}, resonant objects
are not uniformly distributed in azimuth around a star: rather they gather at
specific longitudes relative to the perturbing
planet and subsequently form clumps. This arises from properties
specific to MMRs as a given particle trapped in a MMR with a planet
undergoes conjunctions with the planet at specific locations along its
orbit. The particles tend to gather around the most stable orbital configurations that
ensure that the conjunctions occur at the maximum relative distance. The
clumps, which are the result of the collective effects of resonant particles,
generally corotate with the planet \citep{2003ApJ...588.1110K}, while each of these resonant bodies
has a different period from that of the planet (except for $1$:$1$ resonant
planetesimals): hence the motion of these density waves differs from the
orbital motion of the resonant particles. However,
MMRs are very thin radial structures that usually trap a small
number of particles in a given disk. Therefore, any structure due to
MMRs has a high chance of being totally  hidden by the emission of
the non-resonant particles, as illustrated in Fig. \ref{withoutMigration}. 

For clumps due to MMRs to be observed, the population of resonant
particles must be significantly enhanced by an additional physical
process. Two mechanisms can account for this: Poynting-Robertson (P-R)
drag and
planet migration. Dust particles that are too large to be ejected from the system by
radiation pressure can spiral inward  into the star due to P-R
drag and to some other minor forces like stellar wind drag \citep[e.g.][]{2006A&A...455..987A}. In the
course of its inward migration, a dust particle can be trapped into exterior MMRs with a
planet, hence increasing the contrast of their asymmetric patterns
\citep{2003ApJ...588.1110K,2005ApJ...625..398D}. Planet migration, on the
other hand, involves particles of all sizes, except those 
ejected by radiation pressure. Many particles can be trapped in MMRs by a planet
migrating outward in the
disk. Each non-resonant particle crossing an MMR has a chance trapped and subsequently migrating,
following the resonance \citep{2003ApJ...598.1321W}.

Several authors have studied either the effect of P-R drag, or
of planet migration, on disk structures, using
different methods (analytic, semi-analytic or numerical) and various
planet parameters (mass and orbital eccentricity). A summary of
the main previous studies is provided in Table \ref{previousWorks}. The P-R drag scenario has
been extensively  studied for a wide range of parameters,
while  the migrating planet scenario has been investigated only for a planet
on a circular orbit by  \citet{2003ApJ...598.1321W}. It is thus important
to better characterize the latter scenario in order to distinguish which of the
two dominates the morphology of debris disks. Moreover, a number of
studies \citep{1996Icar..123..168L,2005A&A...433.1007W,2007A&A...462..199K}
have shown that collisions may prevent MMRs from being populated by P-R
drag since the collision timescale in massive debris disks might be
much shorter than the P-R drag migration timescale.

Therefore,  we propose in this paper to extensively study the
planet migration scenario, using numerical modeling, by generating a
synthetic catalog similar to what has been done for the P-R drag scenario using
analytical \citep{2003ApJ...588.1110K} or numerical studies
\citep{2005ApJ...625..398D}. We extend the  pioneering work done by
\citet{2003ApJ...598.1321W} in studying the influence of the
planet eccentricity on the visibility of the resonant patterns. In Section \ref{lowEccOrb},
we discuss the case of a planet migrating on a circular or low-eccentricity orbit. In Section \ref{highEccOrb}, we extend this study to
planets on orbits with eccentricities up to $0.7$. This study is
generalized to various migration rates and disk initial states in
Section \ref{Generalization}, and compared to previous studies in
Section \ref{Comparison}. The limitations
of our approach are discussed in Section \ref{Limitation}. 

\begin{table*}
\centering
\caption{\label{previousWorks}Summary of recent papers on particles trapping in MMRs with a
planet.}
\begin{tabularx}{\textwidth}{c c c c c X}
Authors & Method & \multicolumn{2}{c}{Planet parameters} & Migration &
Notes \\
&& Mass ratio$^a$ & eccentricity \\
\hline\hline
\\
\citet{2003ApJ...598.1321W} & semi-analytic & $0.0003$ to $3$& $0$ & planet & Forced migration\\
\citet{2006ApJ...639.1153W} & semi-analytic &&& none & This work
extends the previous one to smaller particles sensitive to
radiation pressure.\\
\citet{2003ApJ...588.1110K} & analytic & $0.005$ to $15$ & $0$ to $0.6$ & 
particles & Only resonant particles trapped during migration
due to P-R drag. \\
\citet{2005ApJ...625..398D} & numerical & $0.01$ to $3$
& $0$ to $0.7$&
particles & Particles migrate due to P-R drag and solar wind. \\
\citet{2002AJ....124.2305M} & numerical &$0.05$&$0$&
particles & Study of the Kuiper Belt. \\
This paper & numerical & $0.001$ to $3$ & $0$ to $0.7$ & planet & Only
planetesimals disk, forced migration \\
\hline
\end{tabularx}
\begin{list}{}{}
\item[$^a$]Planetary mass in Jovian mass divided by stellar mass
in solar mass.
\end{list}
\end{table*}

\section{Numerical model}
\label{model}
We consider a planetary system consisting of a
star surrounded by a planet and a debris disk. We address the case of
large  particles, which are insensitive to
pressure forces (radiation, stellar wind or gas
pressure). The simulated disk is thus rather a planetesimal
disk than a debris disk and we only consider gravitational
forces. Importantly, we also do not take into account the gravitational interactions
between  planetesimals as they are negligible, nor mutual
collisions. Dynamically speaking, the planetesimals are thus
considered as test particles. A typical configuration for the simulations is a Vega-like
central star ($2.5$ M$_\odot$) and a
planet orbit with a $40$ AU pericenter at the starting time. The initial
planetesimal disk consists of $50\,000$ planetesimals distributed between
$40$ and $75$ AU on circular orbits, with a surface density distribution proportional to
$r^{-1}$. The disk midplane coincides with the orbital plane of
the planet, and the inclinations of the planetesimals are randomly
distributed within $\pm 3^\circ$.

In this model, the planet keeps a Keplerian orbit around the star, or migrates at
a constant rate without modification of its eccentricity.  This basic model is
easy to implement and  to analyze, and corresponds to the case
described by \citet{2003ApJ...598.1321W}. The goal of this
paper is to extend this initial work to a wider range of planet
eccentricities by a numerical study. We start by studying planets on low-eccentricity orbits
($e<0.1$), and then extend our work to larger eccentricities.

To perform our simulations we have used the symplectic package
SWIFT \citep{1991AJ....102.1528W,1994Icar..108...18L}, to which we have added 
planetary migration. To do this, we plugged in the
\citet{2003ApJ...598.1321W} prescription. This method consists of adding an acceleration in the direction
of the orbital motion of the planet, with an intensity equal to:
$\dot{v_p}=0.5\dot{a_p}\sqrt{GM_{\ast}/a_p^3}$, where $G$ is the
gravitational constant, $M_{\ast}$ the stellar mass and $\dot{a_p}$ the
variation rate of the planet semi-major axis $a_p$. This causes a change in the planet semi-major axis
without modifying its eccentricity (for a planet on a low-eccentricity
orbit; for a planet on a higher eccentricity orbit the change is not
significant) or its inclination. We do not discuss here the origin
of the migration. It can be due either to the migration of a large
internal planet, or to the gravitational influence of the planetesimals themselves. The most
important here is that we keep $\dot{a_p}$ constant during each
simulation, generally at $0.5$ AU Myr$^{-1}$, to match the
\citet{2003ApJ...598.1321W} model for the Vega disk. We have used the
RMVS3 version of the SWIFT integrator, in order to have a better
modeling of the close encounters between the planet and the
planetesimals. In all the simulations, the system evolution is followed for $40$ Myr.

\section{Planets on low-eccentricity orbits}
\label{lowEccOrb}

\begin{figure*}
\makebox[\textwidth]{
\includegraphics[angle=-90,width=0.33\textwidth]{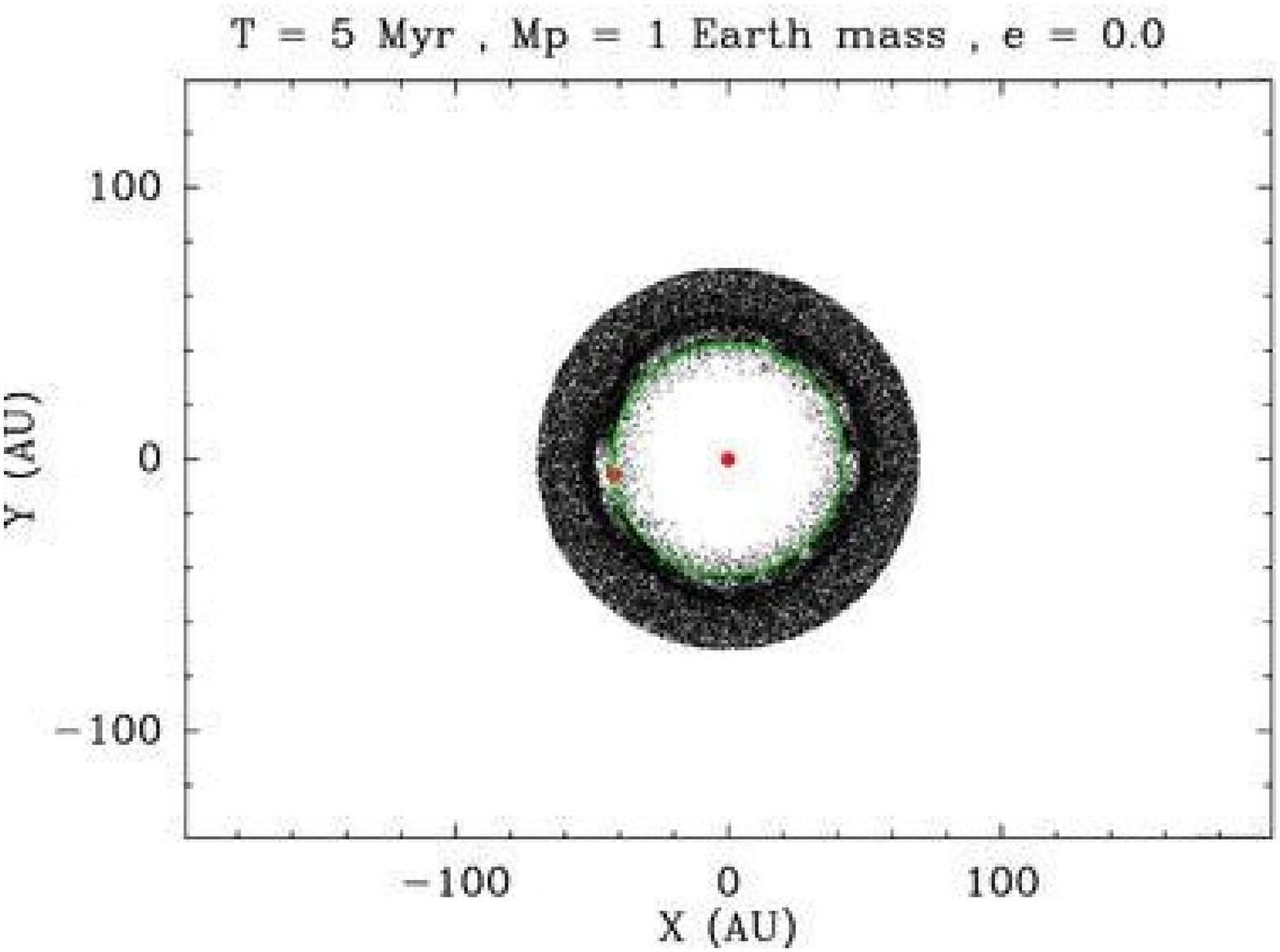}\hfil
\includegraphics[angle=-90,width=0.33\textwidth]{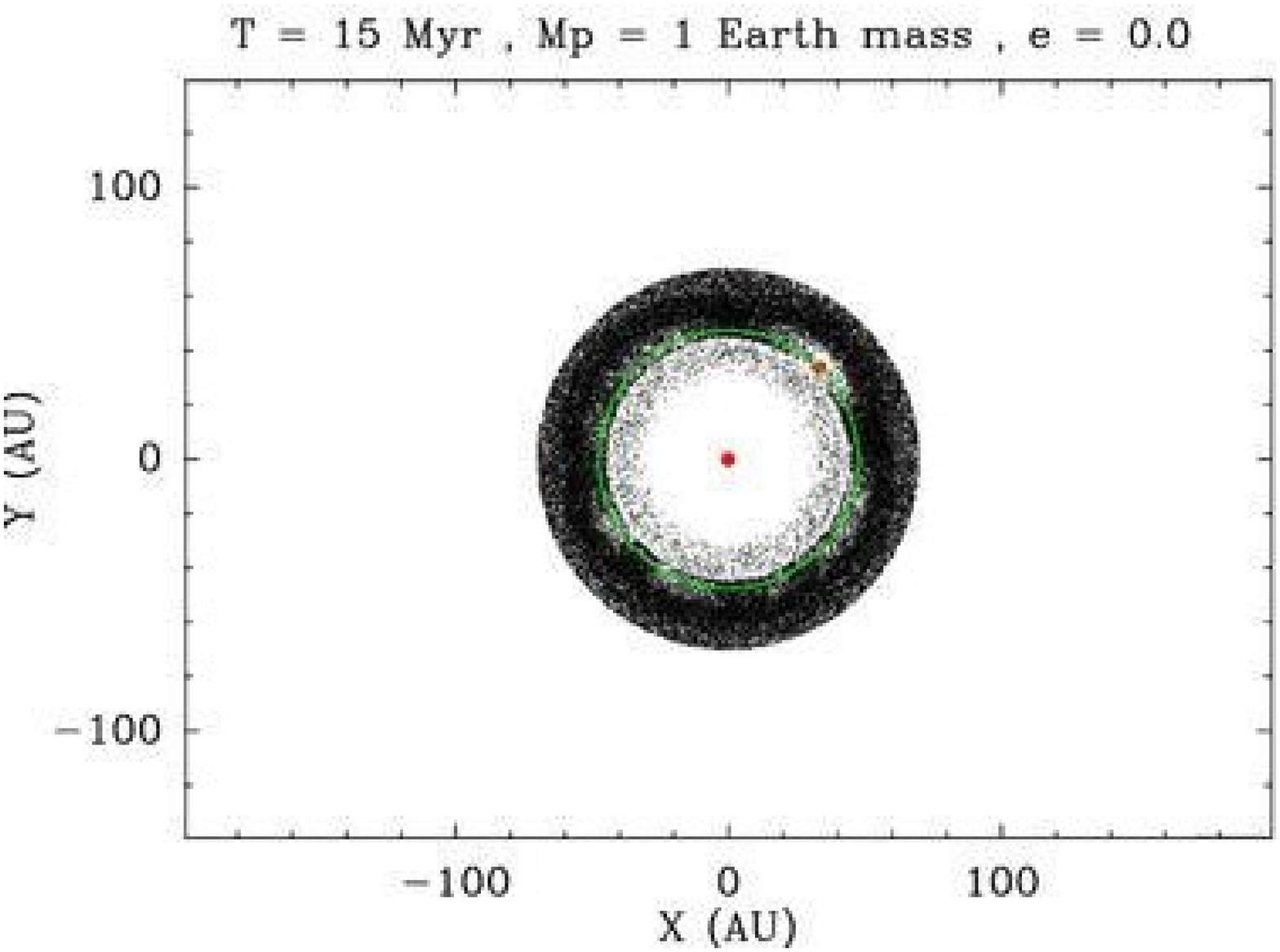}\hfil
\includegraphics[angle=-90,width=0.33\textwidth]{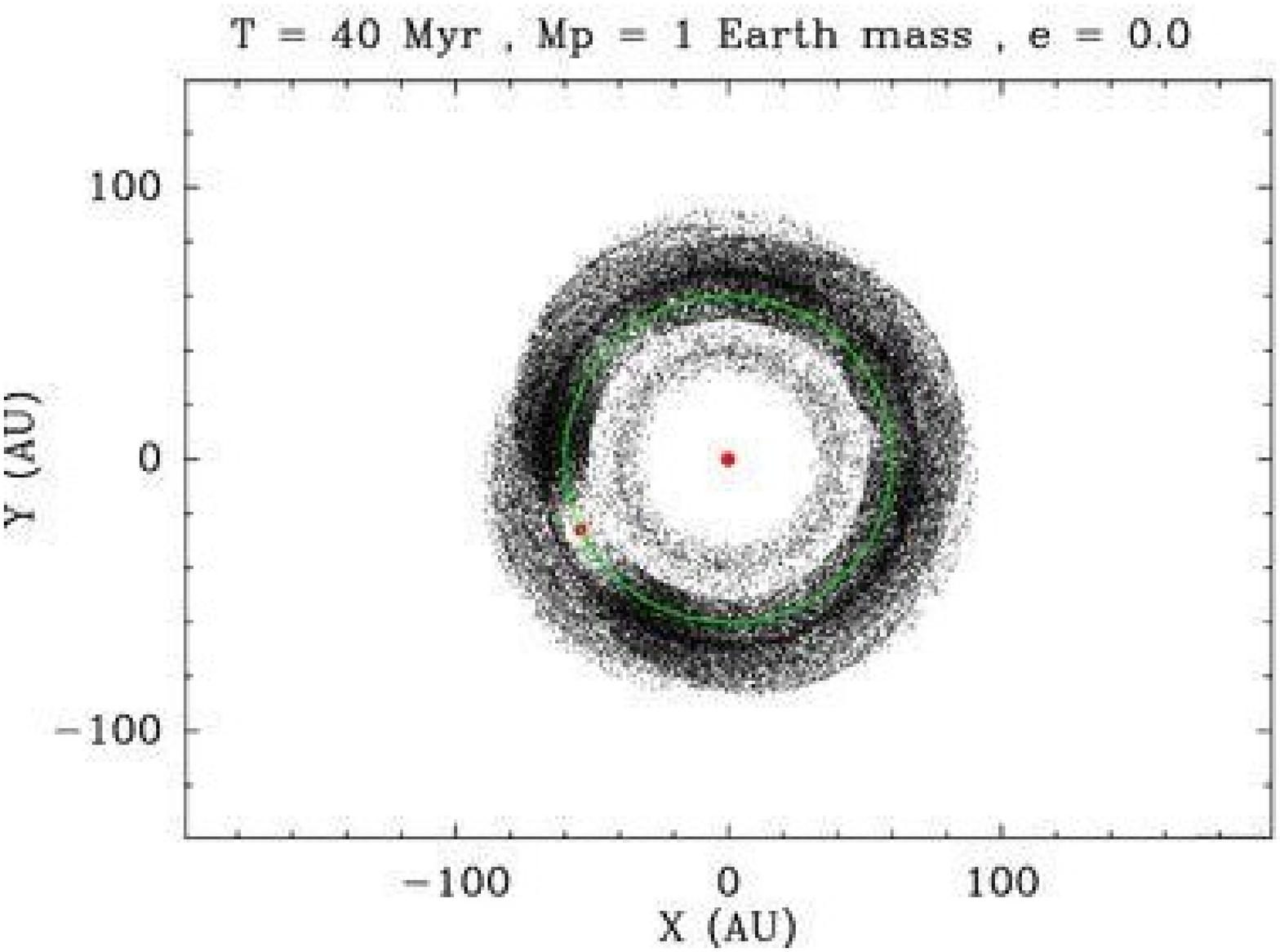}} \\
\makebox[\textwidth]{
\includegraphics[angle=-90,width=0.33\textwidth]{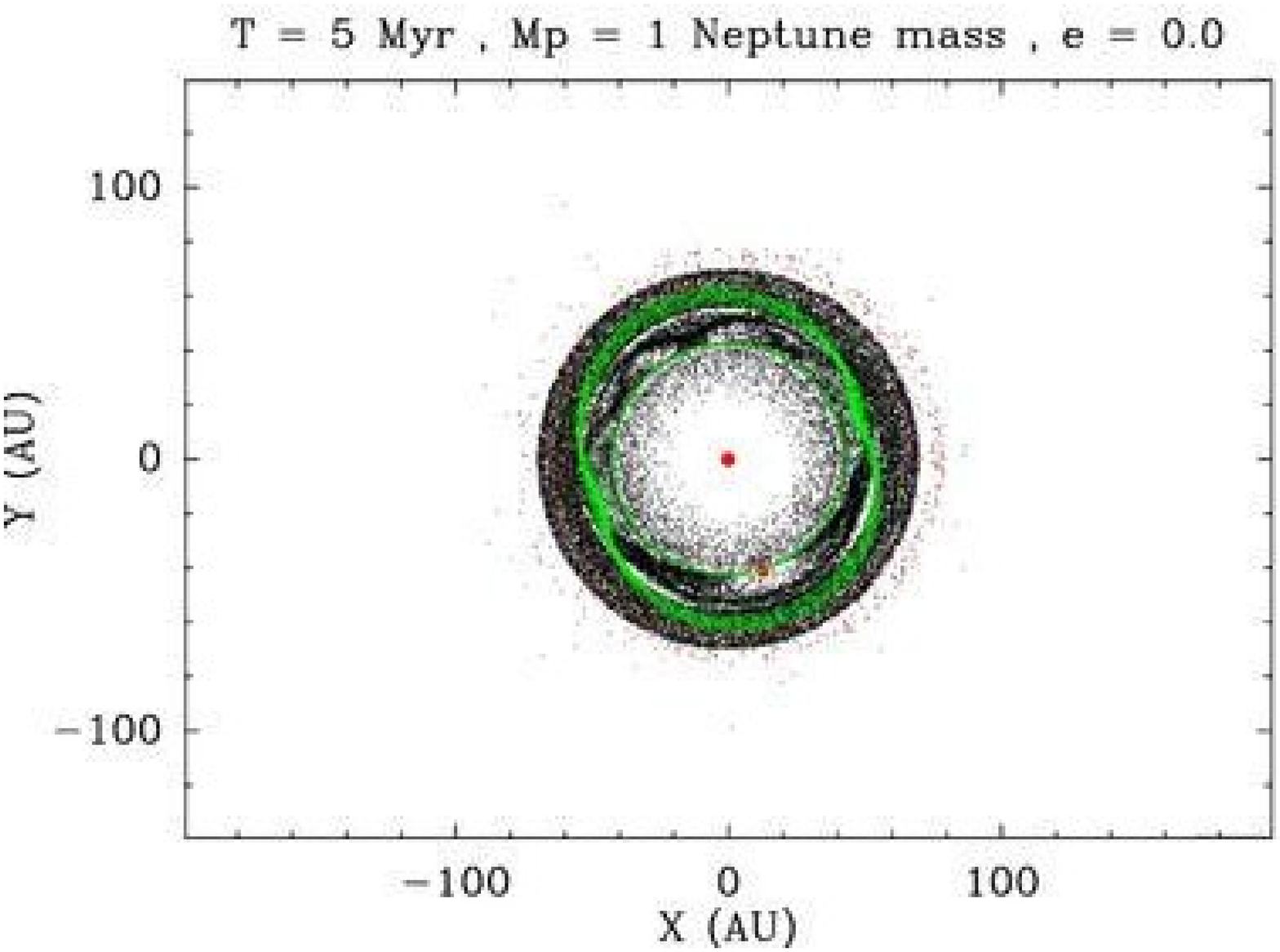}\hfil
\includegraphics[angle=-90,width=0.33\textwidth]{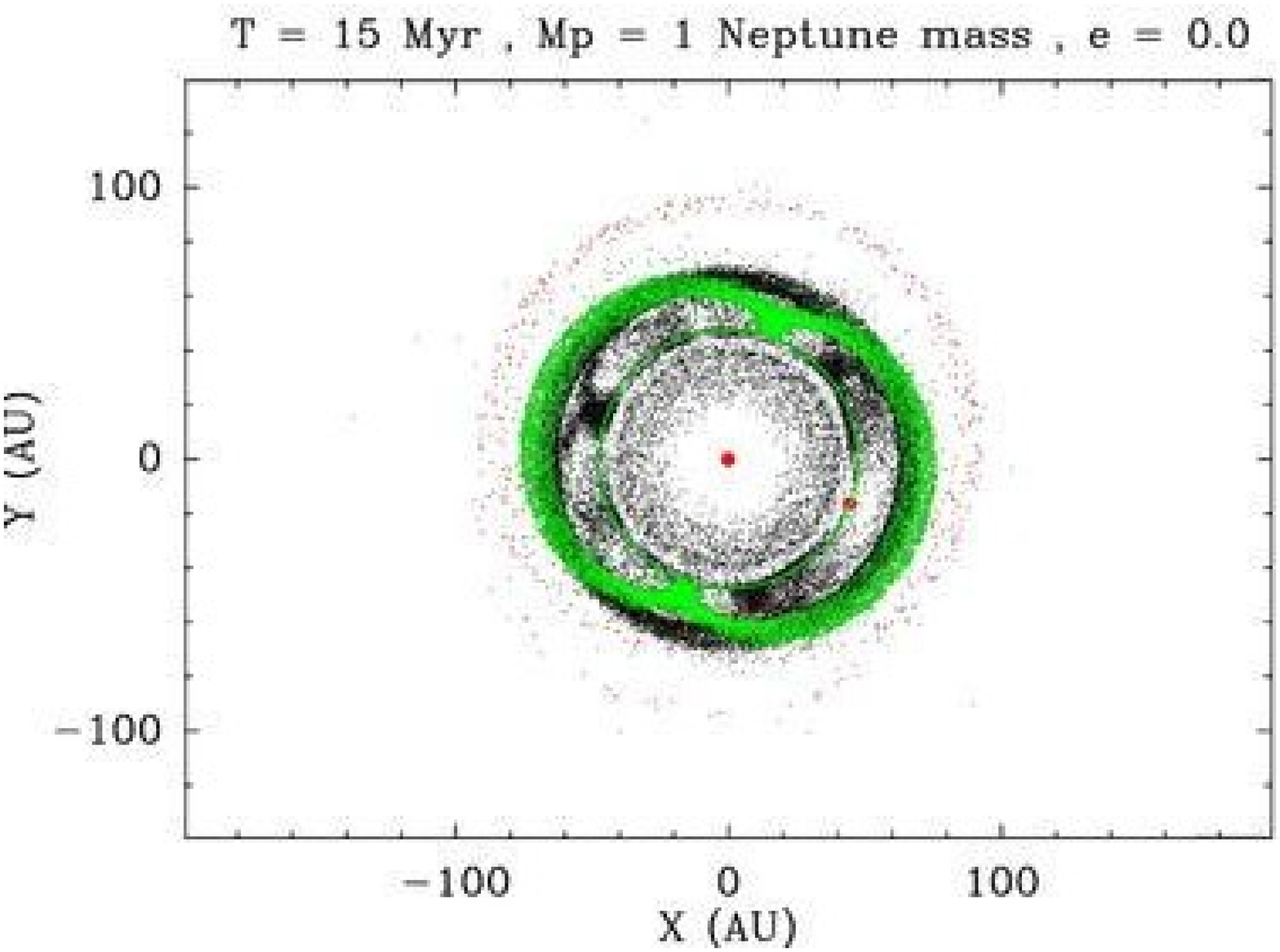}\hfil
\includegraphics[angle=-90,width=0.33\textwidth]{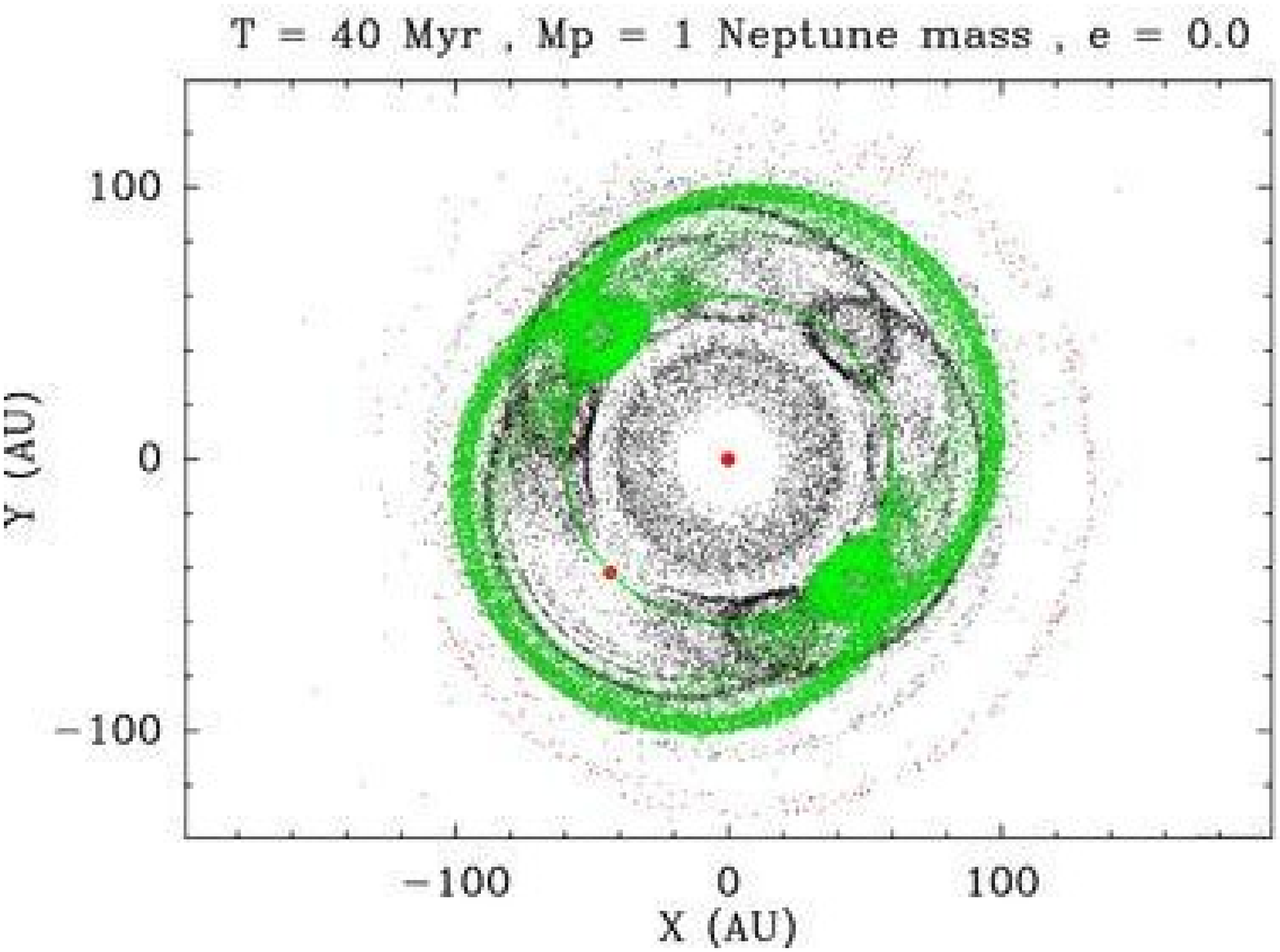}} \\
\makebox[\textwidth]{
\includegraphics[angle=-90,width=0.33\textwidth]{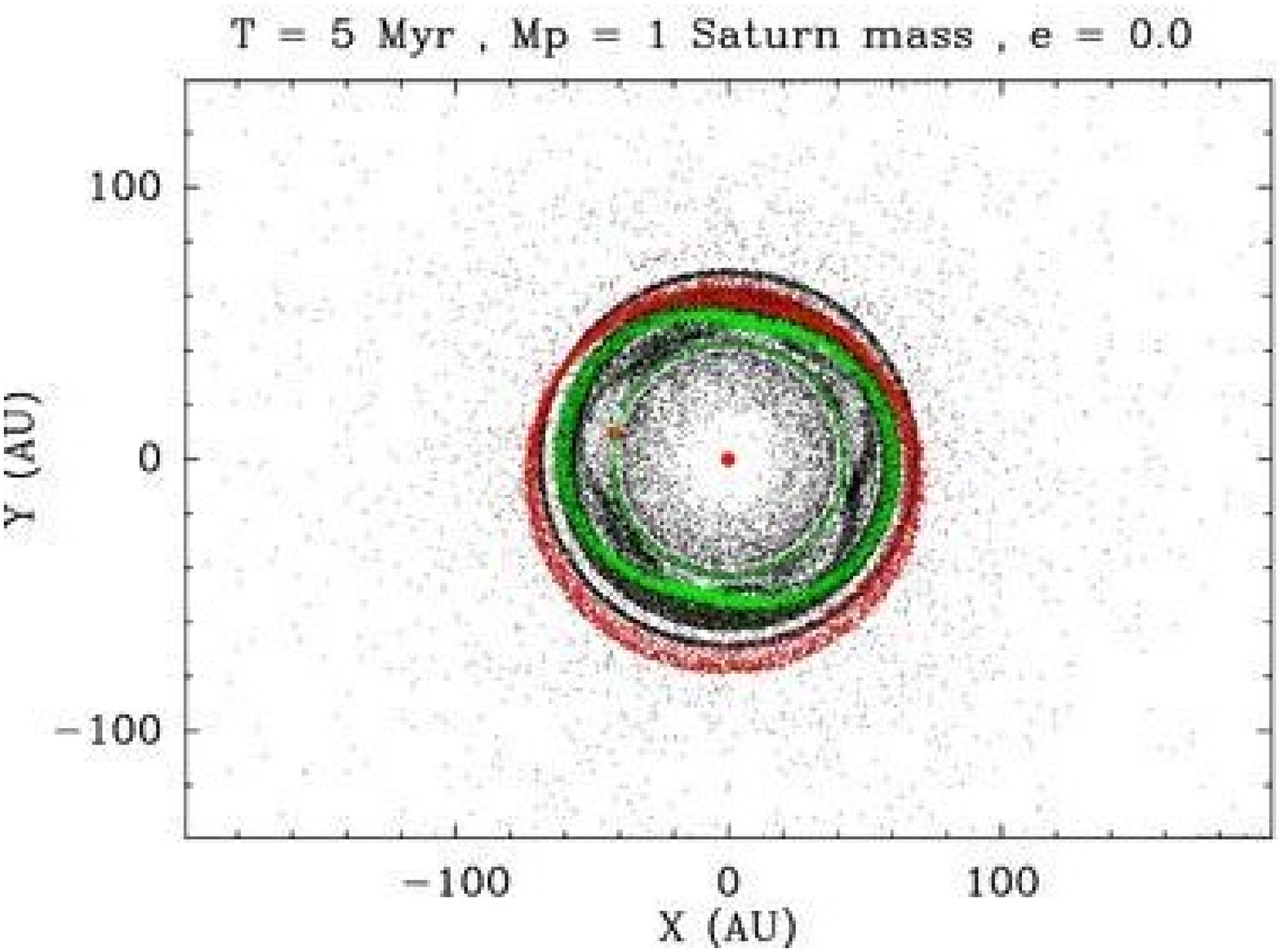} \hfil
\includegraphics[angle=-90,width=0.33\textwidth]{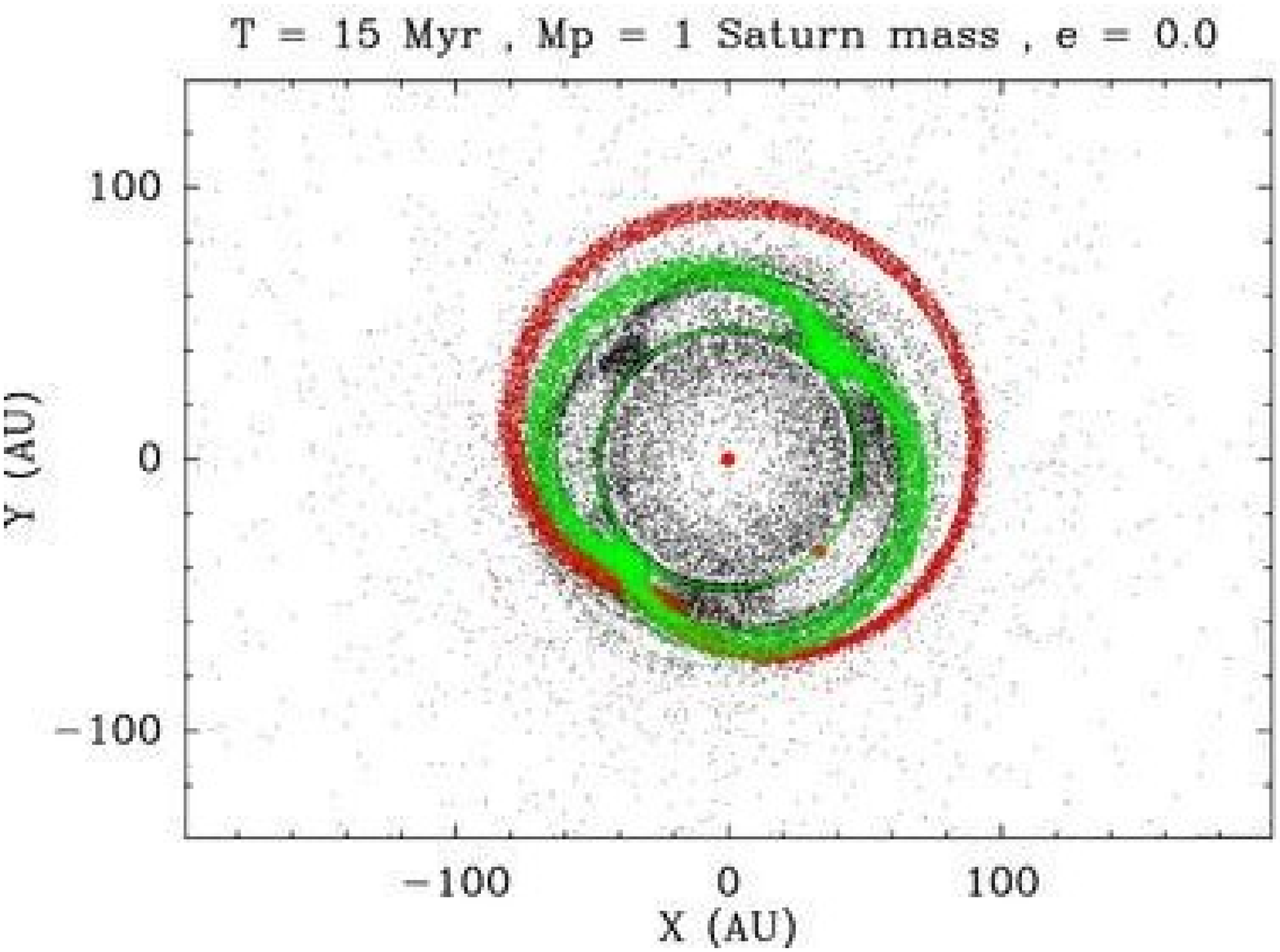} \hfil
\includegraphics[angle=-90,width=0.33\textwidth]{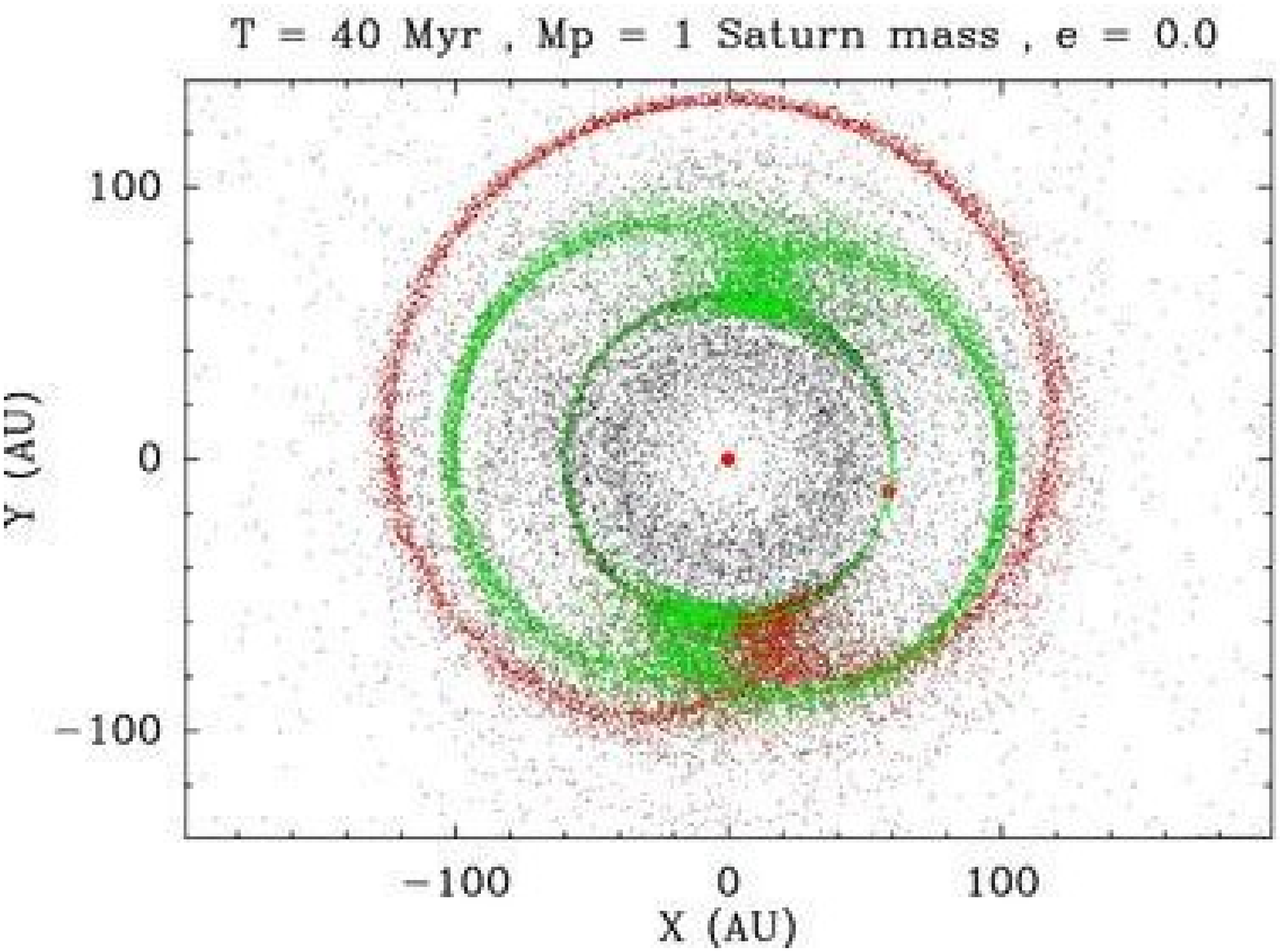}} \\
\makebox[\textwidth]{
\includegraphics[angle=-90,width=0.33\textwidth]{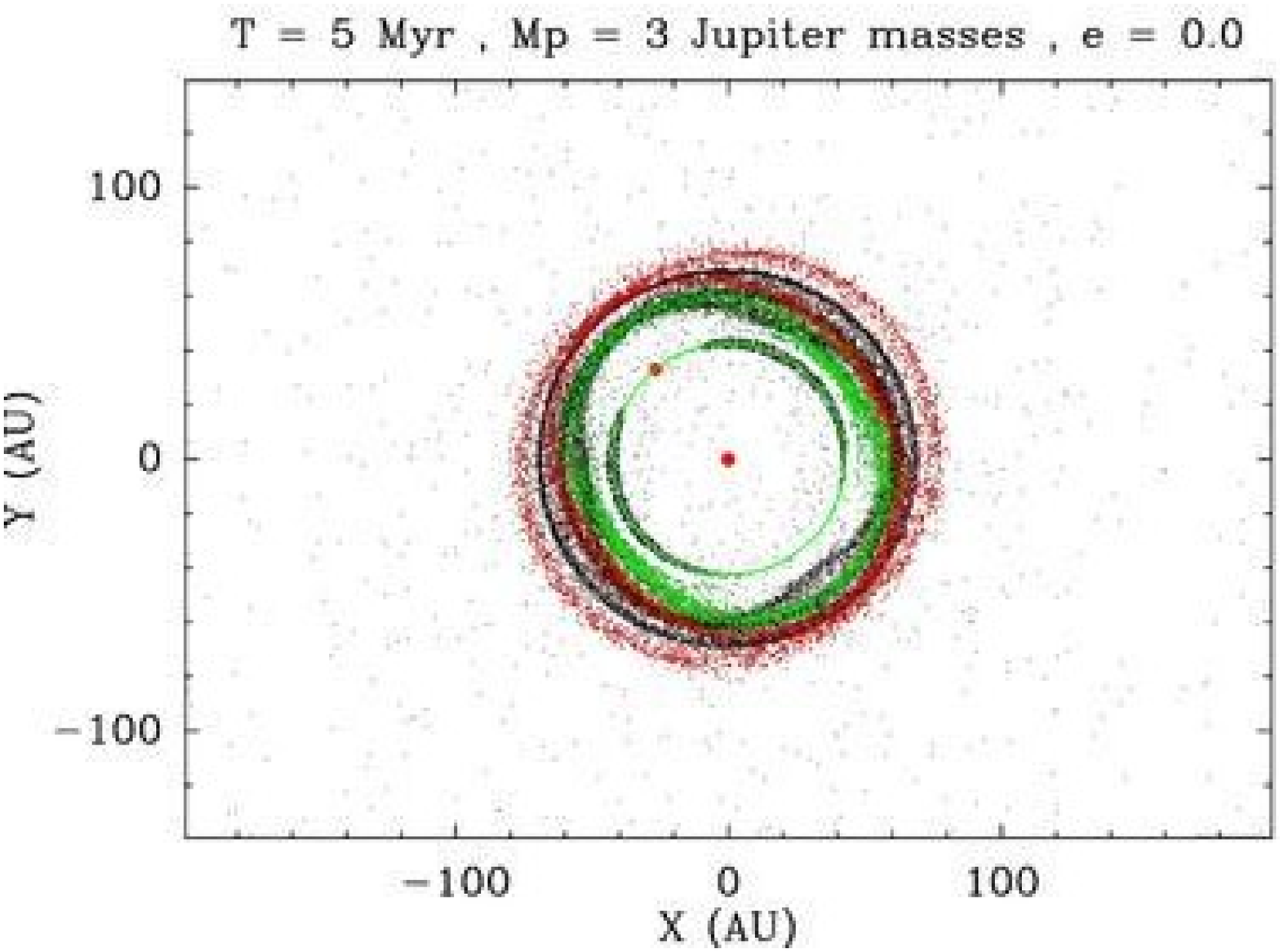}\hfil
\includegraphics[angle=-90,width=0.33\textwidth]{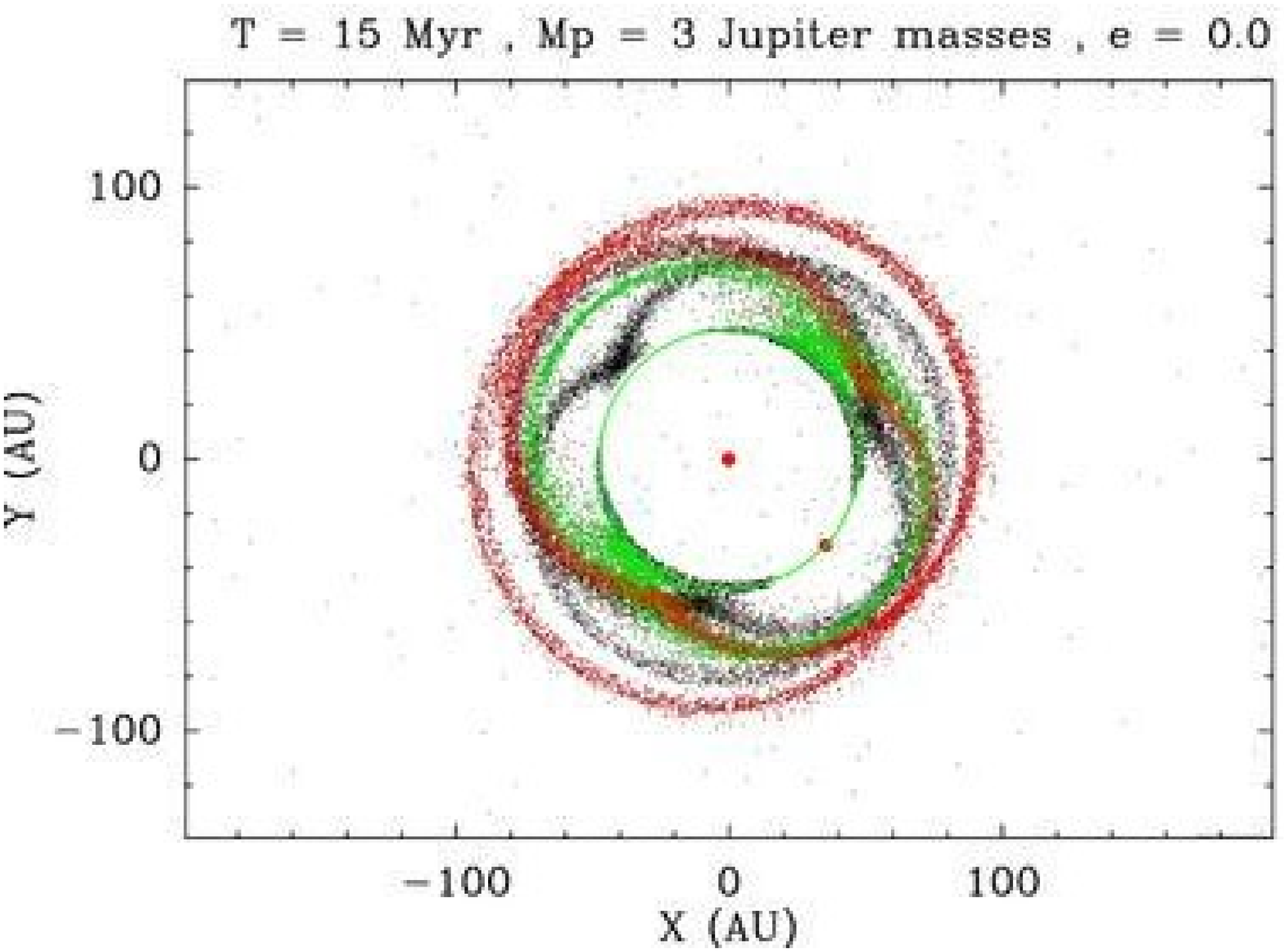} \hfil
\includegraphics[angle=-90,width=0.33\textwidth]{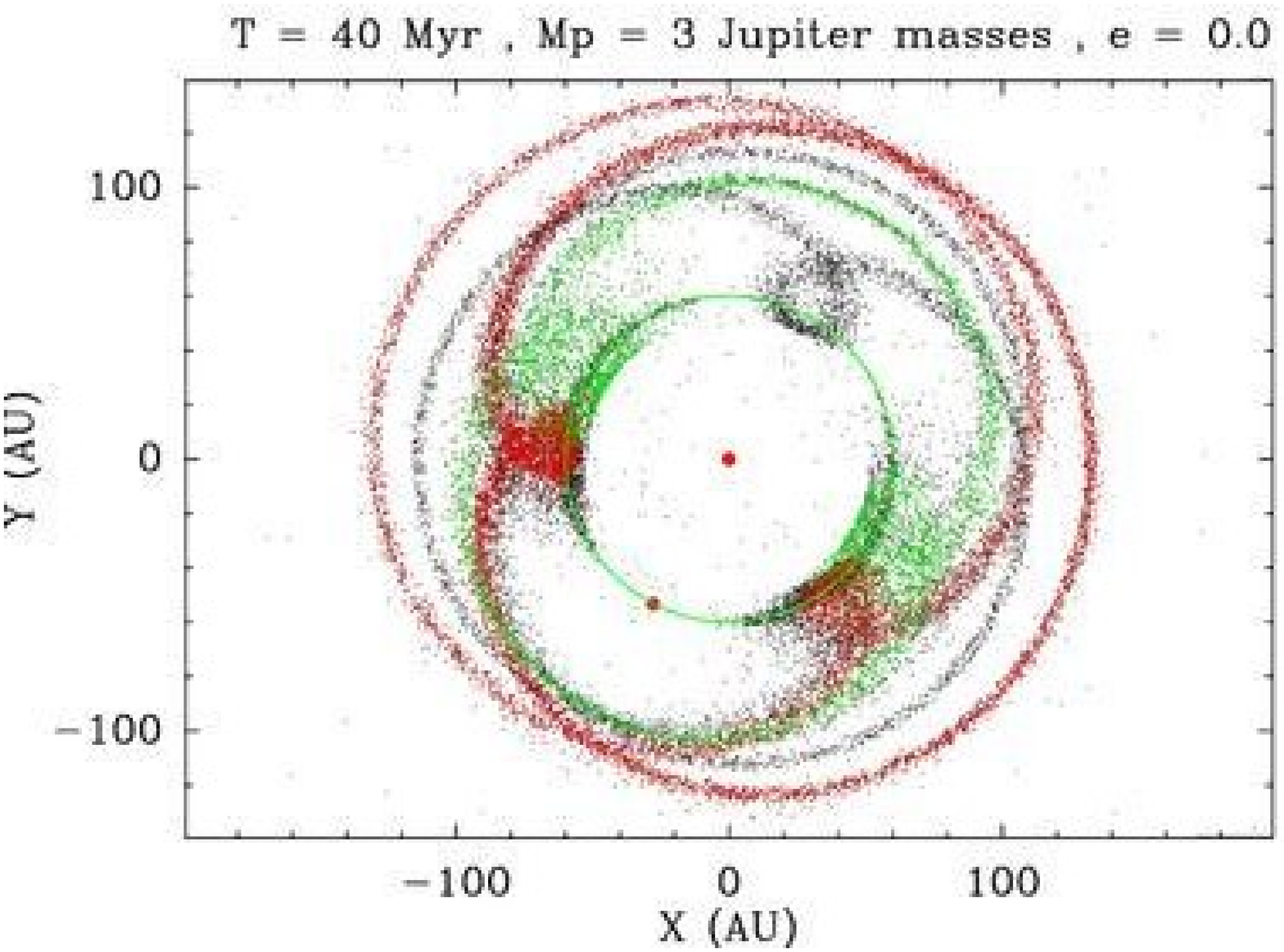}} \\
\caption{\label{figure_e0}Spatial distribution of planetesimals for a planet on
  a strictly circular orbit. The star and planet locations,
  projected onto the orbital plane of the planet, are represented by larger
  red points, and the planet orbit by a thin green line. The $4$
  rows correspond  respectively to Earth mass, Neptune mass, Saturn mass and $3$
  Jupiter mass planets, from top to bottom. The $3$
  columns show the disk after 5, 15  and 40 Myr. The
  initial planetesimal  disk consists of $50\,000$ planetesimals distributed between
  40 and 75 AU, with the surface density distribution proportional to 
  $r^{-1}$. \thanks{See the electronic edition of the
    Journal for a color version of this figure.}}
\end{figure*}

\begin{figure*}
\makebox[\textwidth]{
\includegraphics[angle=-90,width=0.33\textwidth]{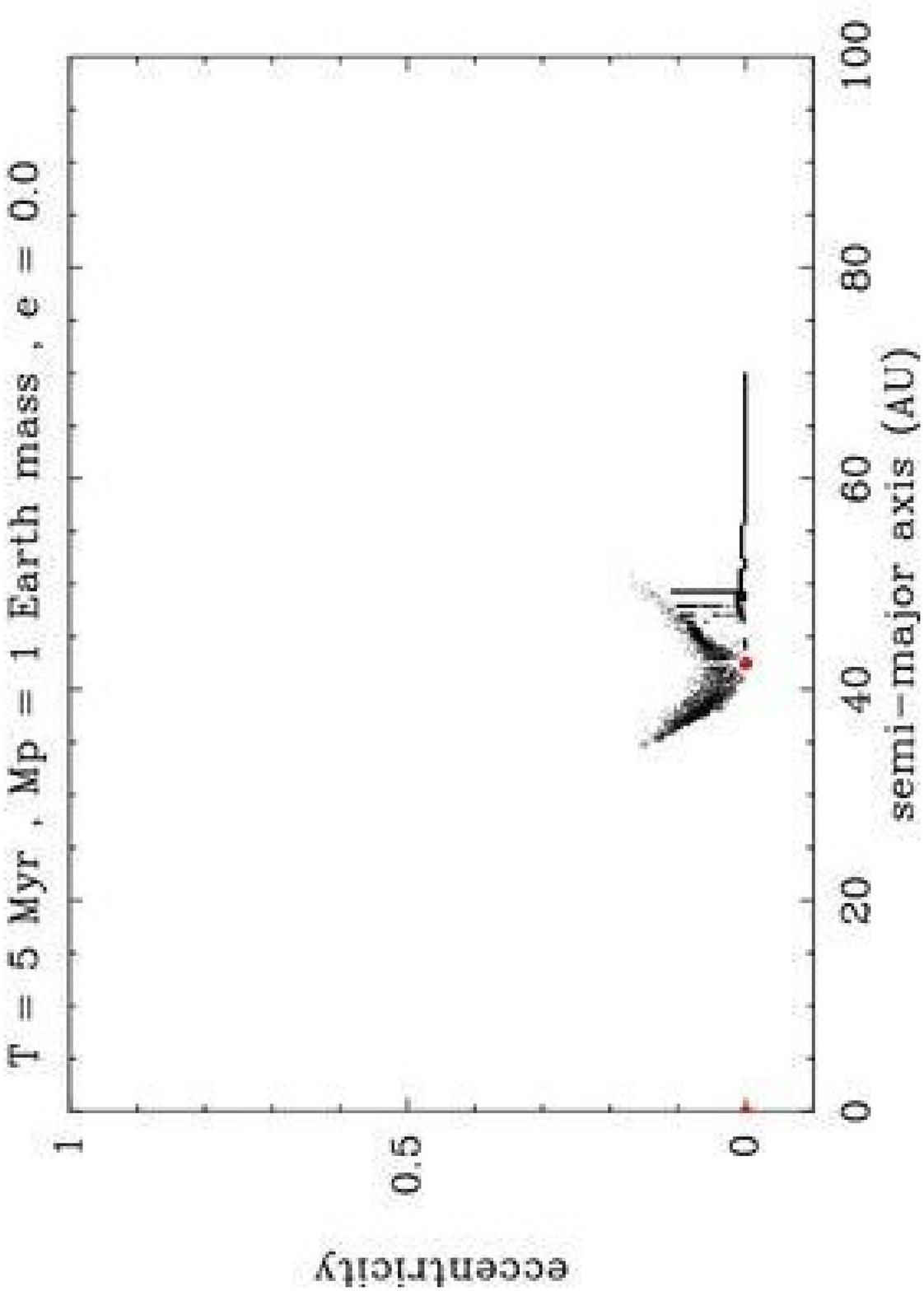} \hfil
\includegraphics[angle=-90,width=0.33\textwidth]{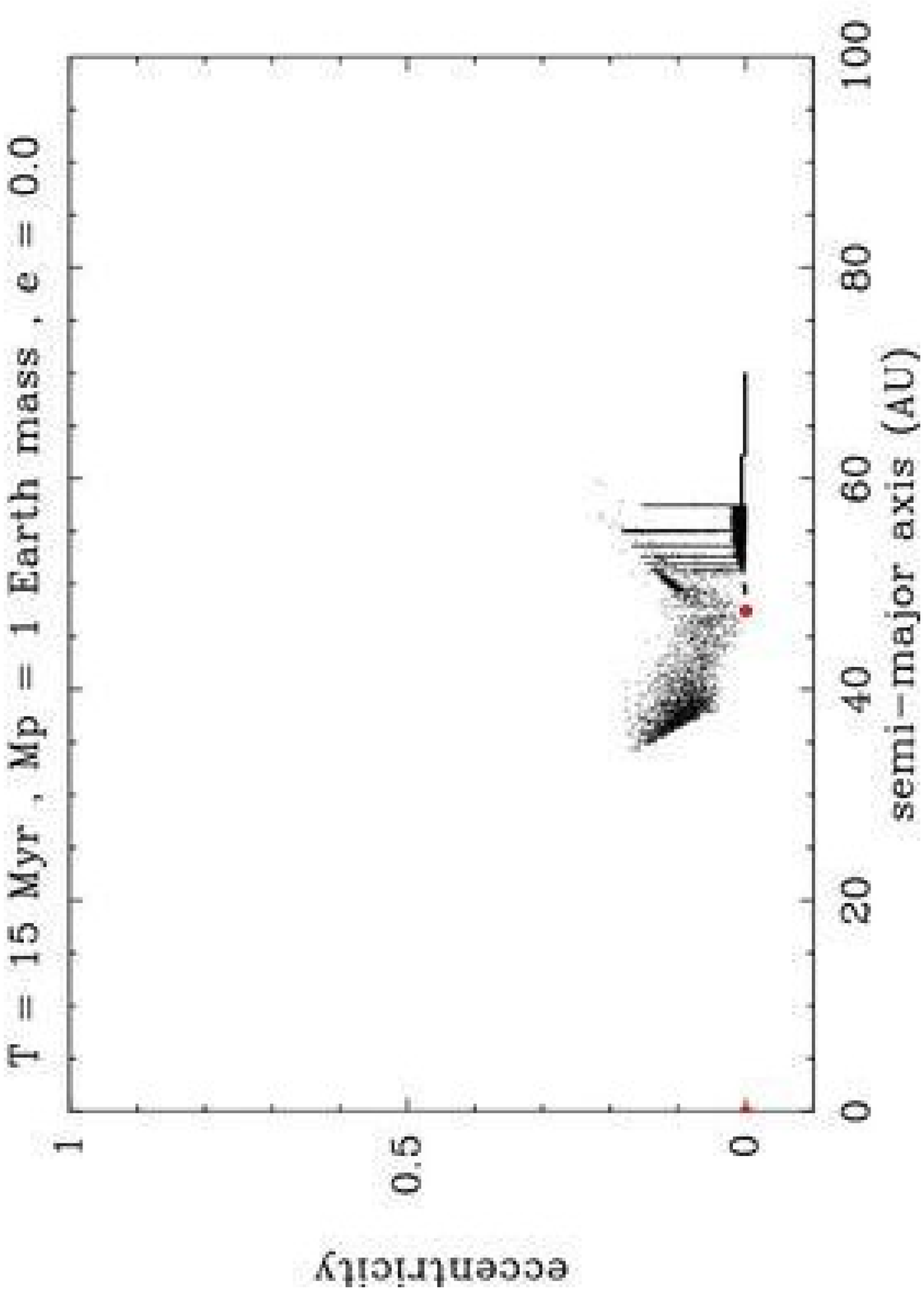} \hfil
\includegraphics[angle=-90,width=0.33\textwidth]{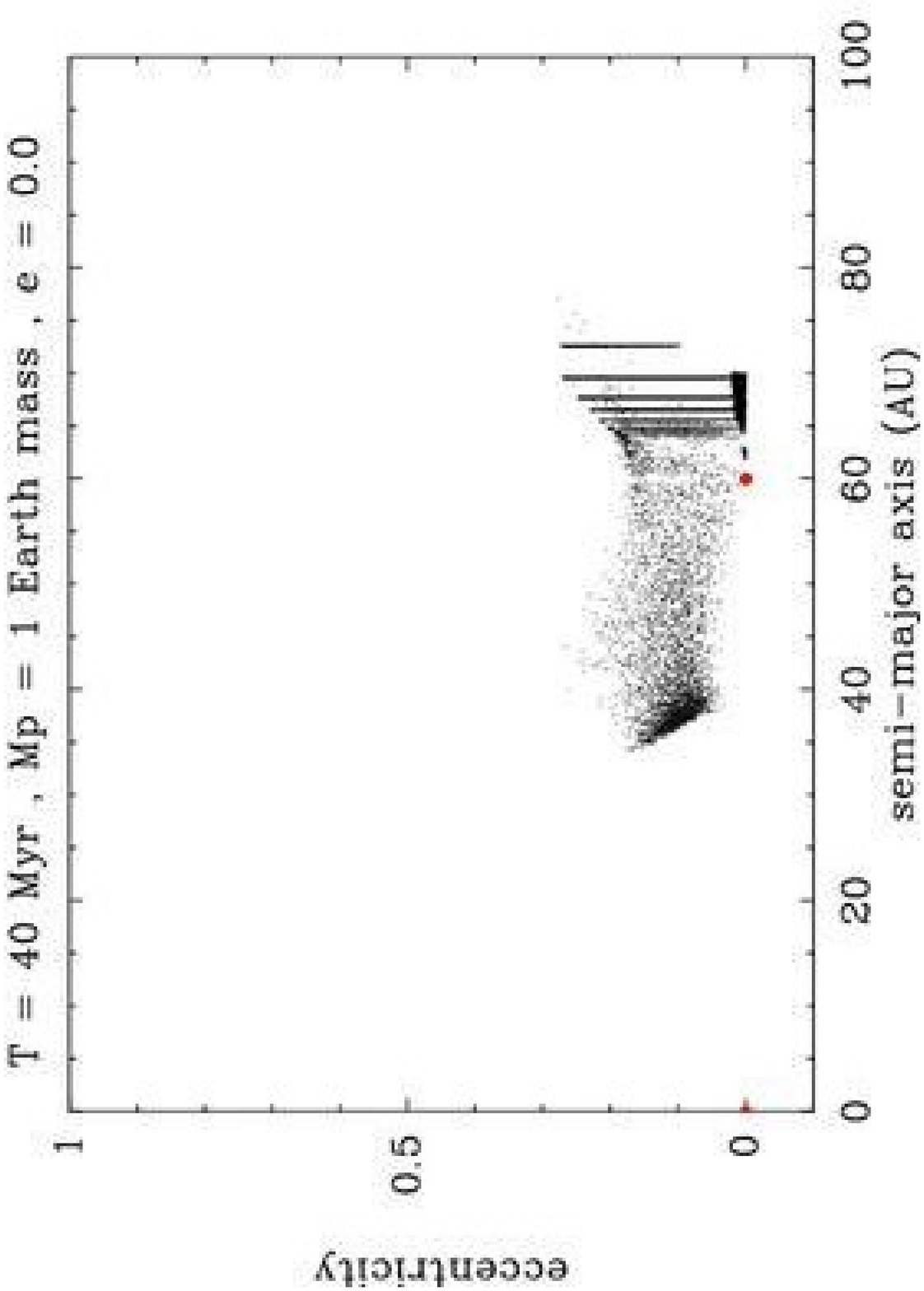}}\\
\makebox[\textwidth]{
\includegraphics[angle=-90,width=0.33\textwidth]{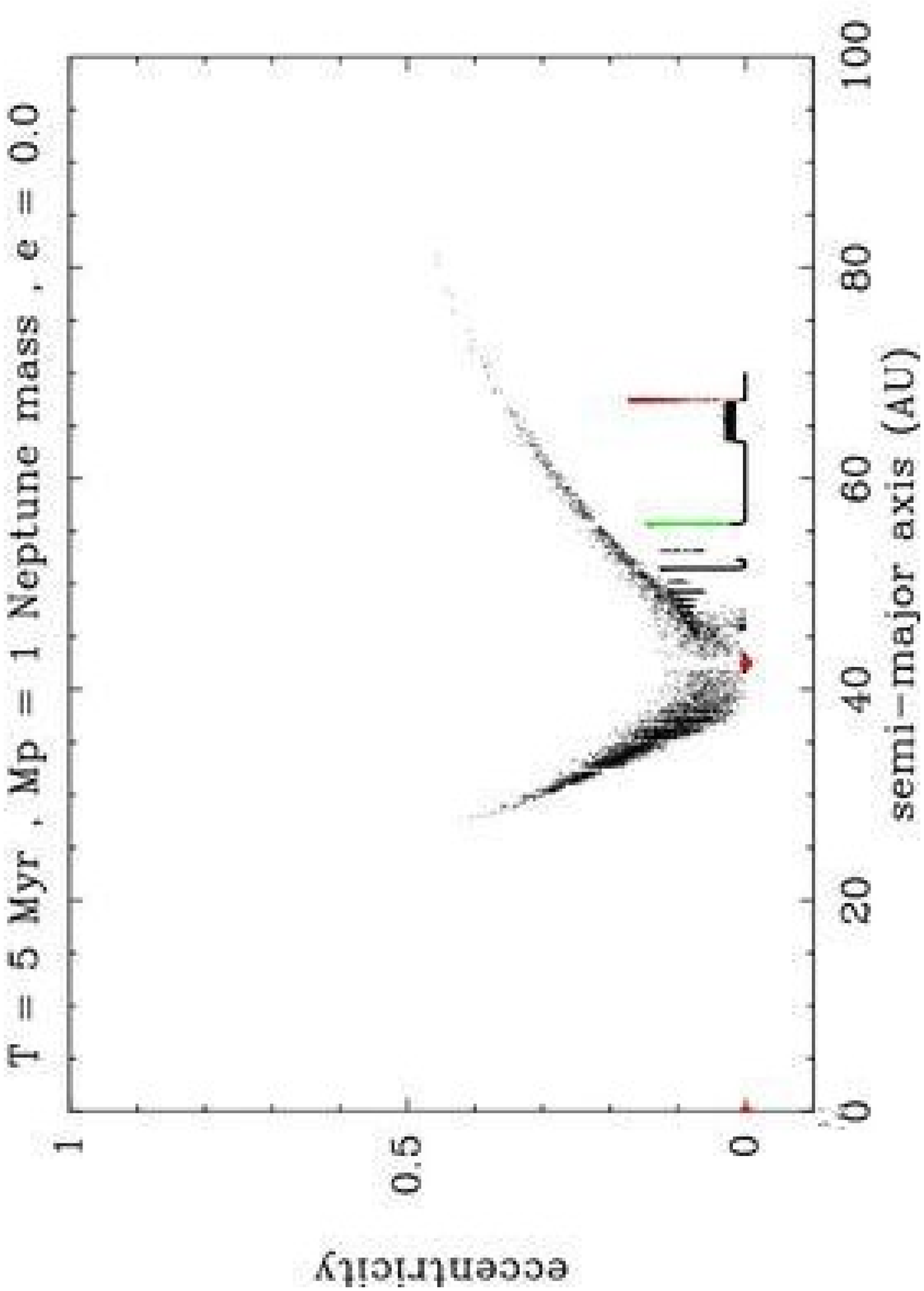} \hfil
\includegraphics[angle=-90,width=0.33\textwidth]{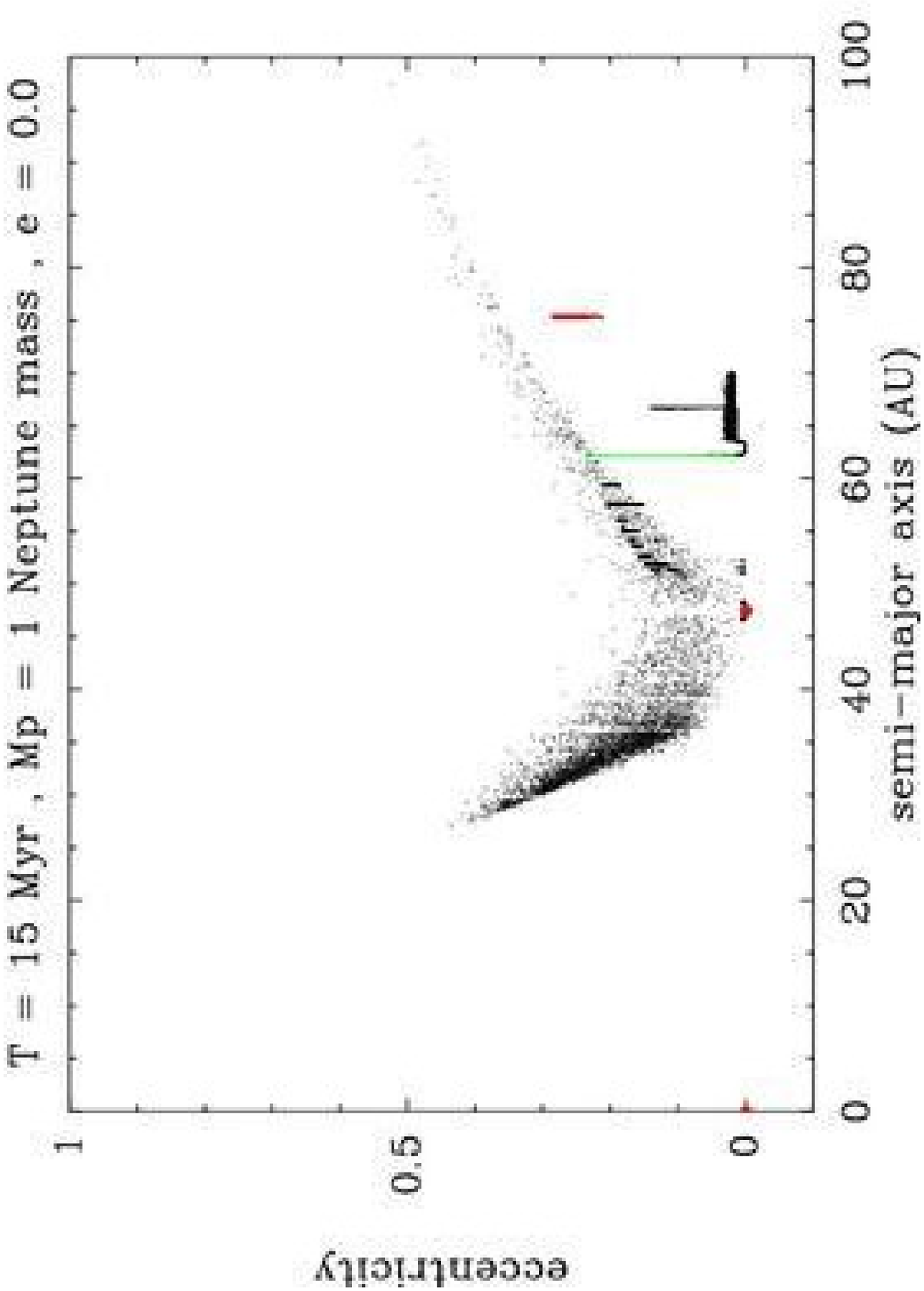} \hfil
\includegraphics[angle=-90,width=0.33\textwidth]{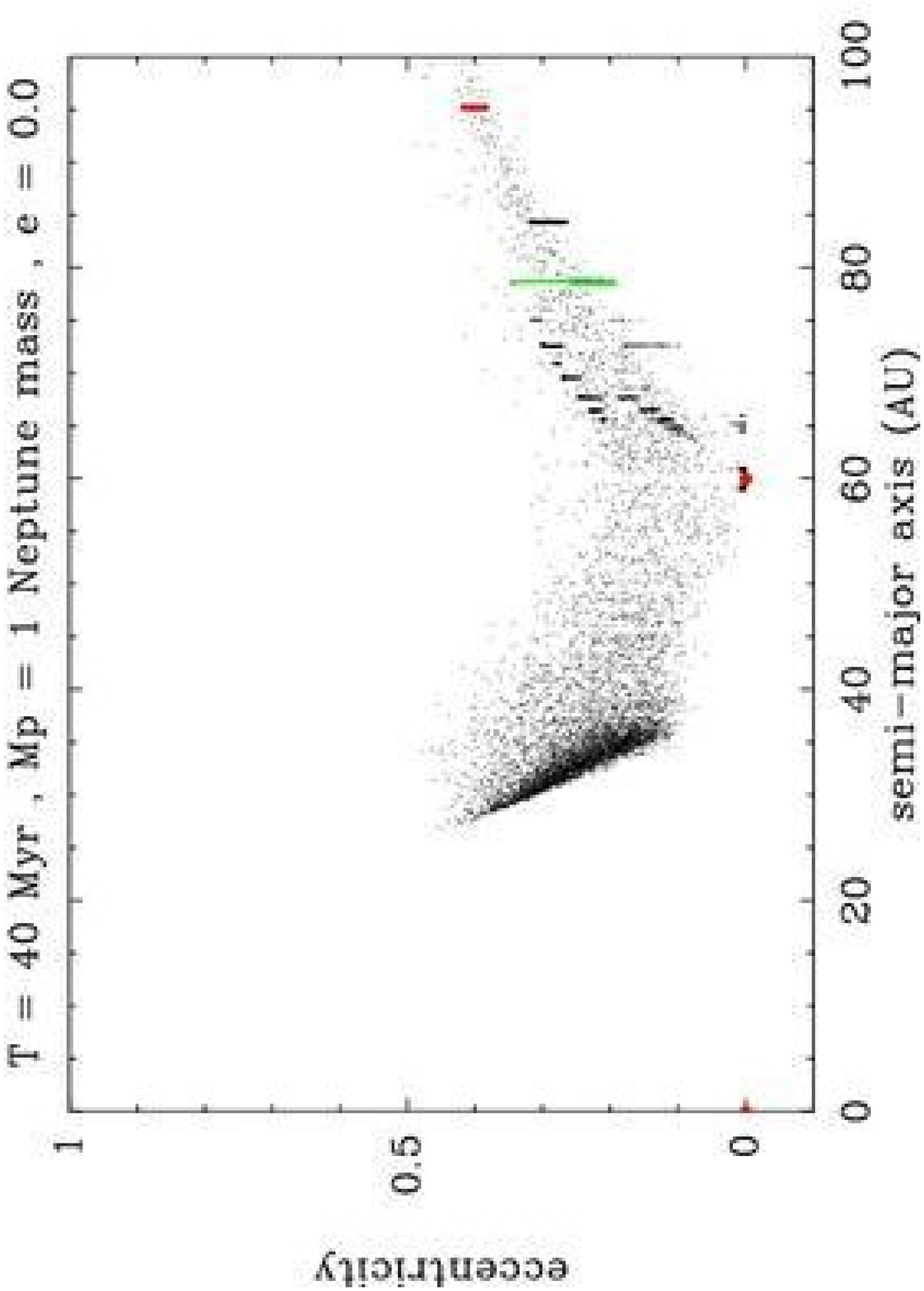}} \\
\makebox[\textwidth]{
\includegraphics[angle=-90,width=0.33\textwidth]{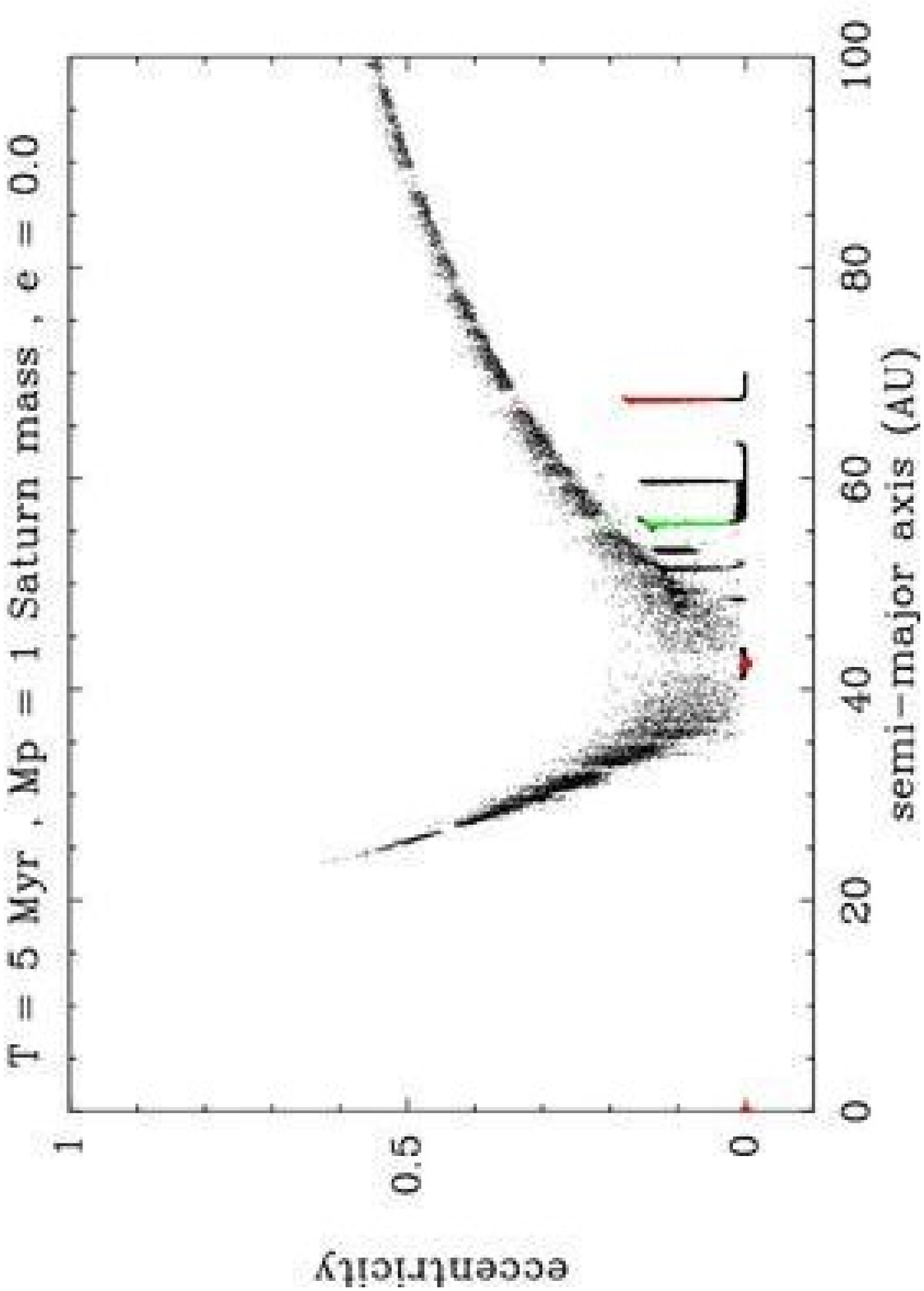} \hfil
\includegraphics[angle=-90,width=0.33\textwidth]{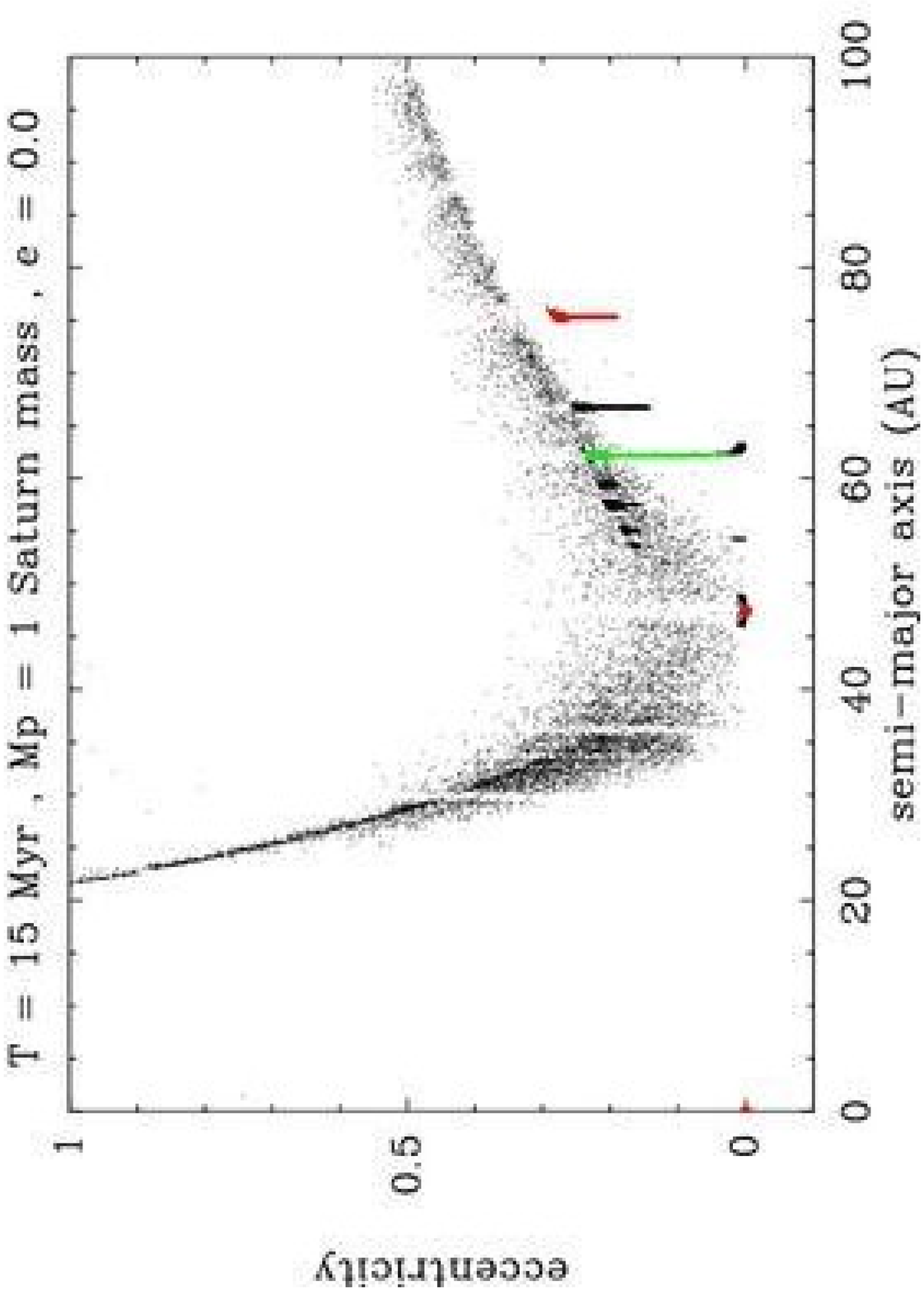} \hfil
\includegraphics[angle=-90,width=0.33\textwidth]{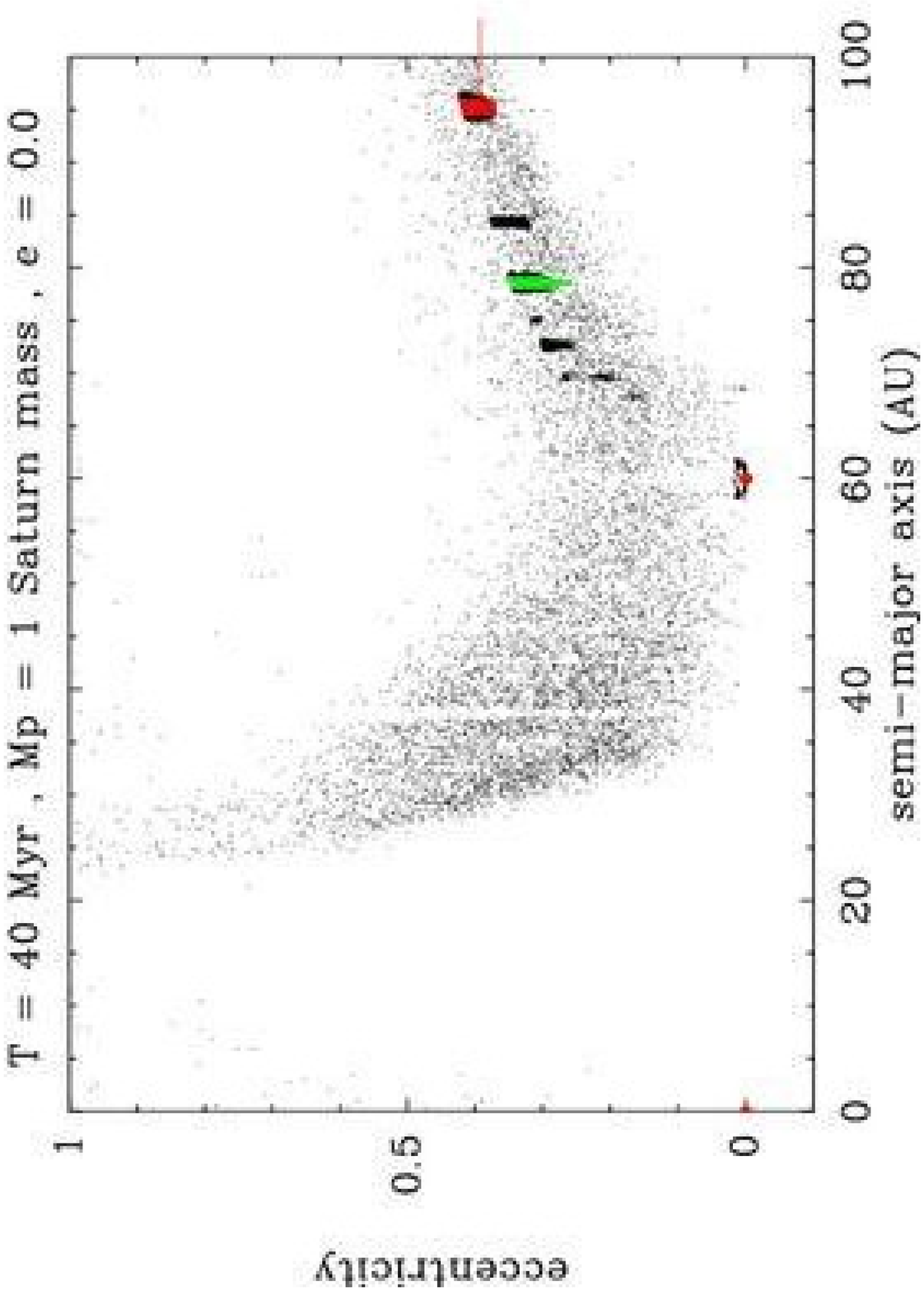}} \\
\makebox[\textwidth]{
\includegraphics[angle=-90,width=0.33\textwidth]{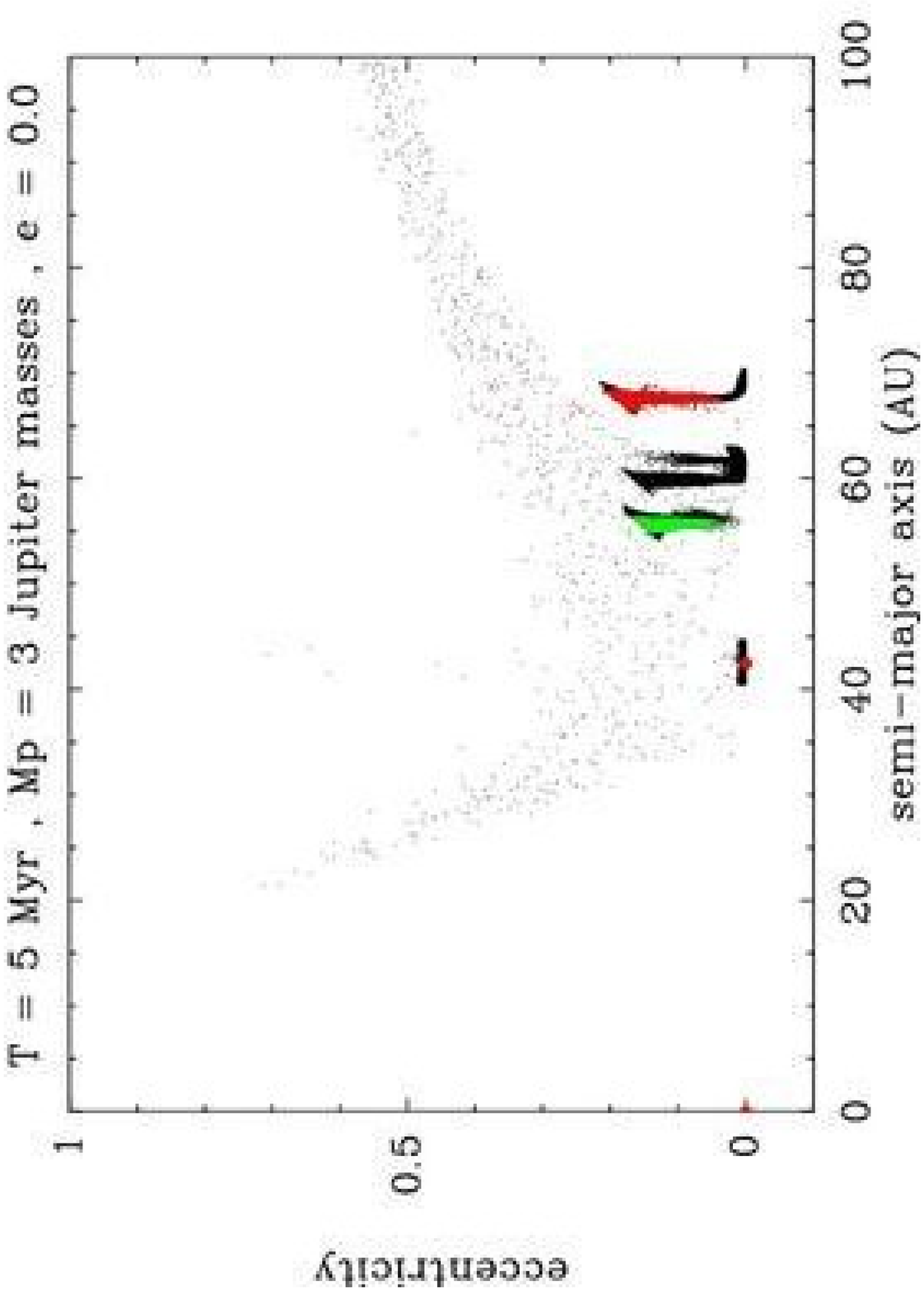} \hfil
\includegraphics[angle=-90,width=0.33\textwidth]{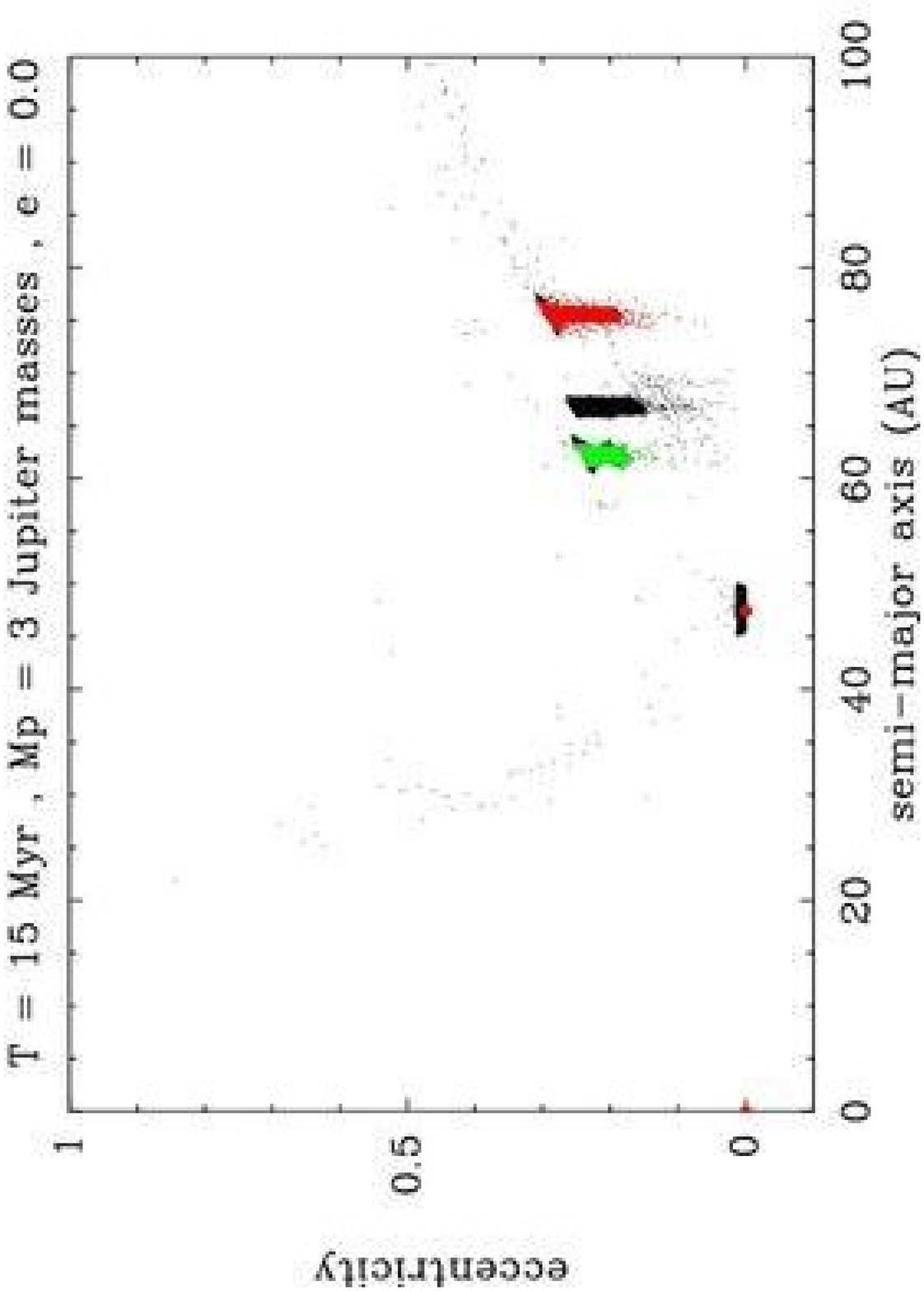} \hfil
\includegraphics[angle=-90,width=0.33\textwidth]{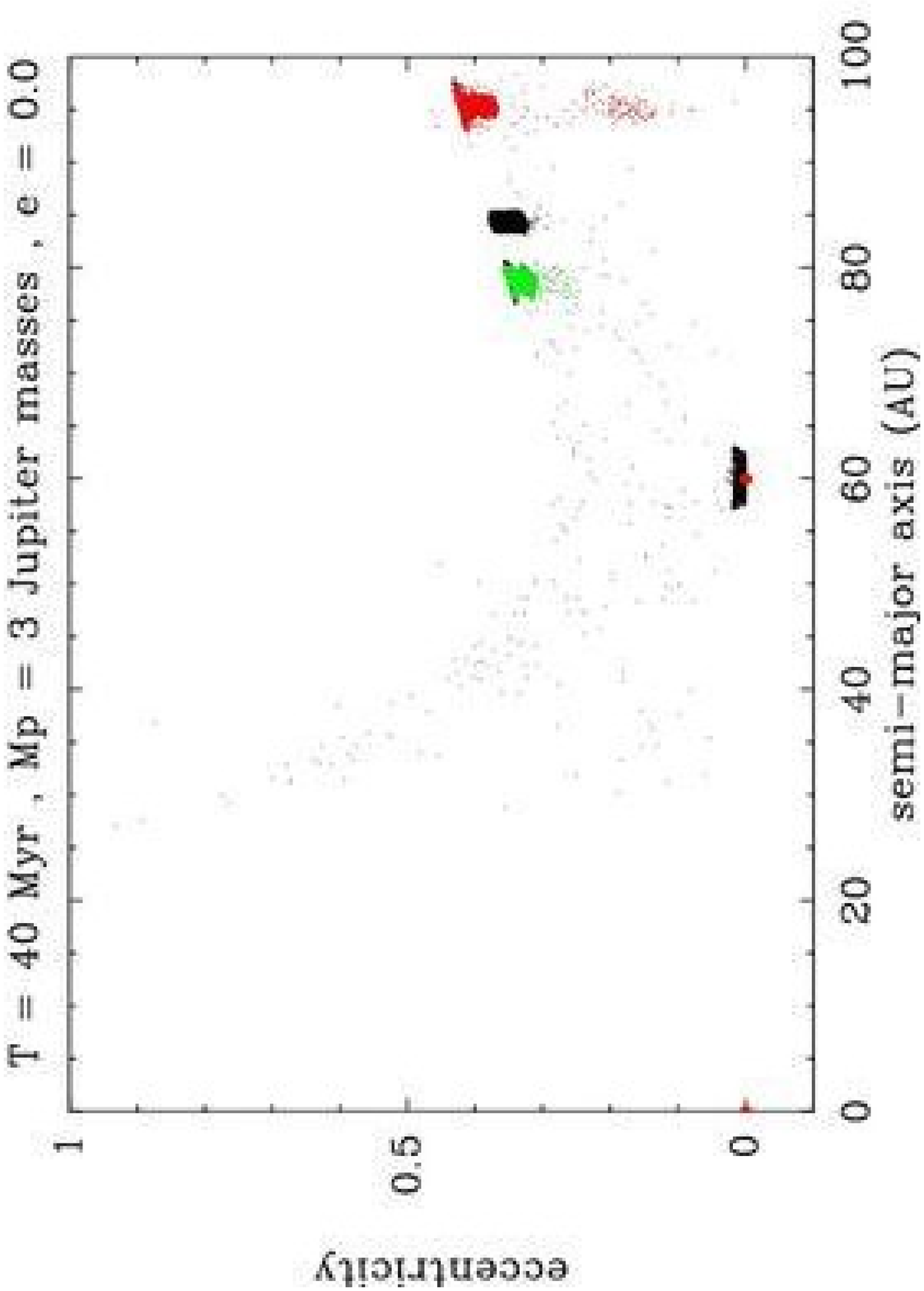}} \\
\caption{ \label{figure_e0_ae} Same simulations as in Fig.
  \ref{figure_e0}, with the planetesimals represented in a
  (semi-major axis, eccentricity) plane. The plotting conventions are
  the same as in  Fig. \ref{figure_e0}.\thanks{See the electronic edition of the
    Journal for a color version of this figure.}}
\end{figure*}

\begin{figure*}
\makebox[\textwidth]{
\includegraphics[angle=-90,width=0.33\textwidth]{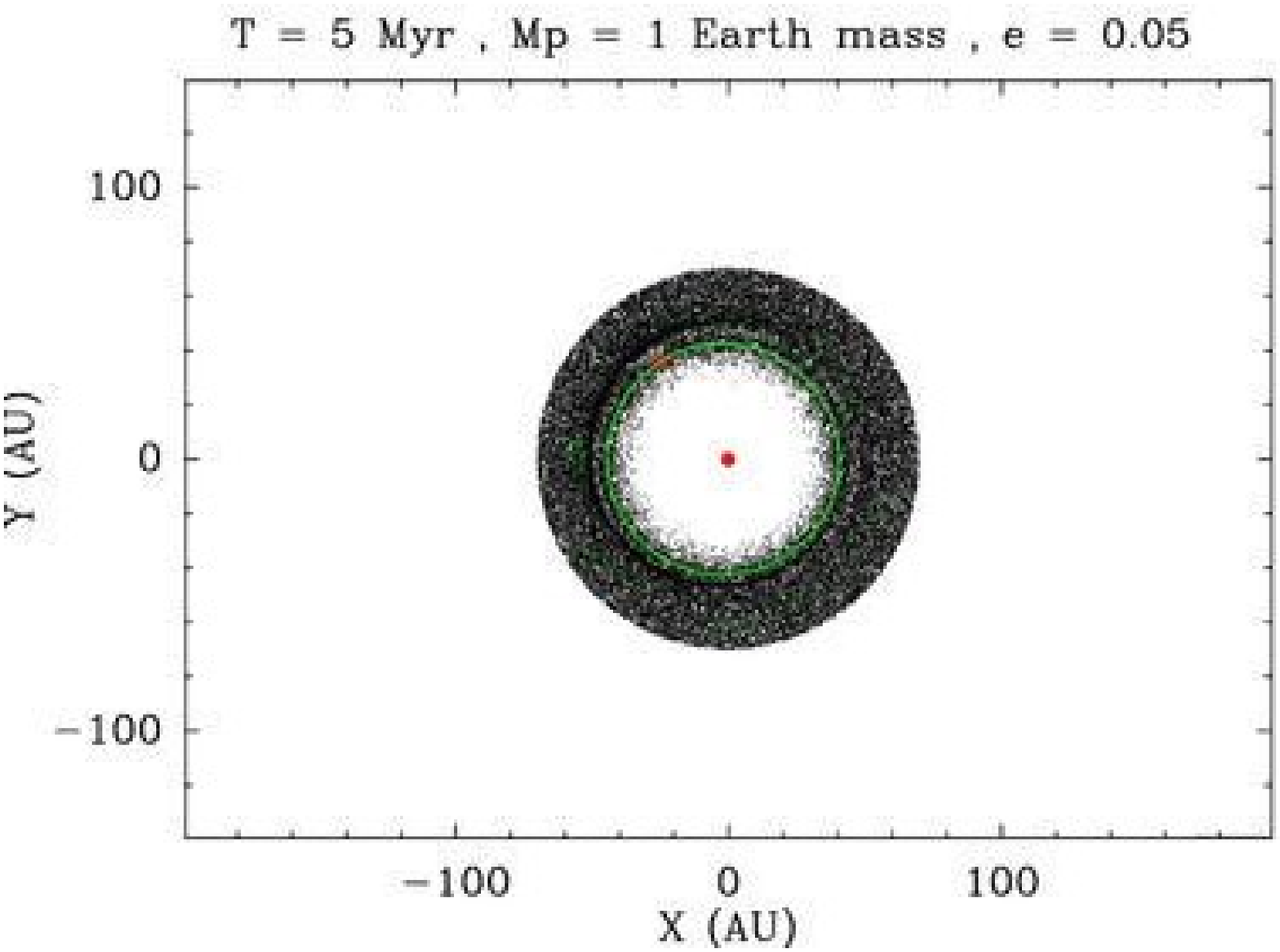} \hfil
\includegraphics[angle=-90,width=0.33\textwidth]{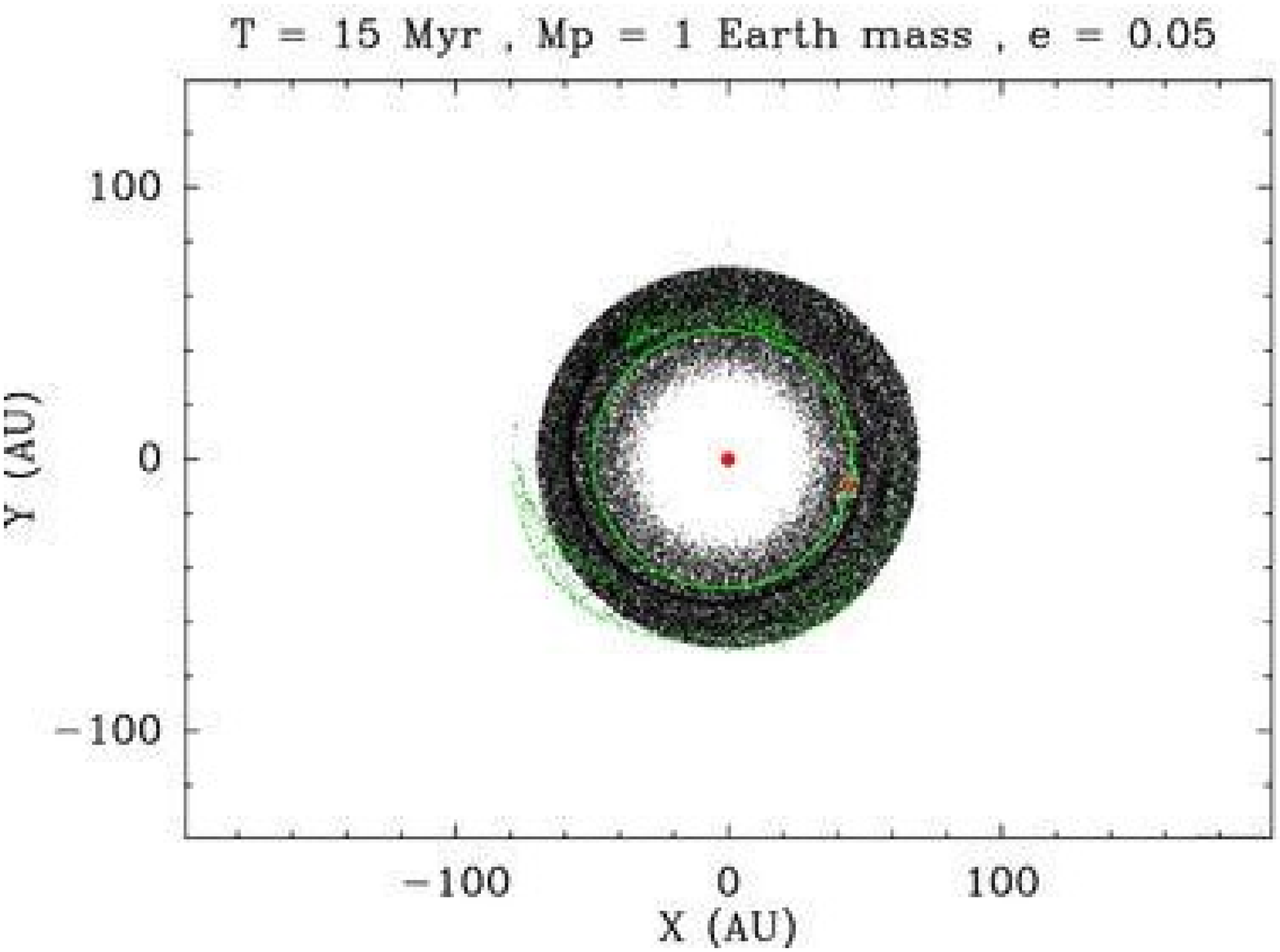} \hfil
\includegraphics[angle=-90,width=0.33\textwidth]{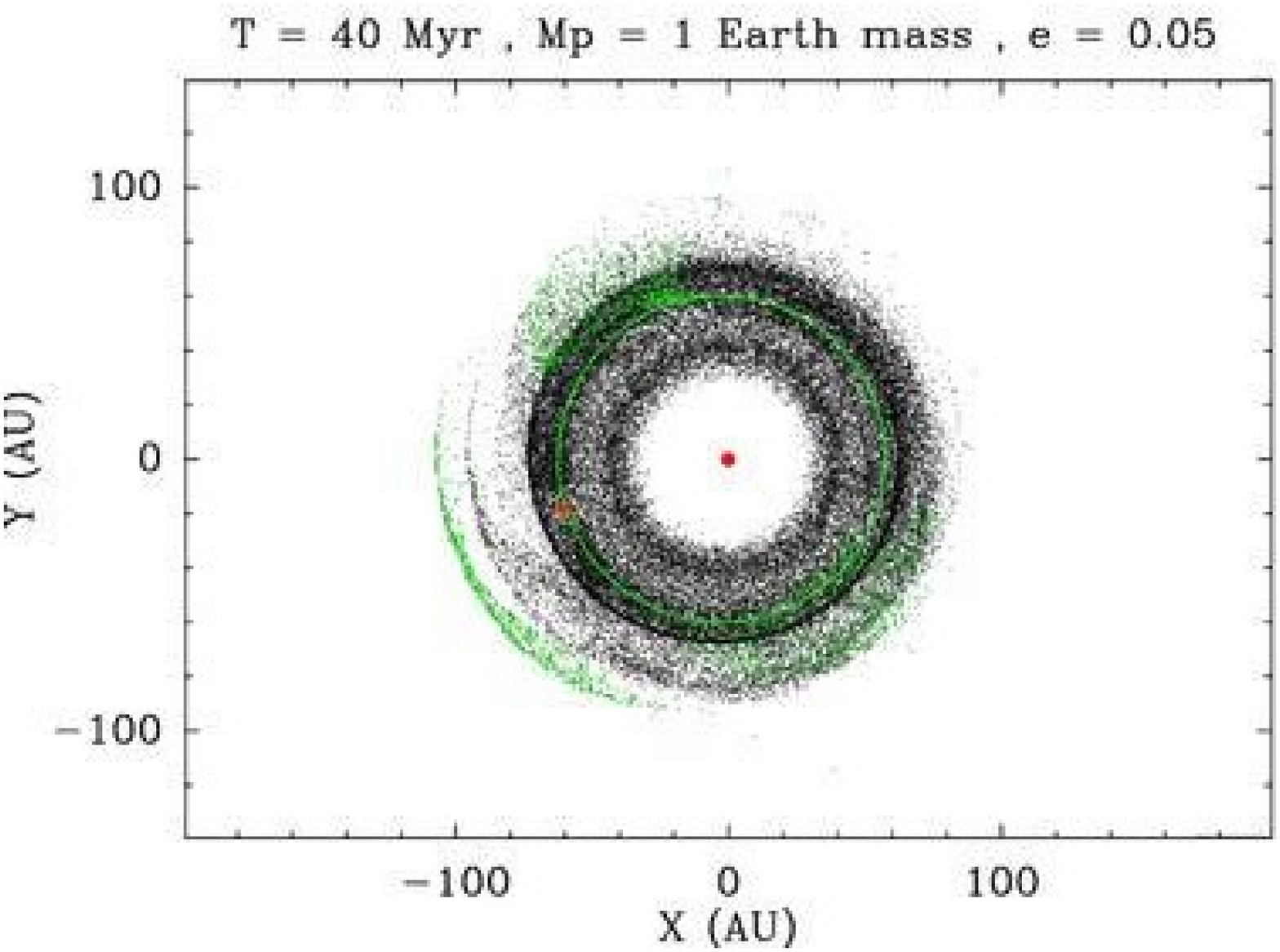}}\\
\makebox[\textwidth]{
\includegraphics[angle=-90,width=0.33\textwidth]{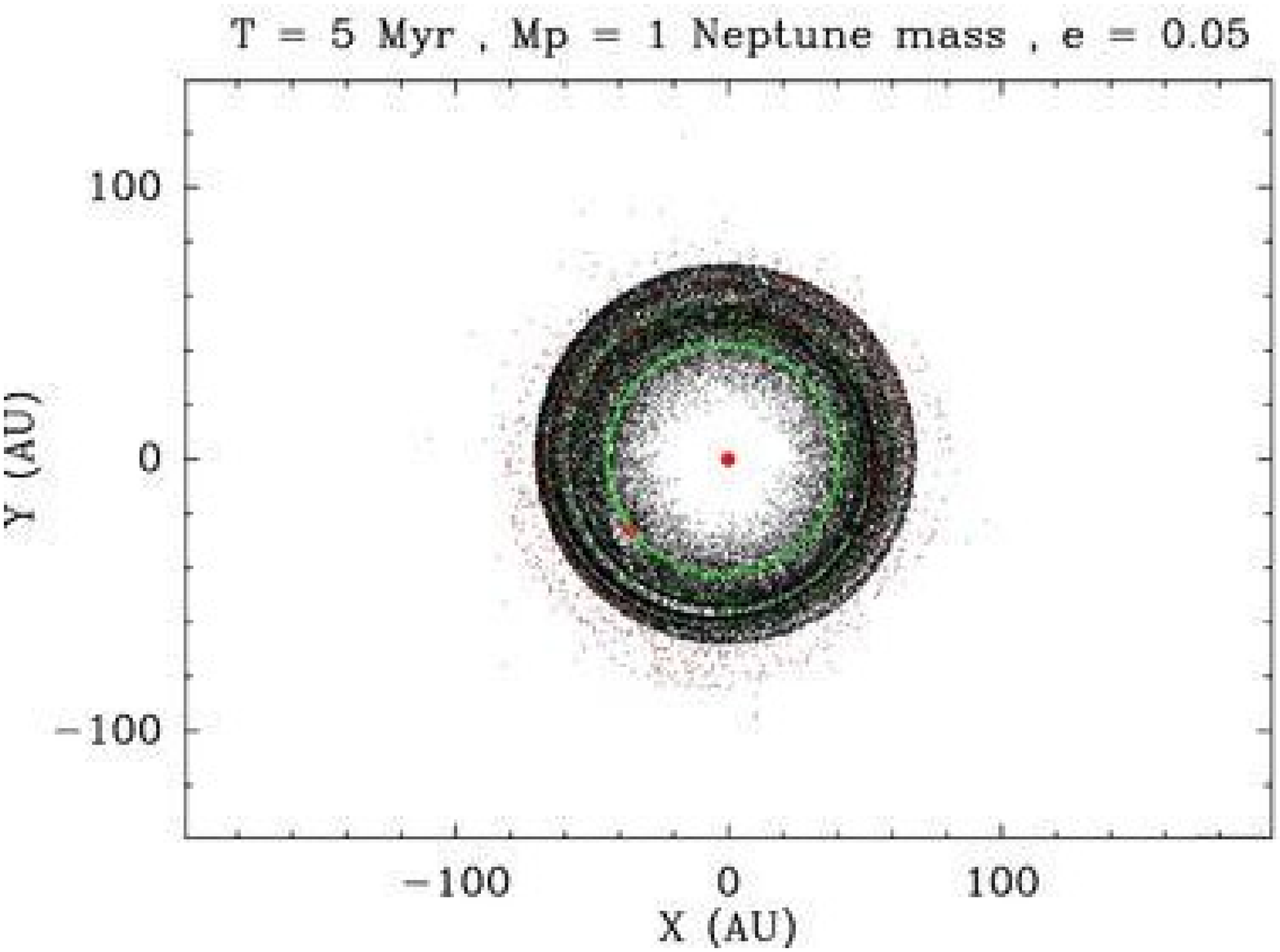} \hfil
\includegraphics[angle=-90,width=0.33\textwidth]{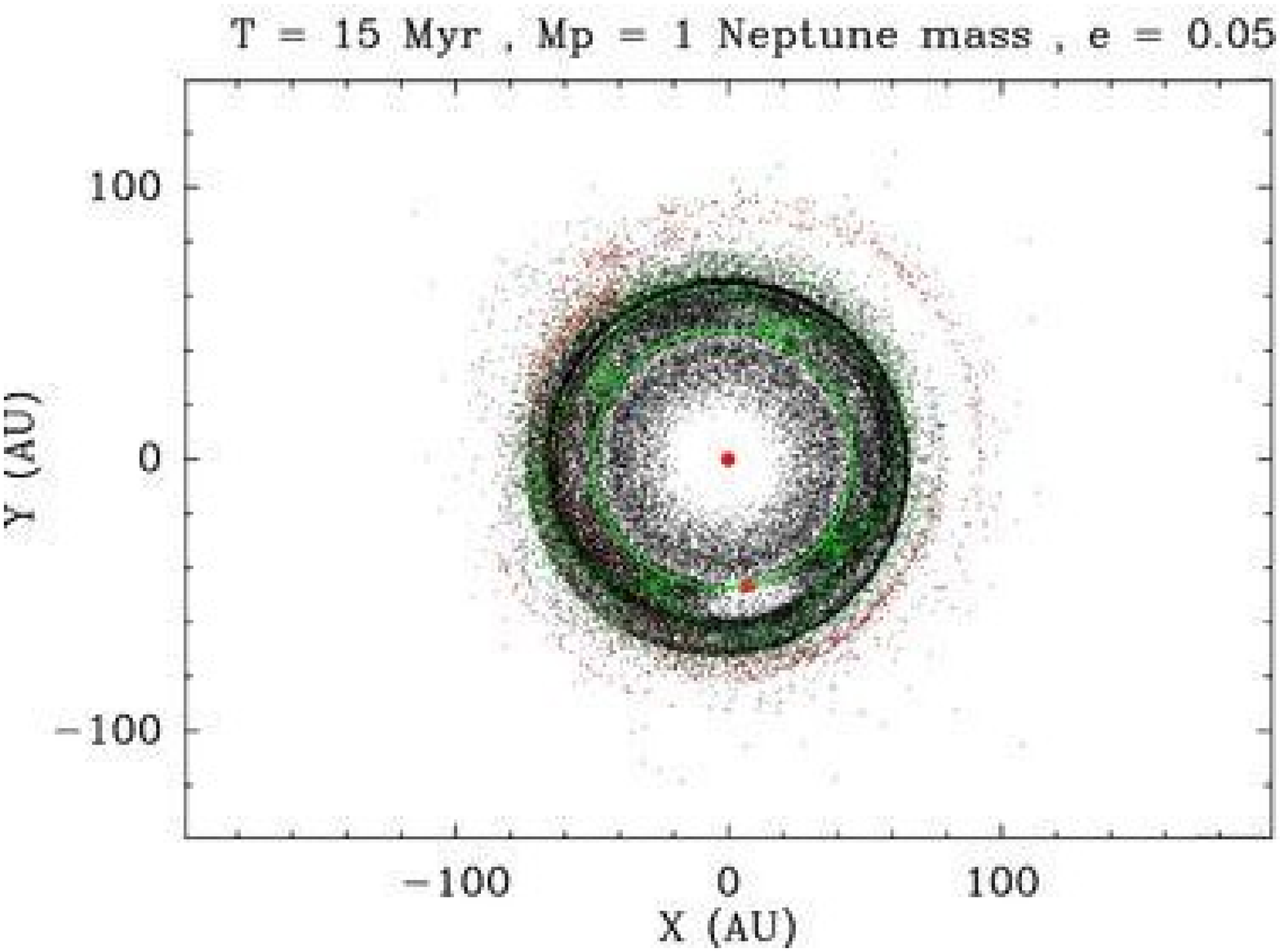} \hfil
\includegraphics[angle=-90,width=0.33\textwidth]{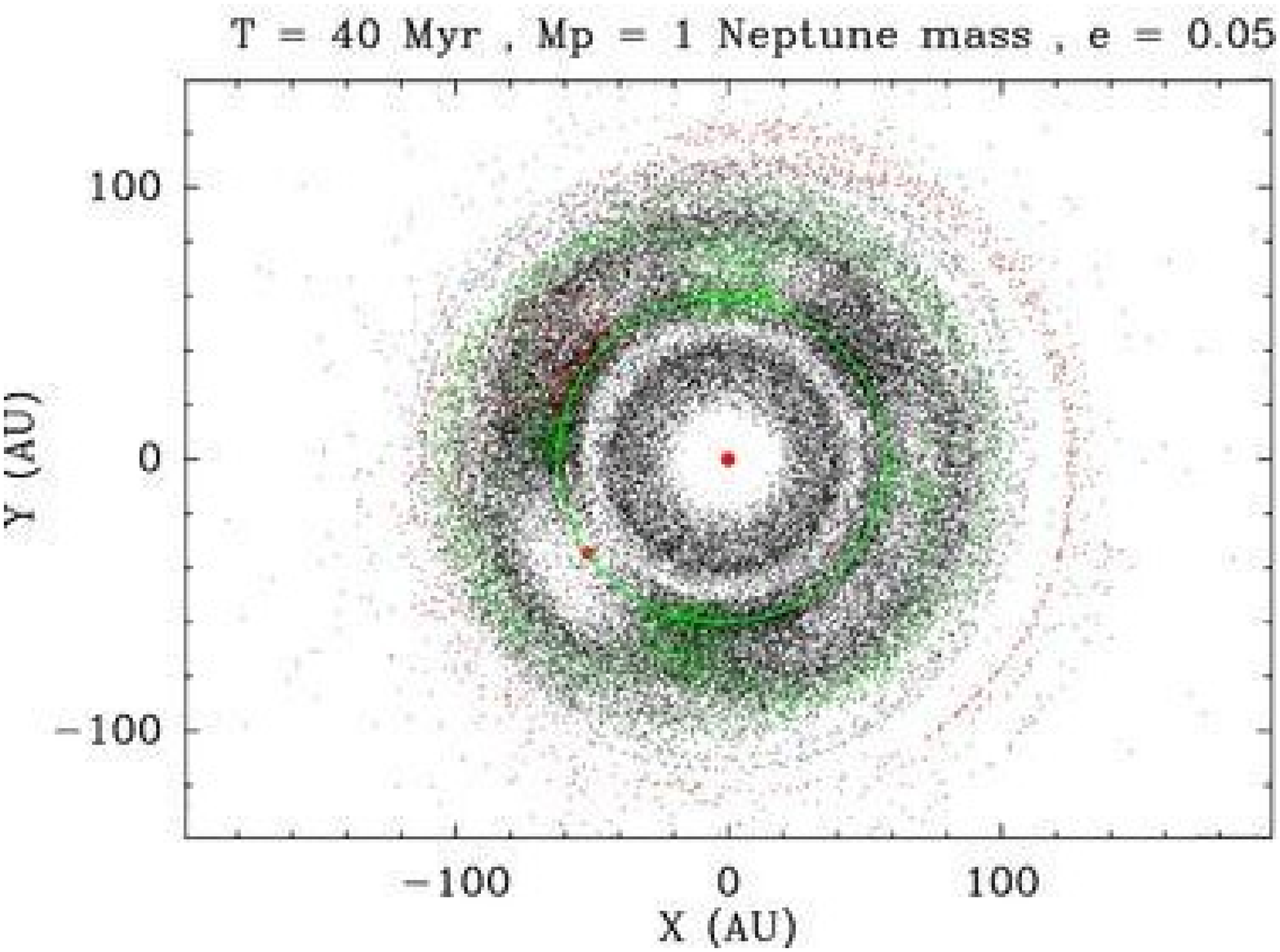}} \\
\makebox[\textwidth]{
\includegraphics[angle=-90,width=0.33\textwidth]{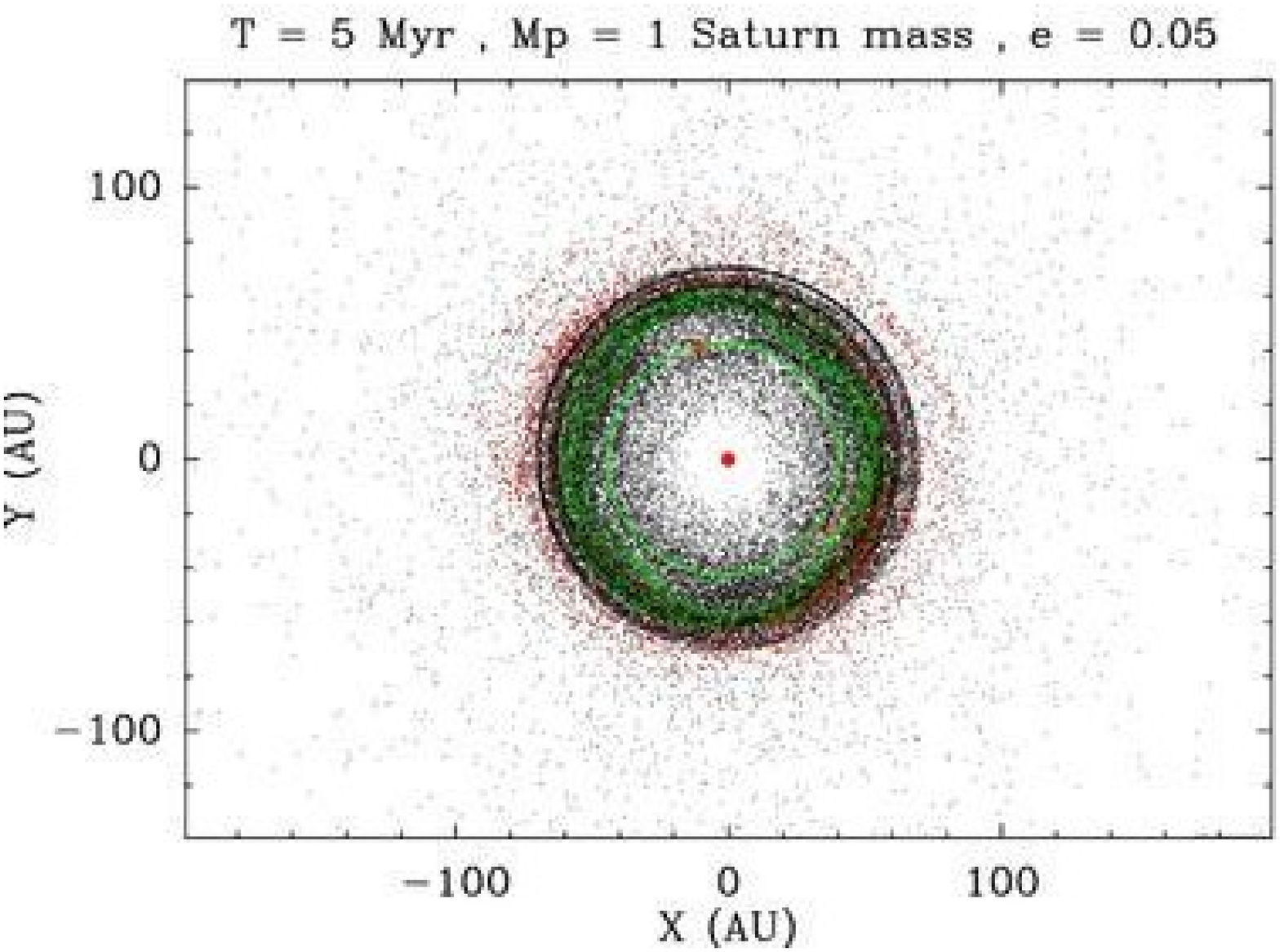} \hfil
\includegraphics[angle=-90,width=0.33\textwidth]{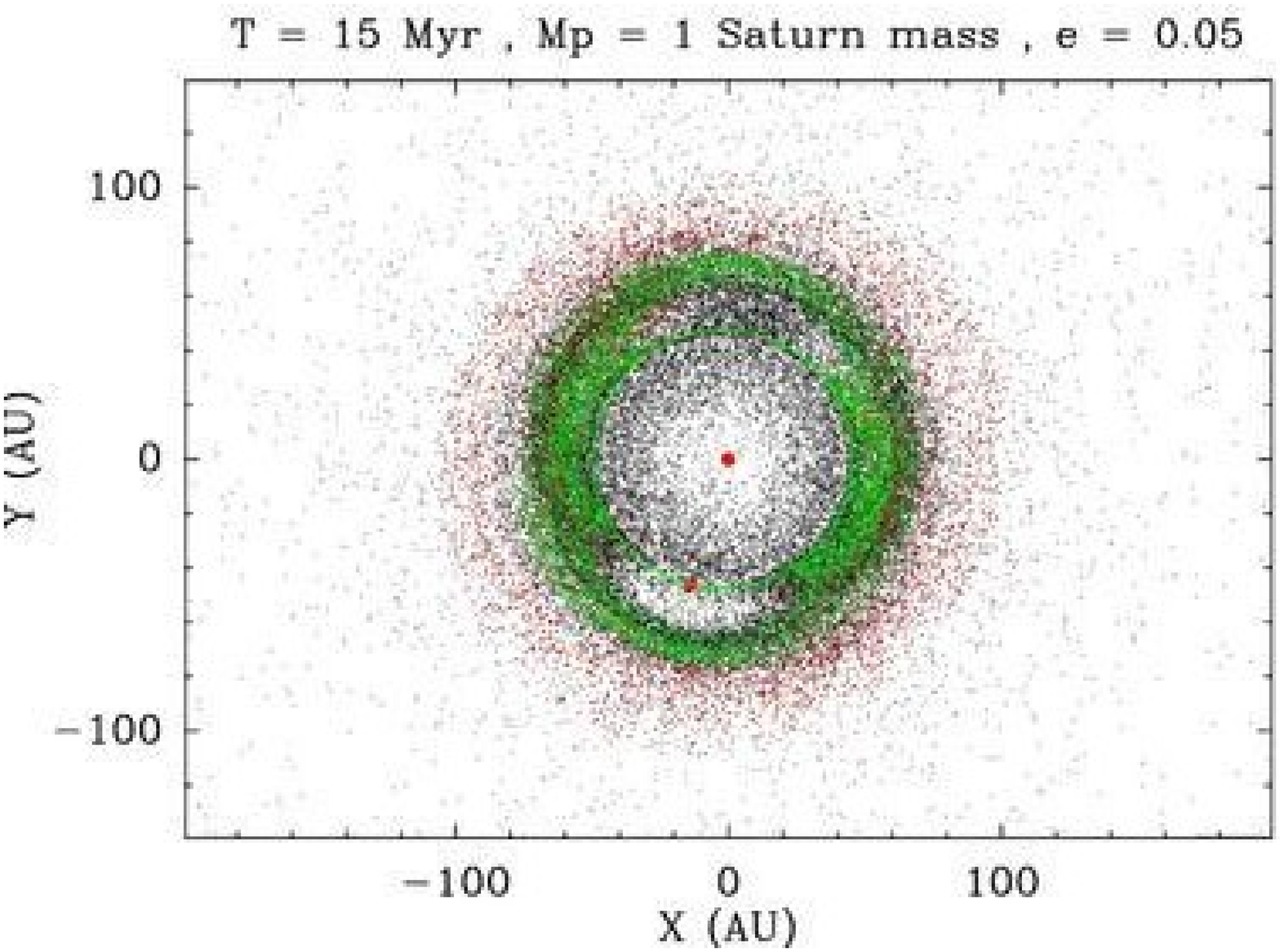} \hfil
\includegraphics[angle=-90,width=0.33\textwidth]{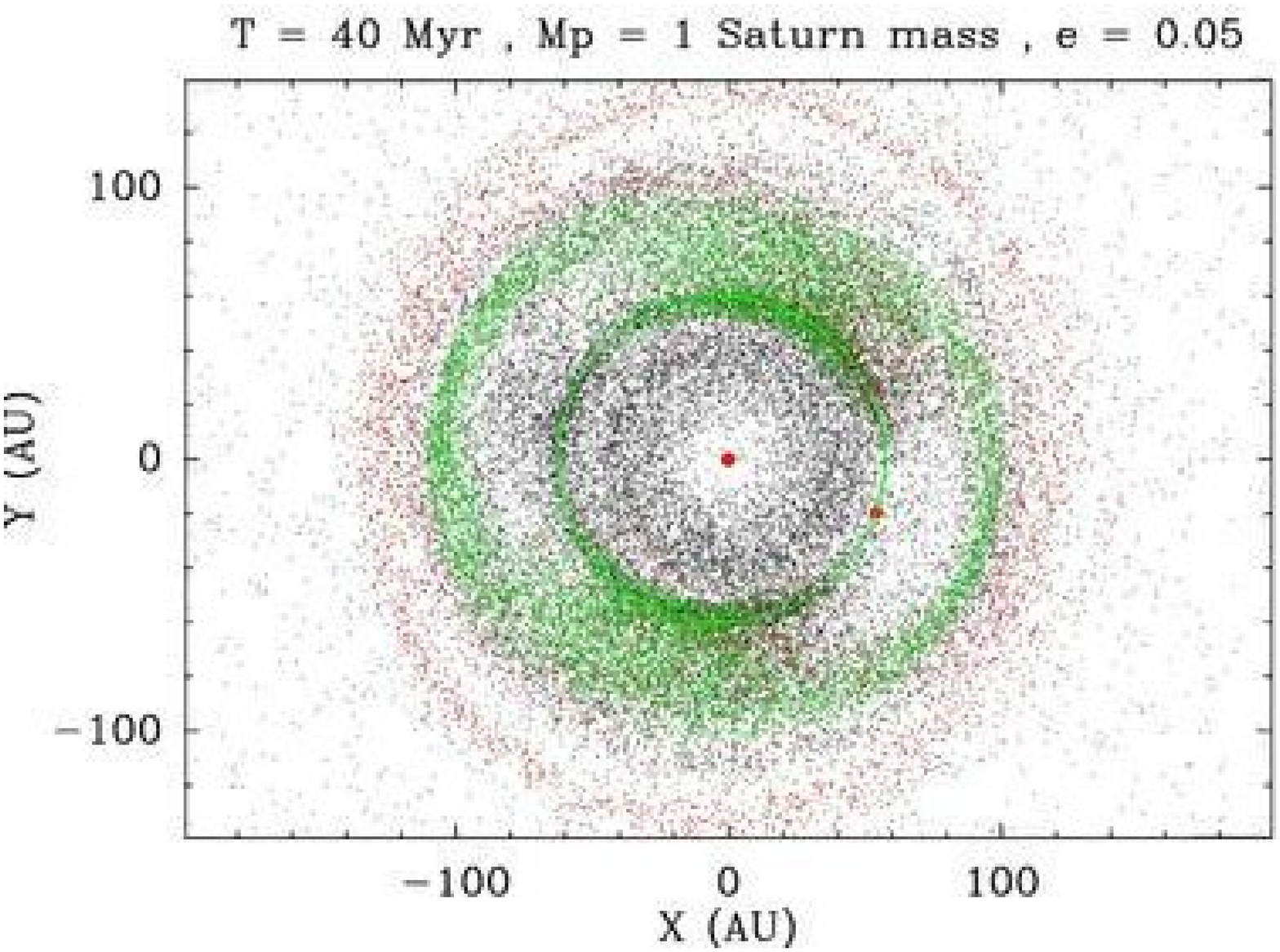}} \\
\makebox[\textwidth]{
\includegraphics[angle=-90,width=0.33\textwidth]{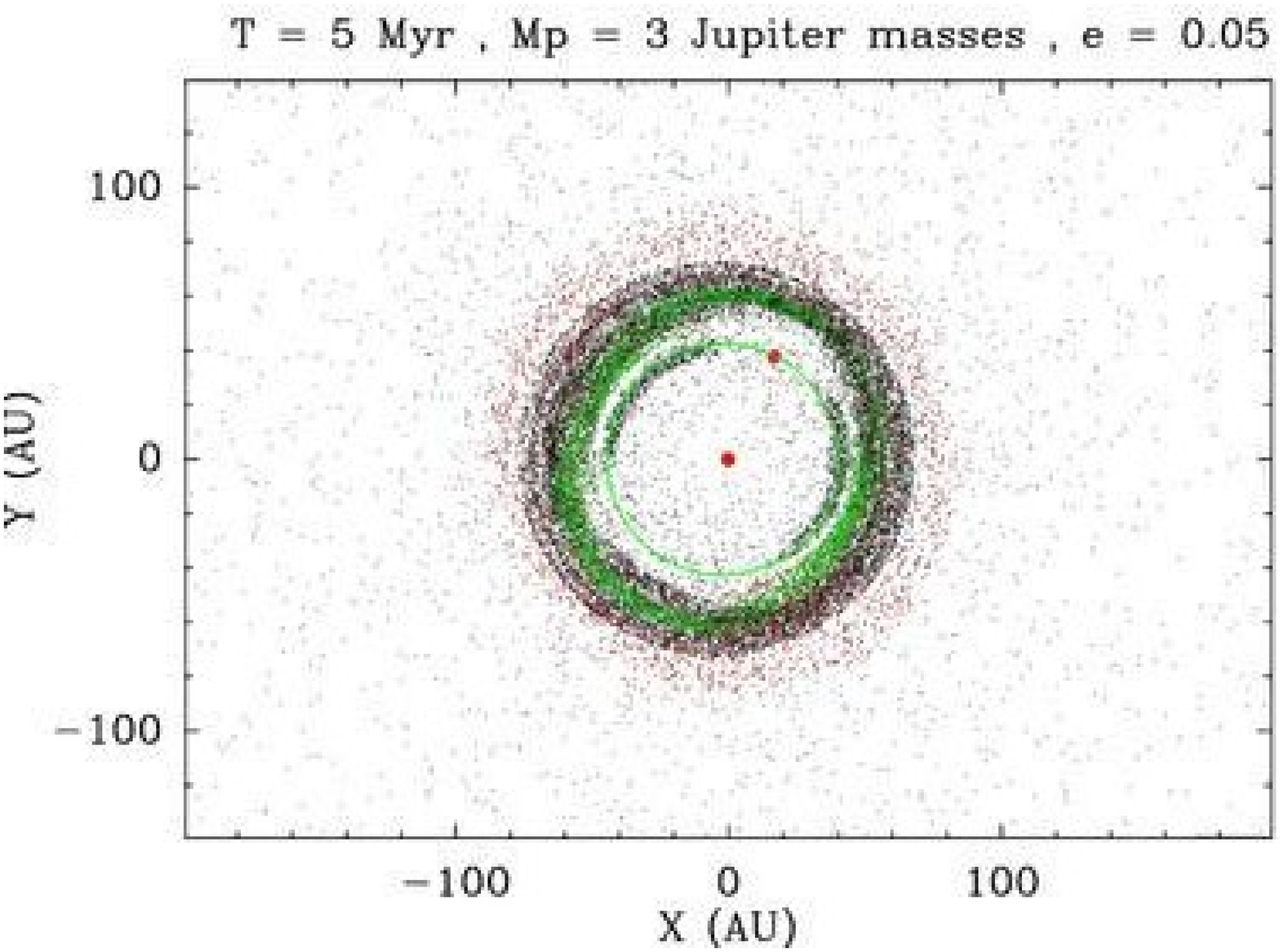} \hfil
\includegraphics[angle=-90,width=0.33\textwidth]{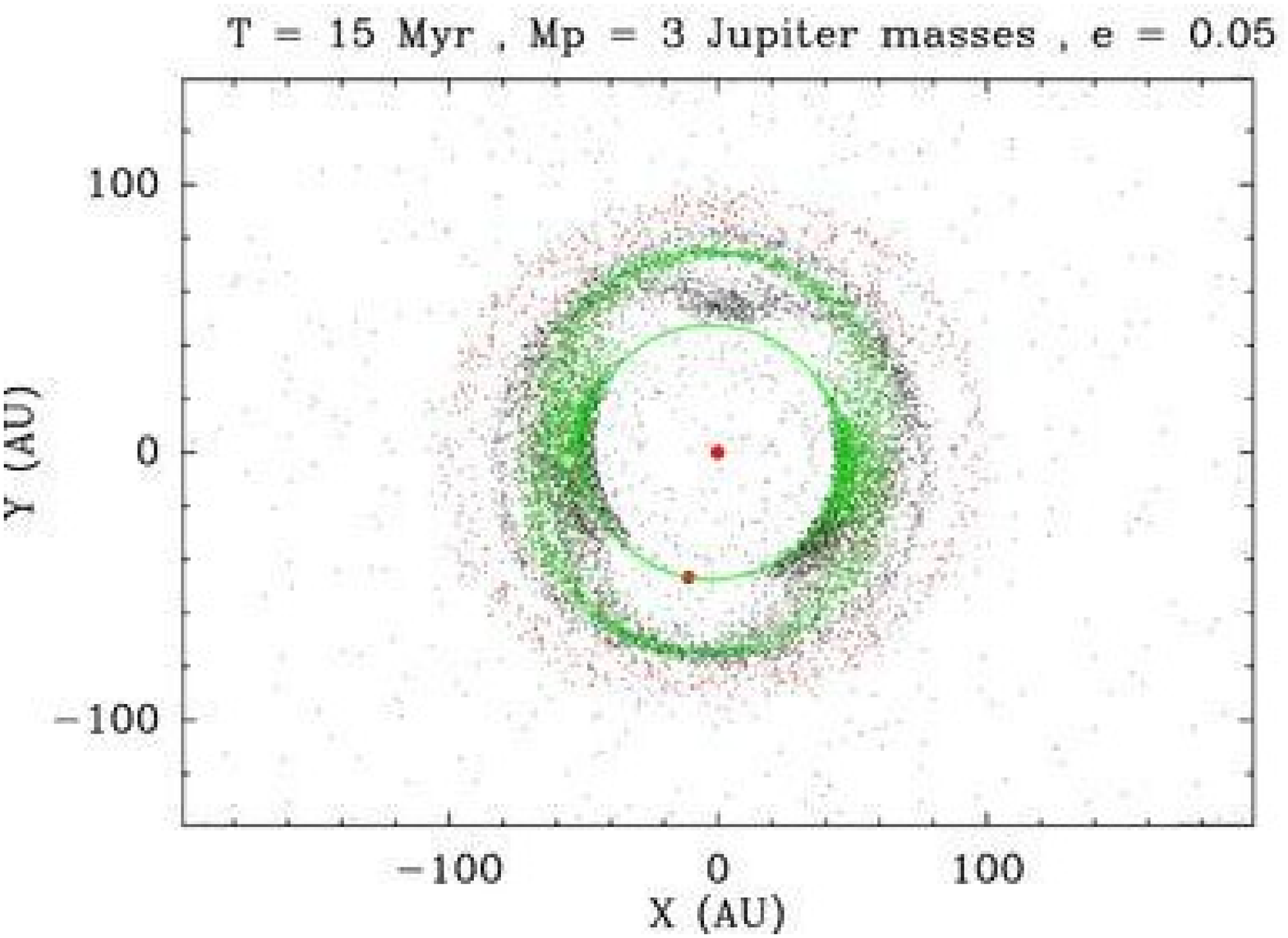} \hfil
\includegraphics[angle=-90,width=0.33\textwidth]{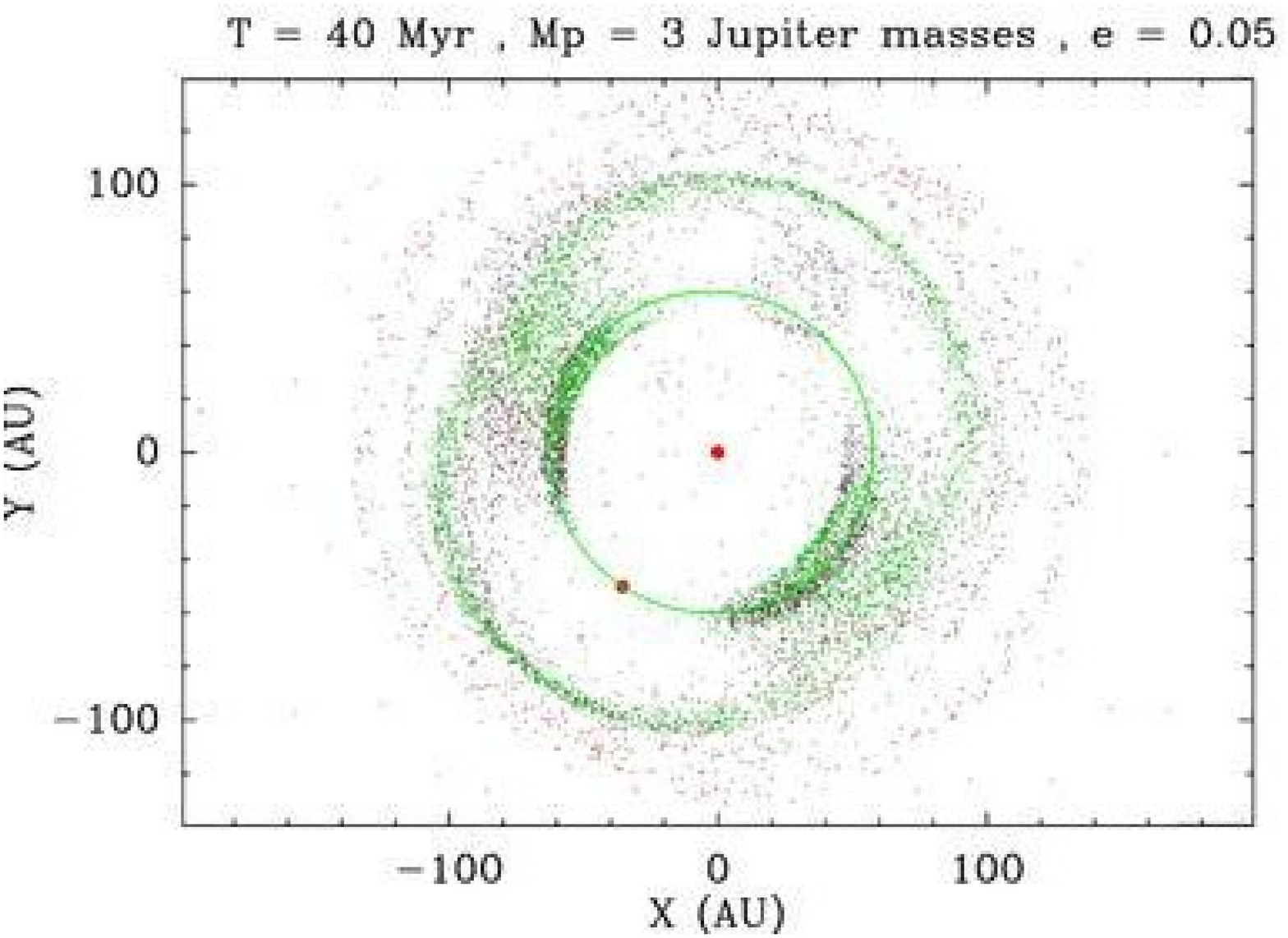}}
\caption{\label{figure_e05} Same as Fig. \ref{figure_e0}, for
  similar planets, but on a low-eccentricity orbit ($e_p=0.05$). \thanks{See the electronic edition of the
    Journal for a color version of this figure.}}
\end{figure*}

\begin{figure*}
\makebox[\textwidth]{
\includegraphics[angle=-90,width=0.33\textwidth]{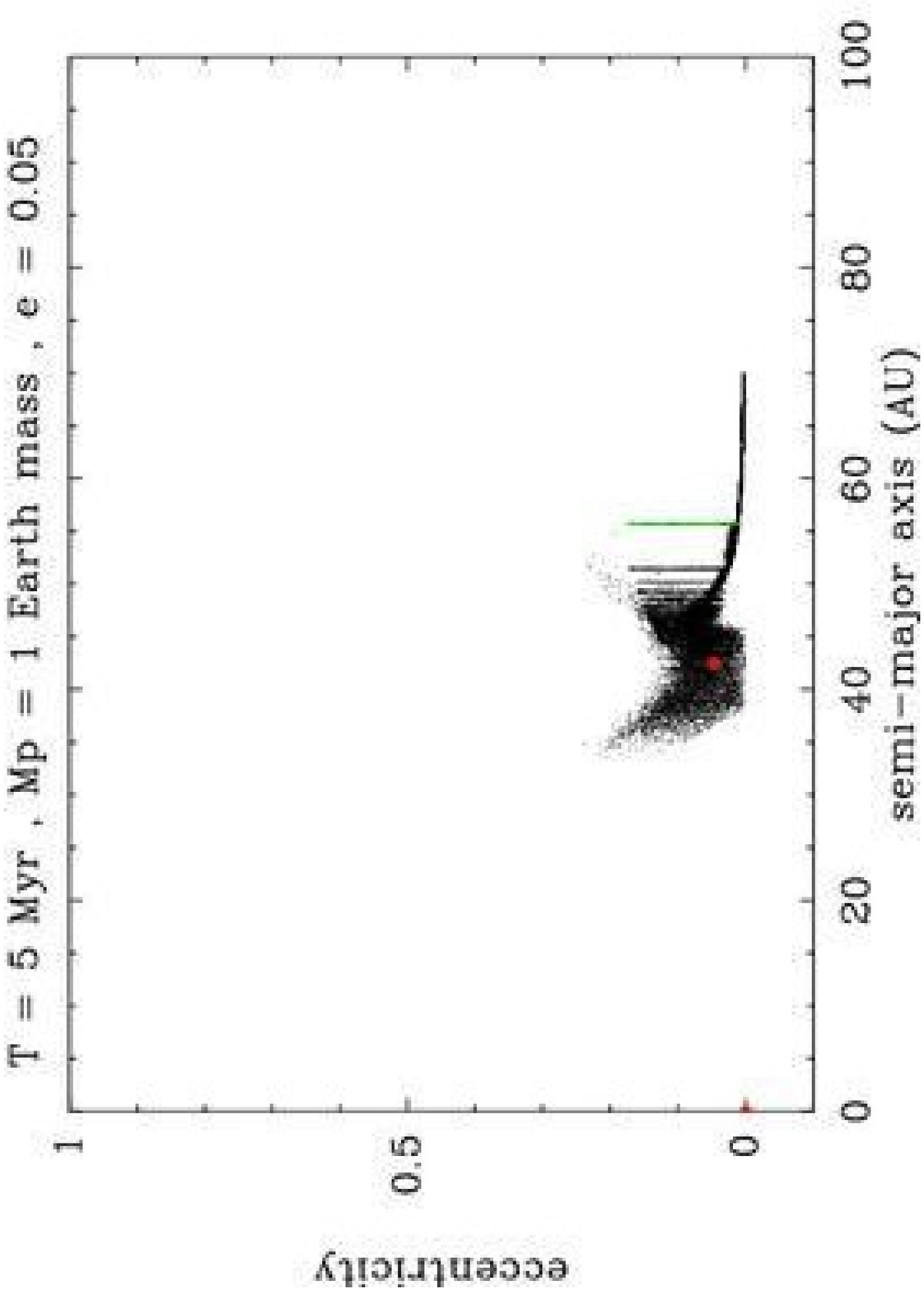} \hfil
\includegraphics[angle=-90,width=0.33\textwidth]{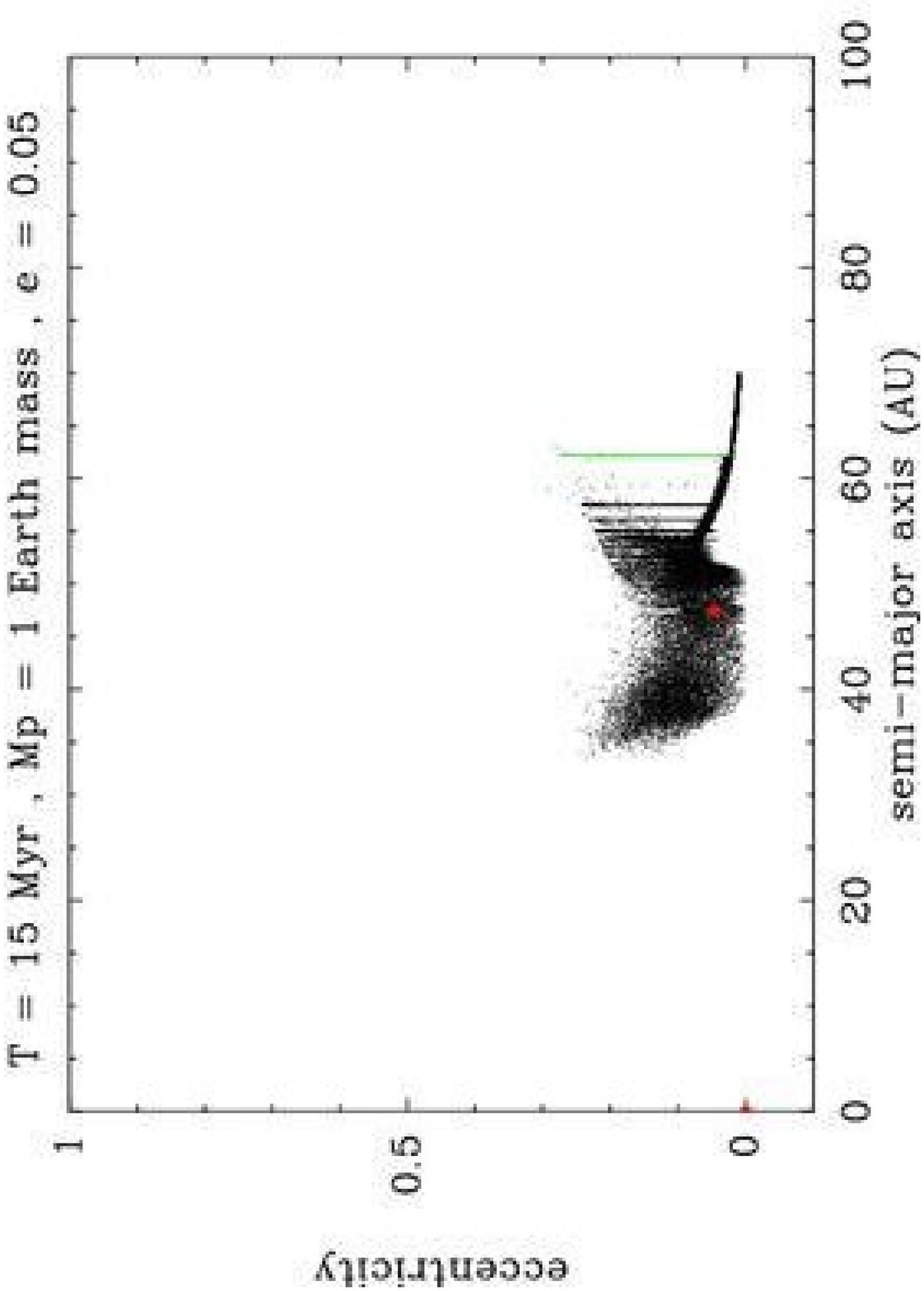} \hfil
\includegraphics[angle=-90,width=0.33\textwidth]{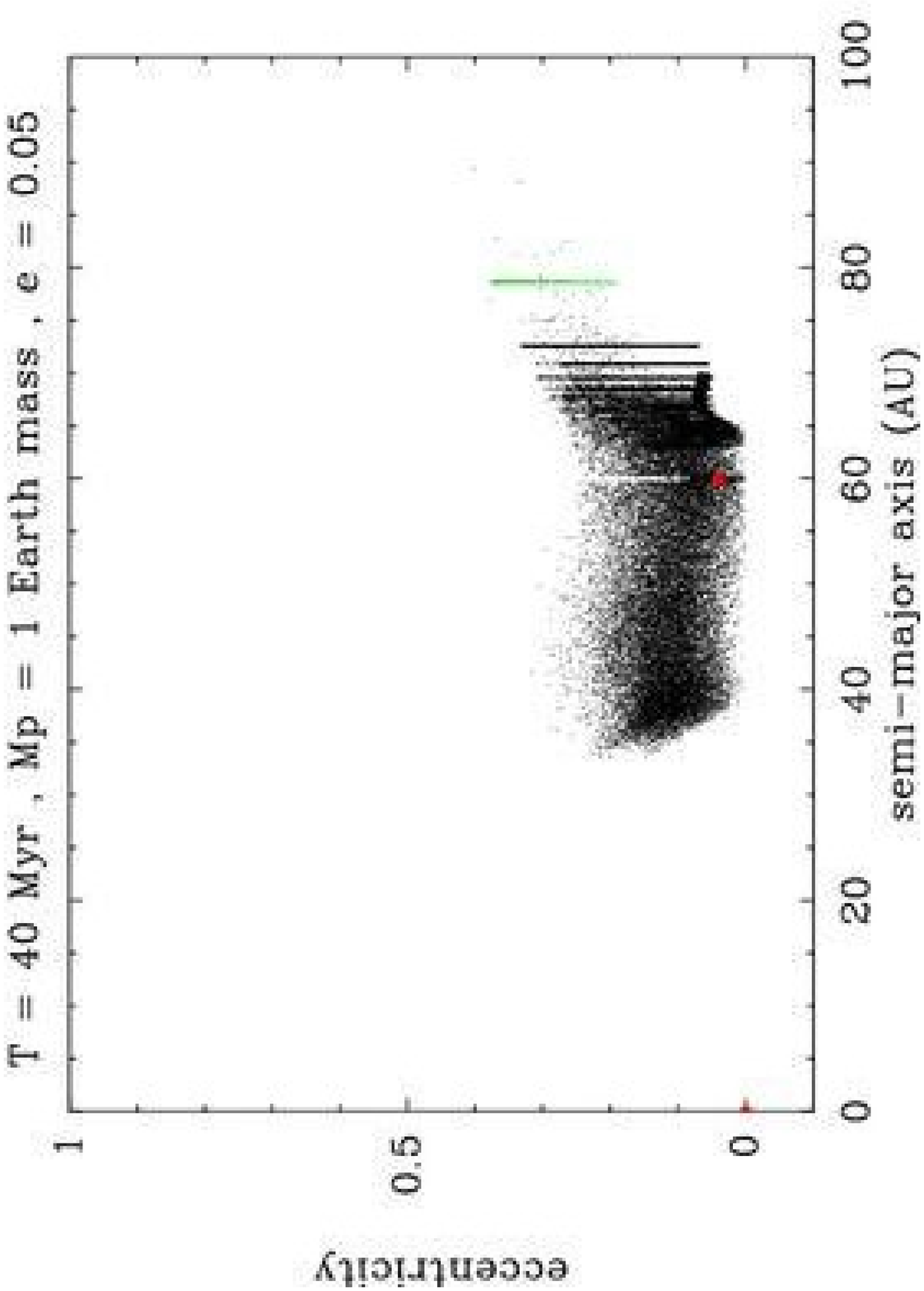}} \\
\makebox[\textwidth]{
\includegraphics[angle=-90,width=0.33\textwidth]{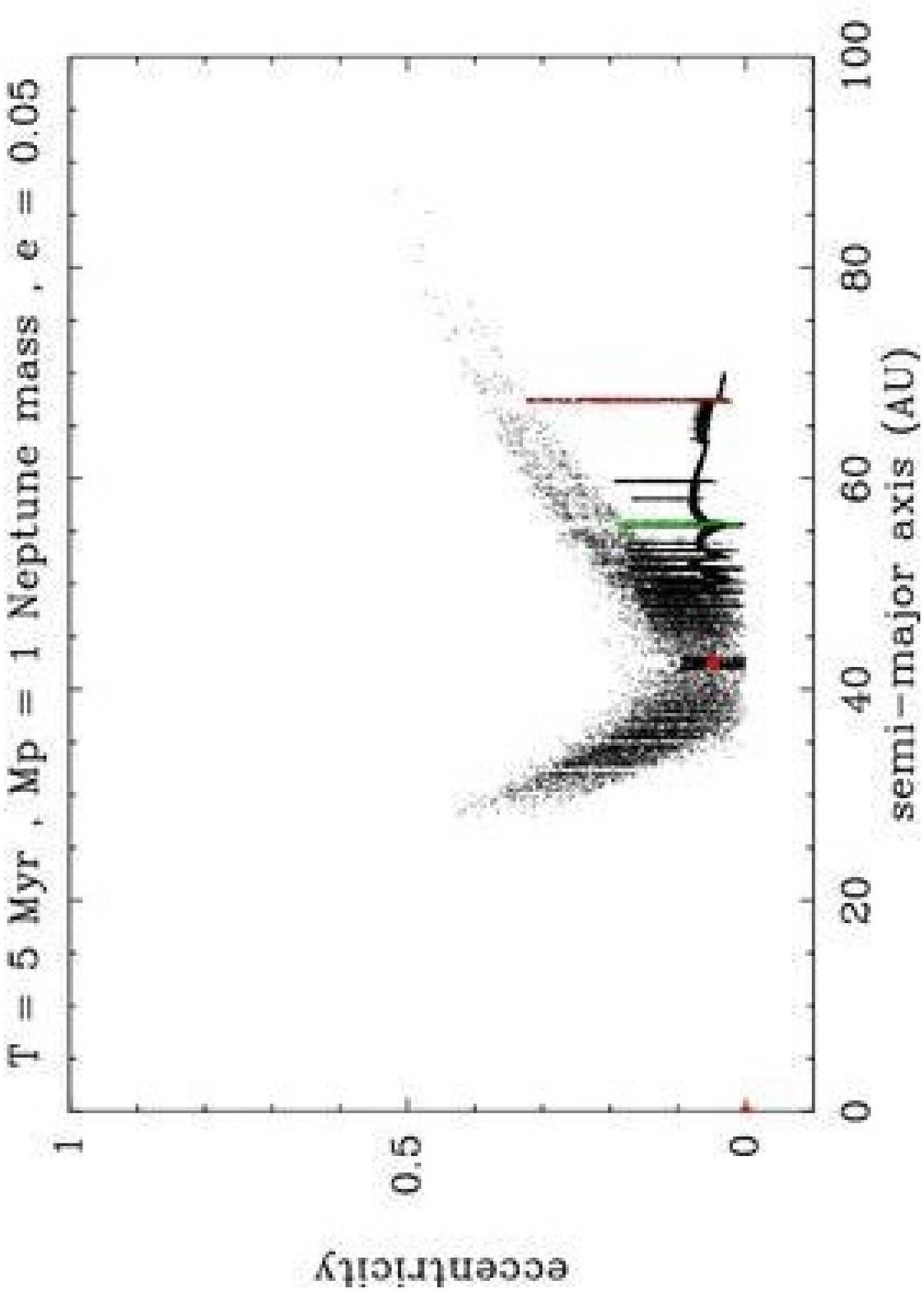} \hfil
\includegraphics[angle=-90,width=0.33\textwidth]{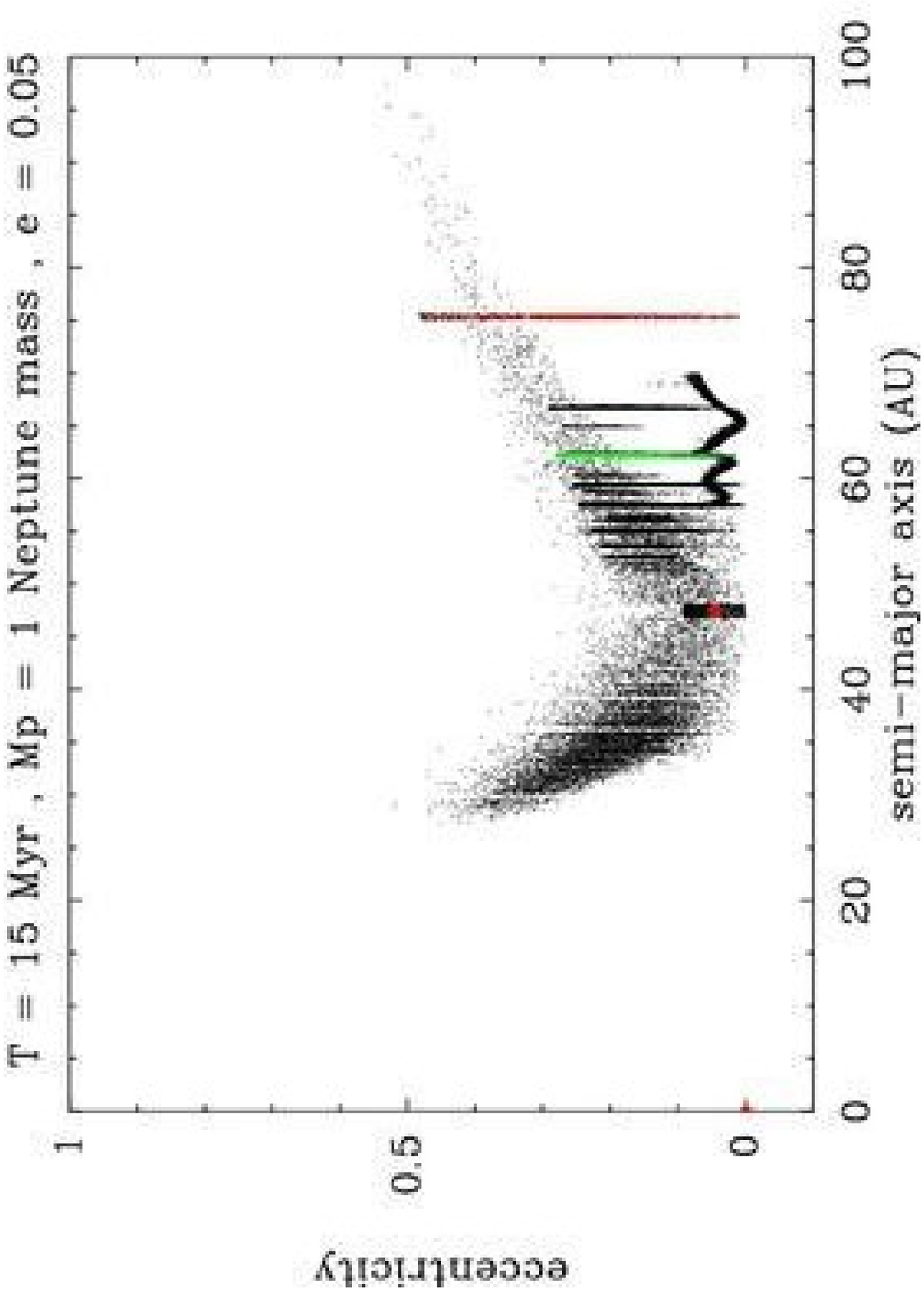} \hfil
\includegraphics[angle=-90,width=0.33\textwidth]{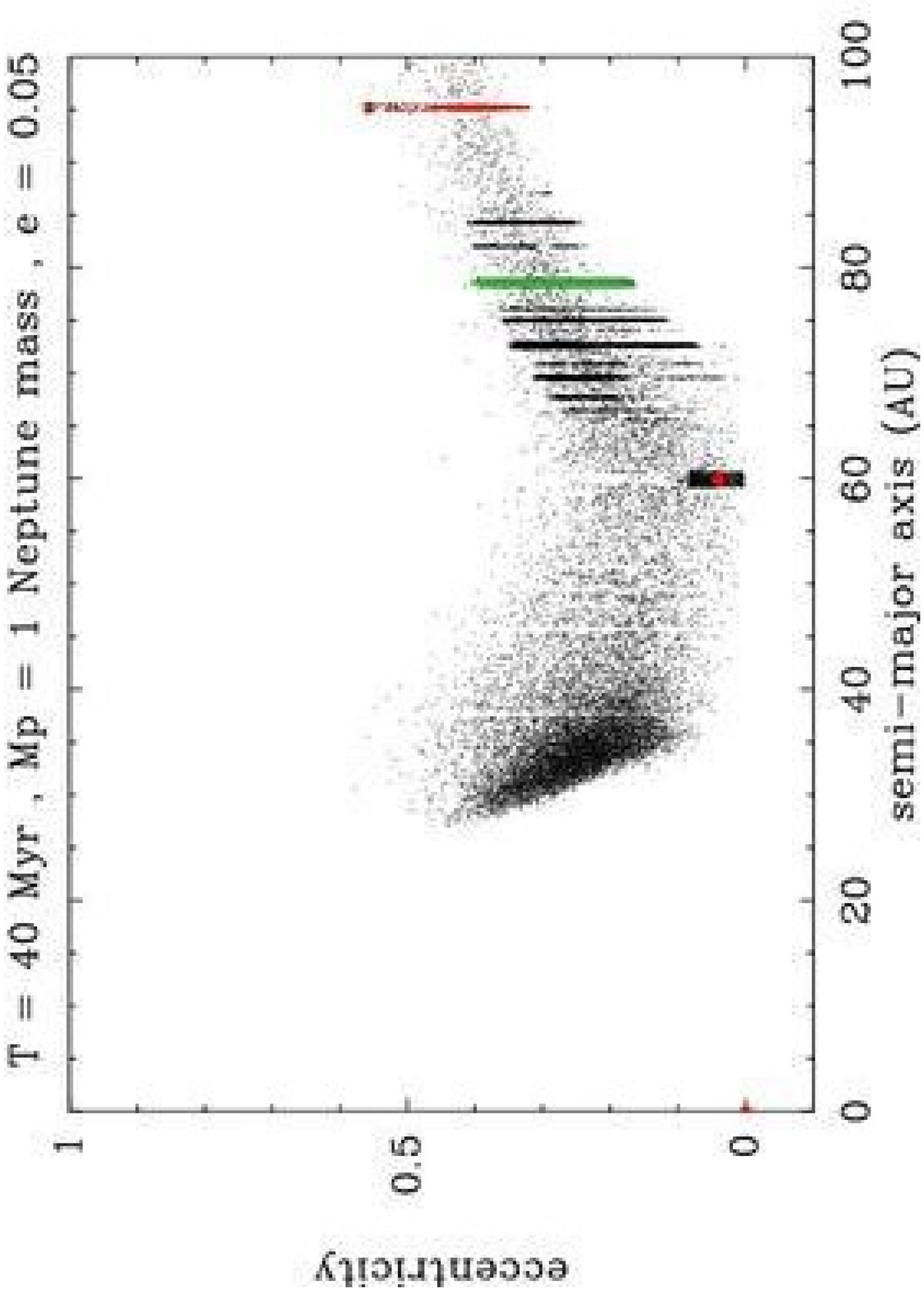}} \\
\makebox[\textwidth]{
\includegraphics[angle=-90,width=0.33\textwidth]{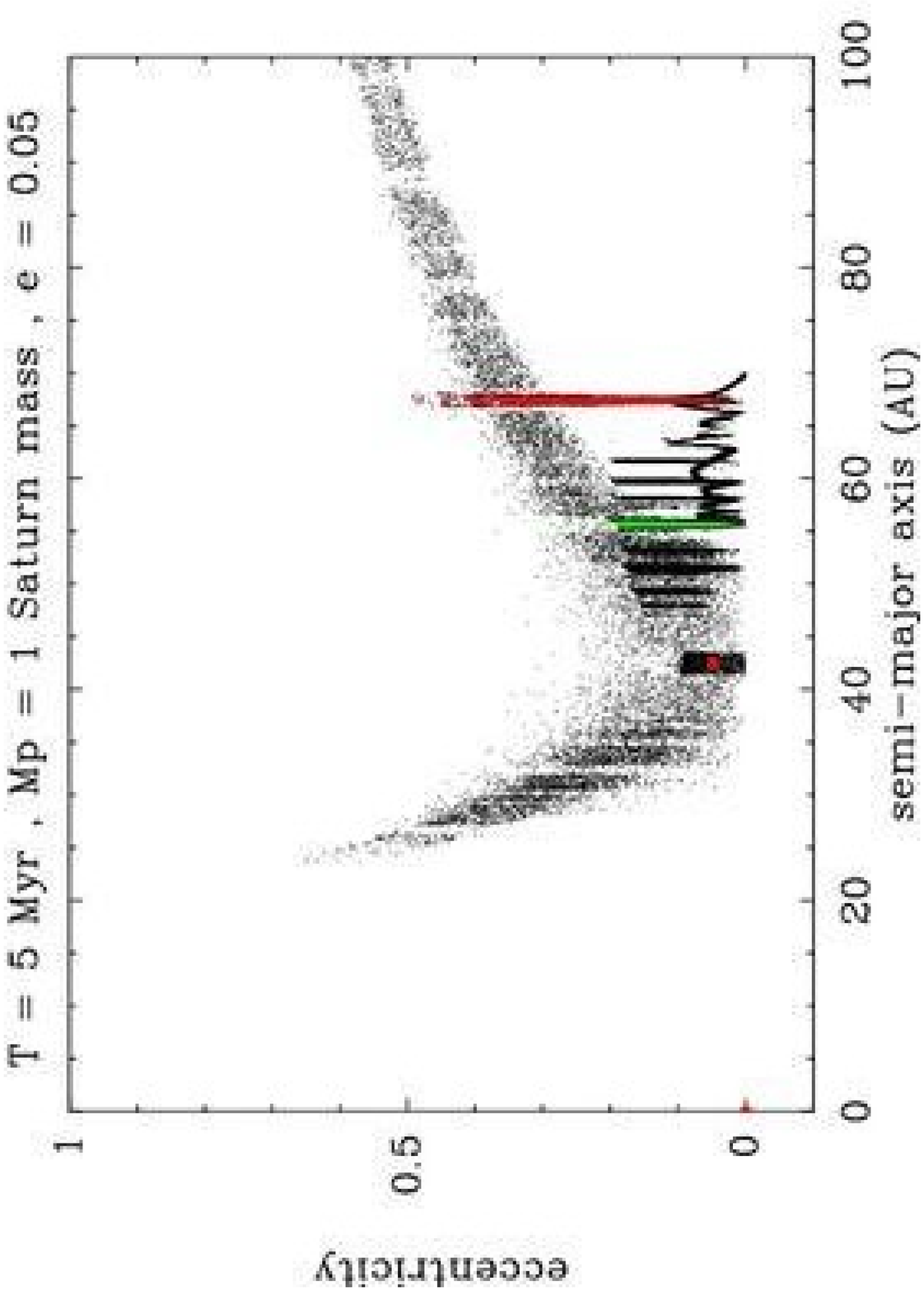} \hfil
\includegraphics[angle=-90,width=0.33\textwidth]{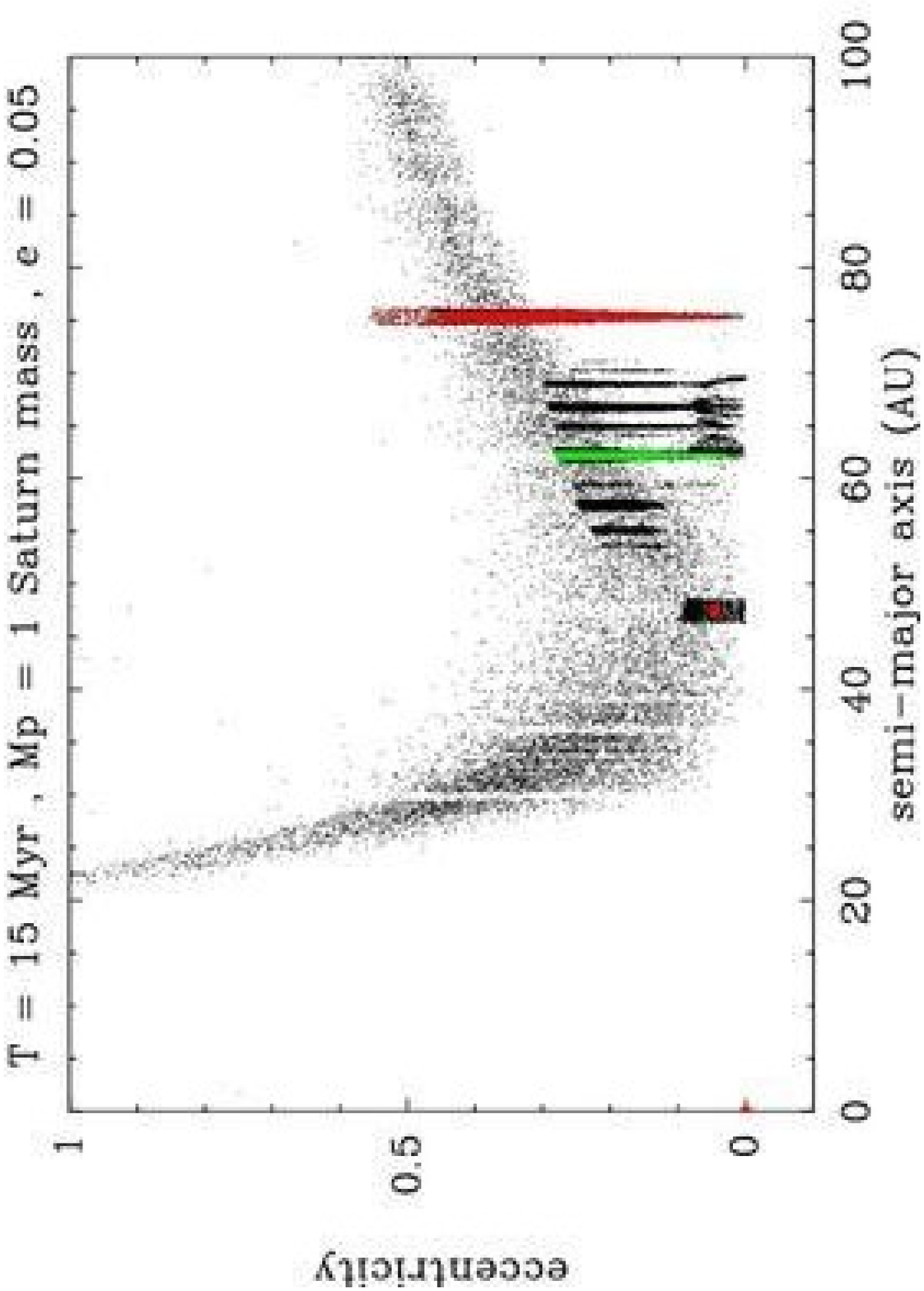} \hfil
\includegraphics[angle=-90,width=0.33\textwidth]{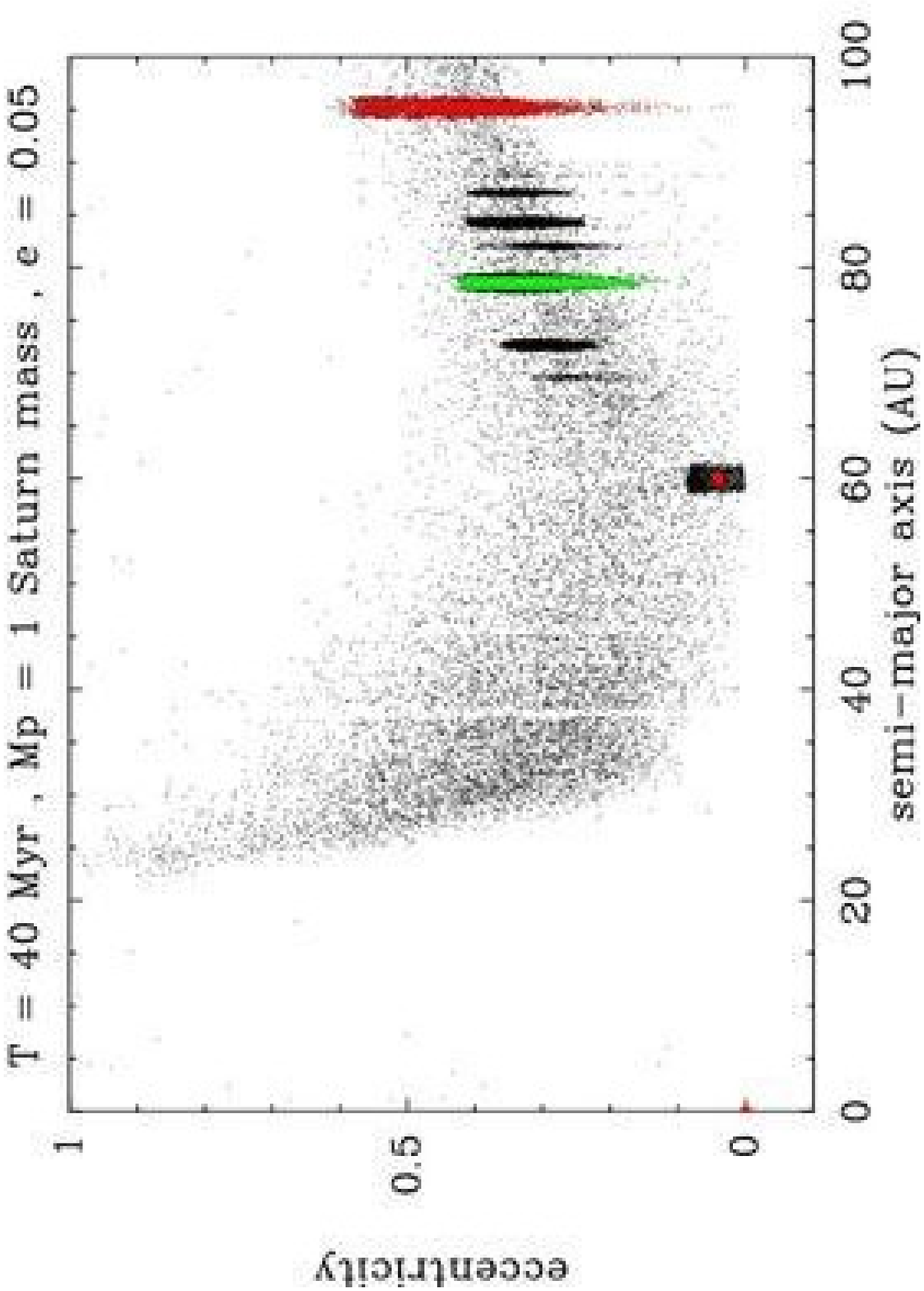}} \\
\makebox[\textwidth]{
\includegraphics[angle=-90,width=0.33\textwidth]{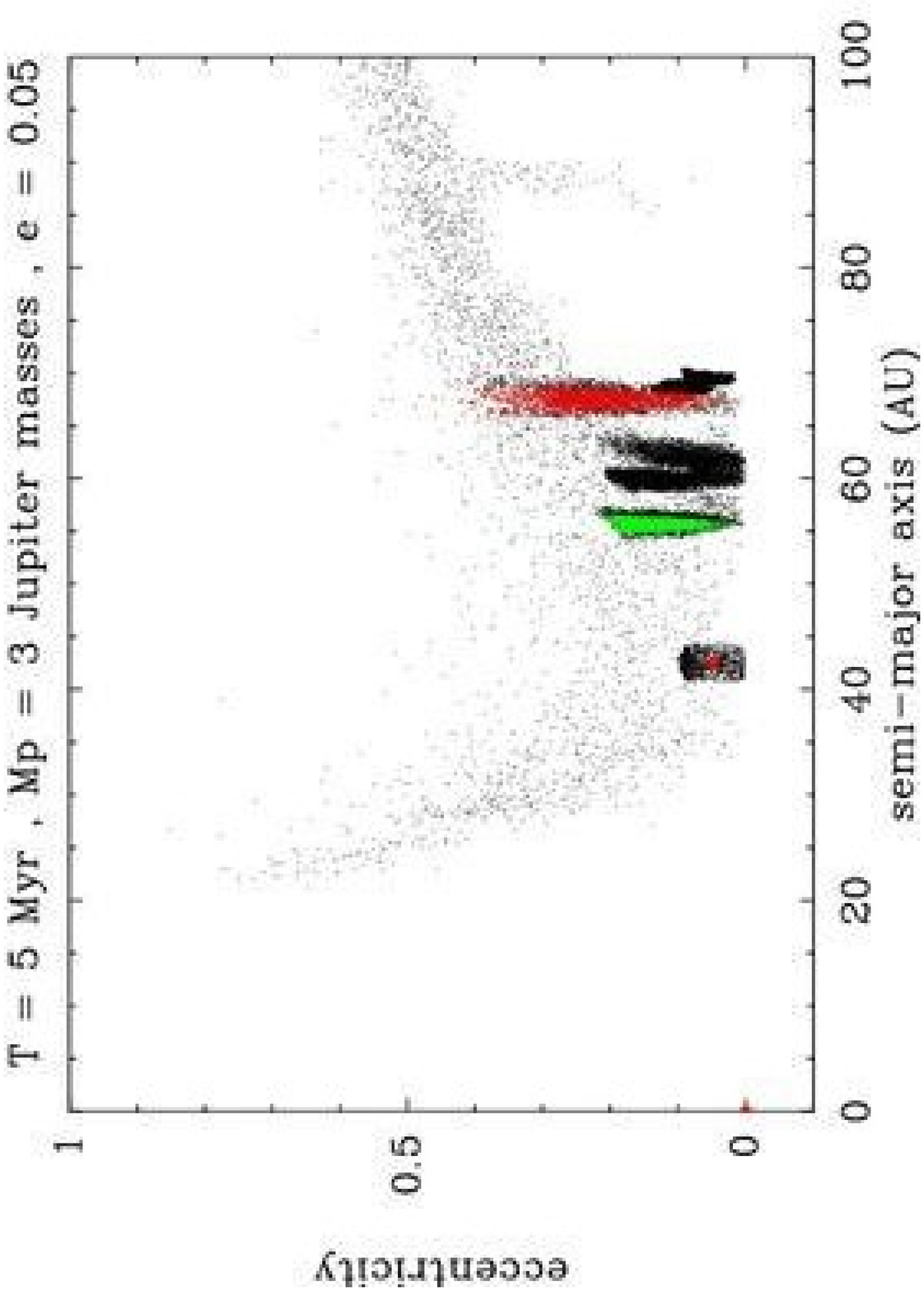} \hfil
\includegraphics[angle=-90,width=0.33\textwidth]{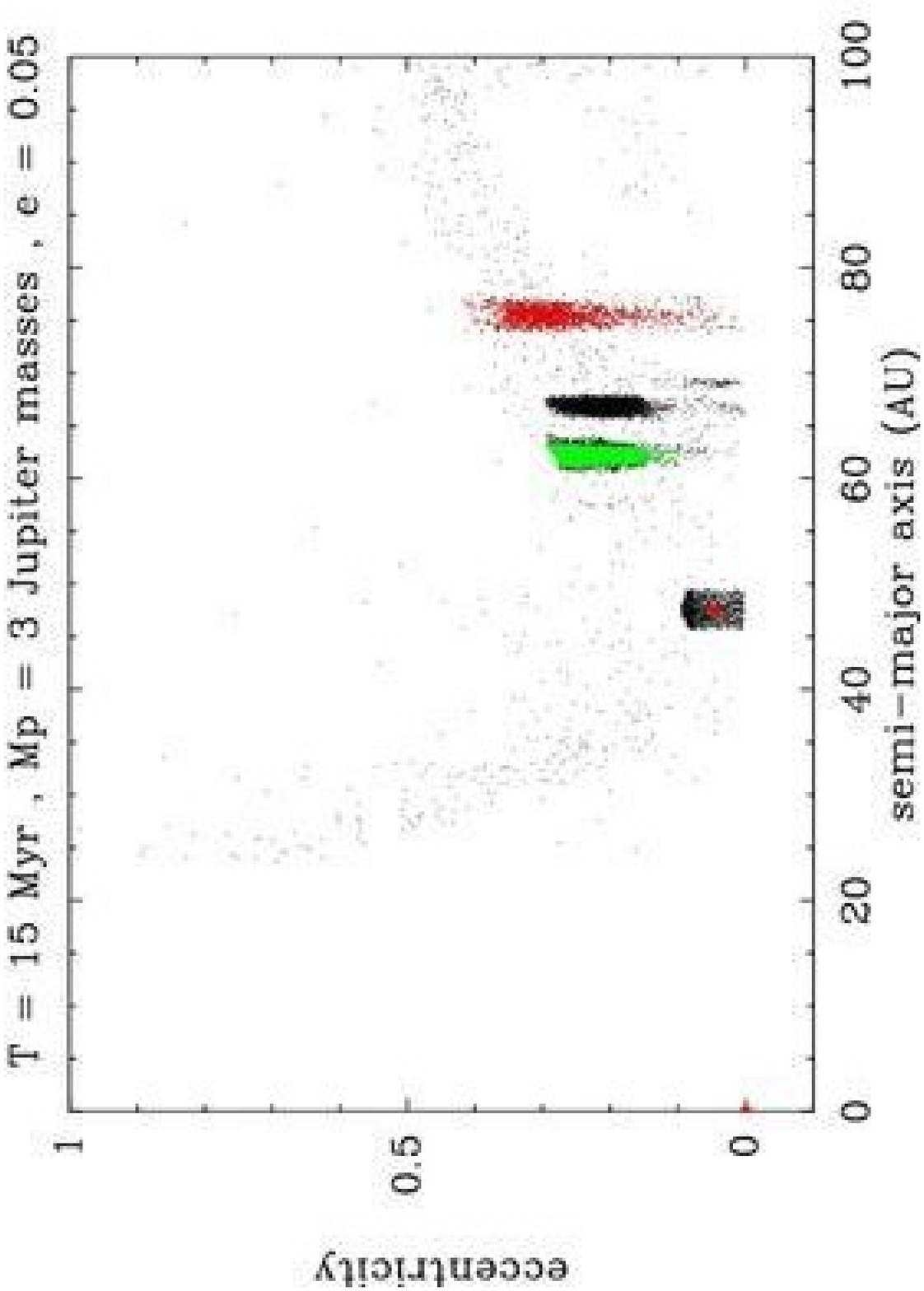} \hfil
\includegraphics[angle=-90,width=0.33\textwidth]{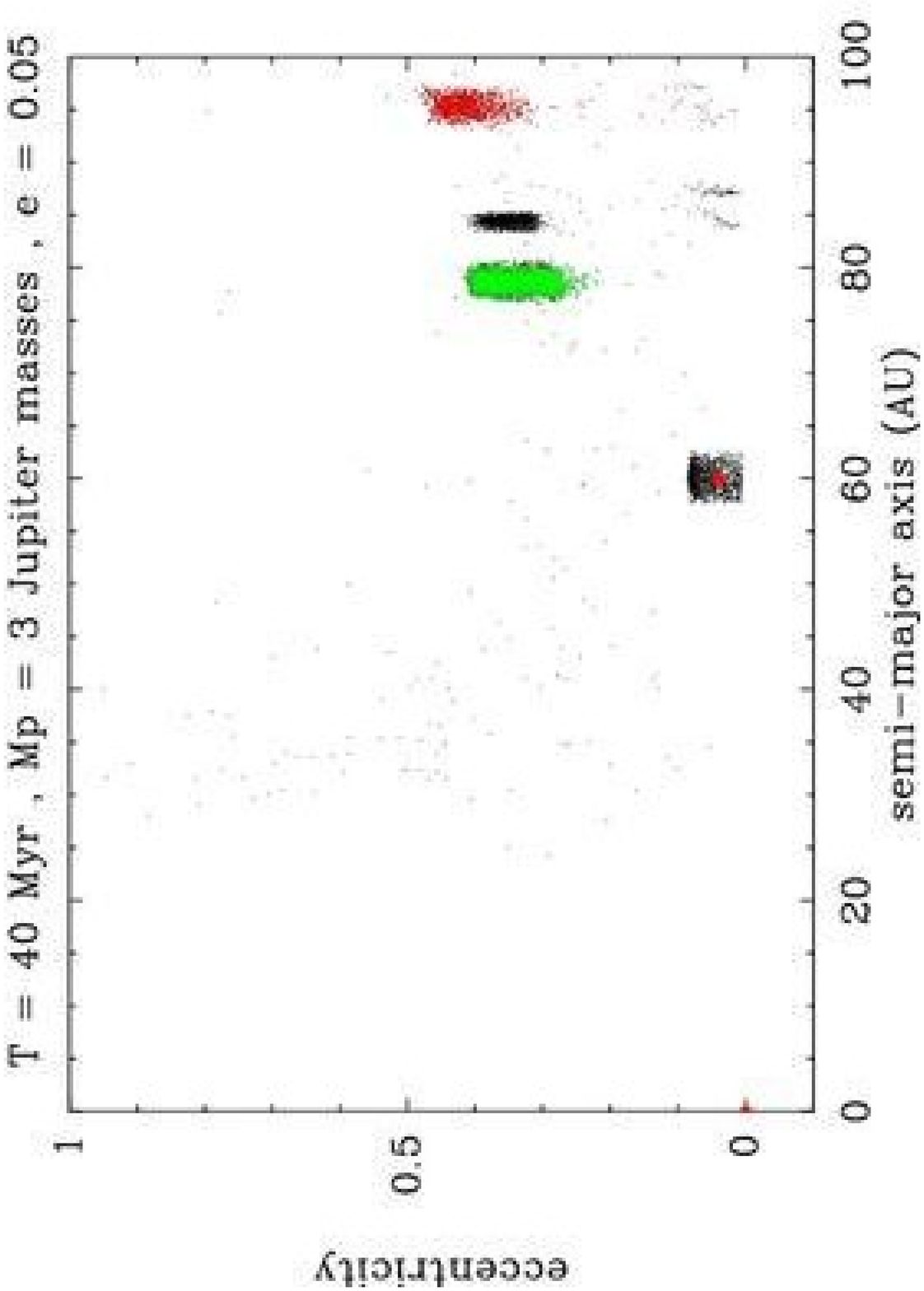}} \\
\caption{\label{figure_e05_ae}. Same as Fig. \ref{figure_e0_ae}, for
  similar planets, but on a low-eccentricity orbit ($e_p=0.05$). \thanks{See the electronic edition of the
    Journal for a color version of this figure.}}
\end{figure*}

The scenario of a planet on a low-eccentricity orbit is the most studied
case (see Table \ref{previousWorks}), for several reasons. First, planets were
originally expected to be on almost circular orbits because, during the protoplanetary
phase, circumstellar gas forces the planets to remain on very low eccentricity
orbits.  A planet on such an orbit therefore corresponds to the ``standard
scenario''. Also, a low or zero planetary eccentricity simplifies an
analytical analysis \citep{2003ApJ...588.1110K}.  Nevertheless, it must be noted that many of the
extrasolar planets detected so far have high
eccentricities\footnote{www.extrasolar.eu}, and we will therefore extend our
study to high eccentricities in Section \ref{highEccOrb}.

Although this standard  scenario has already been well studied, all its aspects have not yet been investigated. \citet{2003ApJ...598.1321W}
studied the  case of an outward migrating planet on a strictly
circular orbit, while \citet{2003ApJ...588.1110K} and \citet{2005ApJ...625..398D}
studied the case of planets on fixed low-eccentricity orbits,
considering only inward dust
migration due to P-R drag. We propose in this section to numerically
study a migrating planet on a circular or low-eccentricity orbit to
search for possible differences with respect to previous studies. 

\subsection{General trends}

Figures~\ref{figure_e0} to \ref{figure_e05_ae} show examples of results
obtained with our numerical model. It appears that, with a planet on a
low-eccentricity orbit, the planetesimals trapped in MMRs are
numerous and dominate the shape of the disk. Four important factors must be taken into account to
determine which
resonances govern the aspect of the structures in the disk:

\begin{itemize}
\item External MMRs at large $a/a_p$ are weaker and less able to trap
  numerous planetesimals than closer MMRs
\citep{2003ApJ...588.1110K}. First-order resonances (i.e. $m=n+1$)
with large $m$ values are the closest and hence the strongest MMRs. However, MMRs near
the planet are located close to each other and compete. A
resonance overlap criterion
\citep{1980AJ.....85.1122W,1989Icar...82..402D} predicts that first-order resonances
become completely chaotic when  $m>0.45\mu^{-2/7}+1$, where
$\mu=m_p/M_{\ast}$. This places a limit on the nearest resonance that can
be populated. In the Solar System, the first completely overlapped MMR
is the $17$:$18$ for the Earth and the $4$:$5$ in the case of Jupiter.
\item \citet{2003ApJ...598.1321W} showed that the planetary mass acts as a threshold for the probability of
capture in first-order external MMRs: for a fixed migration rate,
 the probability drops quickly to $0$ below a certain planetary mass, while above
this mass the probability grows quickly to $1$. For MMRs of higher
order, the transition is less sharp.
\item An MMR that traps all the planetesimals at a given
  semi-major axis $a_r$ stops the growth of any other
  resonance that reaches this $a_r$ afterward. Resonances like
  $3$:$2$ or $5$:$3$  can thus trap a large number of
  planetesimals thanks to a large enough spatial separation, while
  resonances near the planet are too close to each other to trap large
  populations. This shows that any modeling of structure generation
with this process must be done globally, as the various MMRs
compete with each other to be filled. 
\item The long term evolution of planetesimals trapped in MMRs globally results
  in an increase of their eccentricity. If this eccentricity
  becomes high enough, planetesimals can become planet-crossing. Thanks
  to the resonance, they are nevertheless phase-protected against close
  encounters with the planet. But if the eccentricity grows too high,
  this phase-protection does not hold any longer, due to strong modulations of
  the angular velocity. Hence MMRs are limited in eccentricity
  \citep{1993CeMDA..57..373S}, with $e_{max}=(2/5m_p)^{1/2}$. This
  concerns more specifically the MMRs that are close to the planet, as the
  planetesimals may easily become planet crossing. We thus expect the
  closest MMRs to lose planetesimals when they reach a given eccentricity
  limit. This is illustrated in Fig. \ref{figure_e0_ae} where the limits in
  eccentricity of the resonant populations are clearly visible.
\end{itemize}

In conclusion, only a few resonances, namely the most external of the
first or second order resonances ($4$:$3$, $3$:$2$, $5$:$3$ and $2$:$1$
resonance) capture most of the planetesimals. However,
  depending on the planet mass, the planetesimals that are not
  trapped in resonances can, or not, change the shape of the
  disk. In our simulations, planets with a mass above $1$ Jupiter mass
  eject almost all the non-resonant planetesimals: the MMR structures
  therefore appear clearly. Below $1$ Jupiter mass, planets cannot eject
  all the non-resonant planetesimals, which can then partly hide the MMR structures.

\subsection{Circular orbits}
\label{circularOrbits}
In the circular orbit case (Fig.  \ref{figure_e0} and
\ref{figure_e0_ae}), differences appear between the simulations,
depending on the planet mass: 
\begin{itemize}
\item For an Earth mass planet, non resonant planetesimals are still
  bound to the system and hide all the resonant
  structures. However, a small hole at the planet location can be observed.  
\item For a Neptune mass planet, two clumps of equal density are generated by the
  $3$:$2$ resonance and they are located in opposition with respect to
  the star. It also appears that the planet is not massive enough to capture many
  planetesimals in the $2$:$1$ resonance and thus does not  generate thin
  rings at large distances like more massive
  planets. Non-resonant planetesimals create a ring inside the
    planet orbit.
\item For a Saturn mass planet, the $2$:$1$ resonant pattern appears in
  addition to the $3$:$2$ one. The $2$:$1$ MMR produces only one clump near
  one of the two generated by the $3$:$2$ resonance: the global structure
  thus becomes asymmetric. But, as in the previous case, the
    non-resonant planetesimals are still numerous in the inner part of
    the disk and partly hide the resonant structures. 
\item For a $3$ Jupiter mass planet, the $2$:$1$ resonant pattern changes
  and generates two clumps of equal density near those of the $3$:$2$
  resonance. The global structure is symmetric but the two clumps are
  no longer in opposition, because the libration centers of the $2$:$1$
  resonance are separated by less than $180^\circ$ in longitude. The
  change in the $2$:$1$ pattern is discussed in several papers
  \citep{2002AJ....124.3430C,2003ApJ...598.1321W,2005ApJ...619..623M}: this resonance has
  two libration centers but they do not have the same trapping
  probability and only massive planets can populate the second libration
  center.
\end{itemize}

\subsection{Low-eccentricity orbits}

Most of the structures discussed in Section
  \ref{circularOrbits} disappear as the
planet eccentricity increases (Fig.  \ref{figure_e05}). The libration
amplitude of resonant planetesimals indeed
increases, smoothing the density waves along the orbit. For eccentricities between $0.05$ and $0.1$,
the disk looks like a ring with a hole at the location of the
planet, or no longer shows structures for the lower mass
planets. The rings are not only due to the $1$:$1$
resonant planetesimals corotating with the planet, but are also
populated by other major resonances (e.g., $2$:$1$,$3$:$2$). Massive planets (last row of Fig.
\ref{figure_e05}) are less sensitive to this effect because they can
more efficiently eject planetesimals during close encounters, even if they
are in the MMRs. Only resonant planetesimals with low libration
amplitudes can survive and the disk remains structured as in the strictly circular case.

\section{Planets on eccentric orbits}
\label{highEccOrb}

\begin{figure*}
\makebox[\textwidth]{
\includegraphics[angle=-90,width=0.33\textwidth]{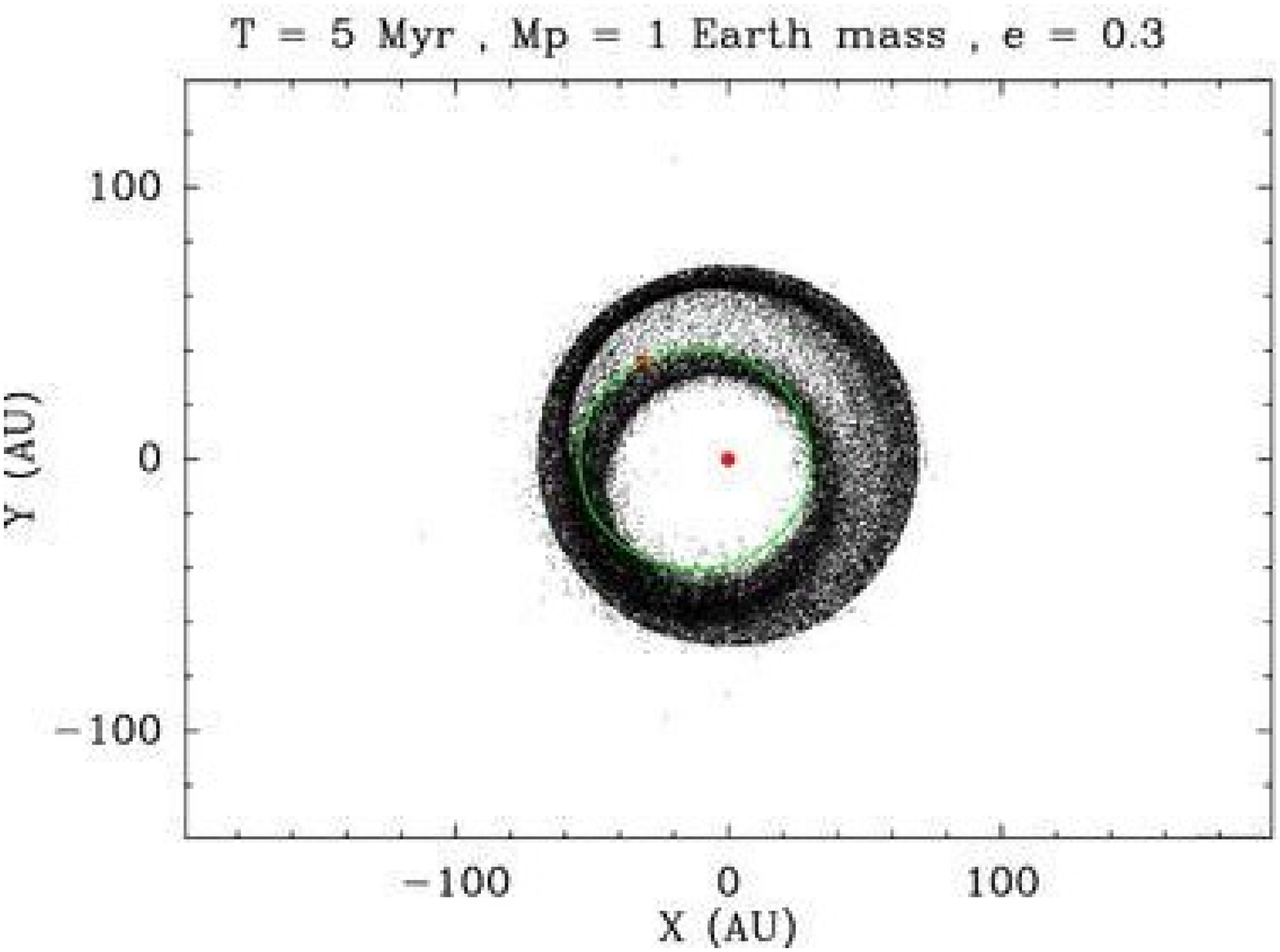} \hfil
\includegraphics[angle=-90,width=0.33\textwidth]{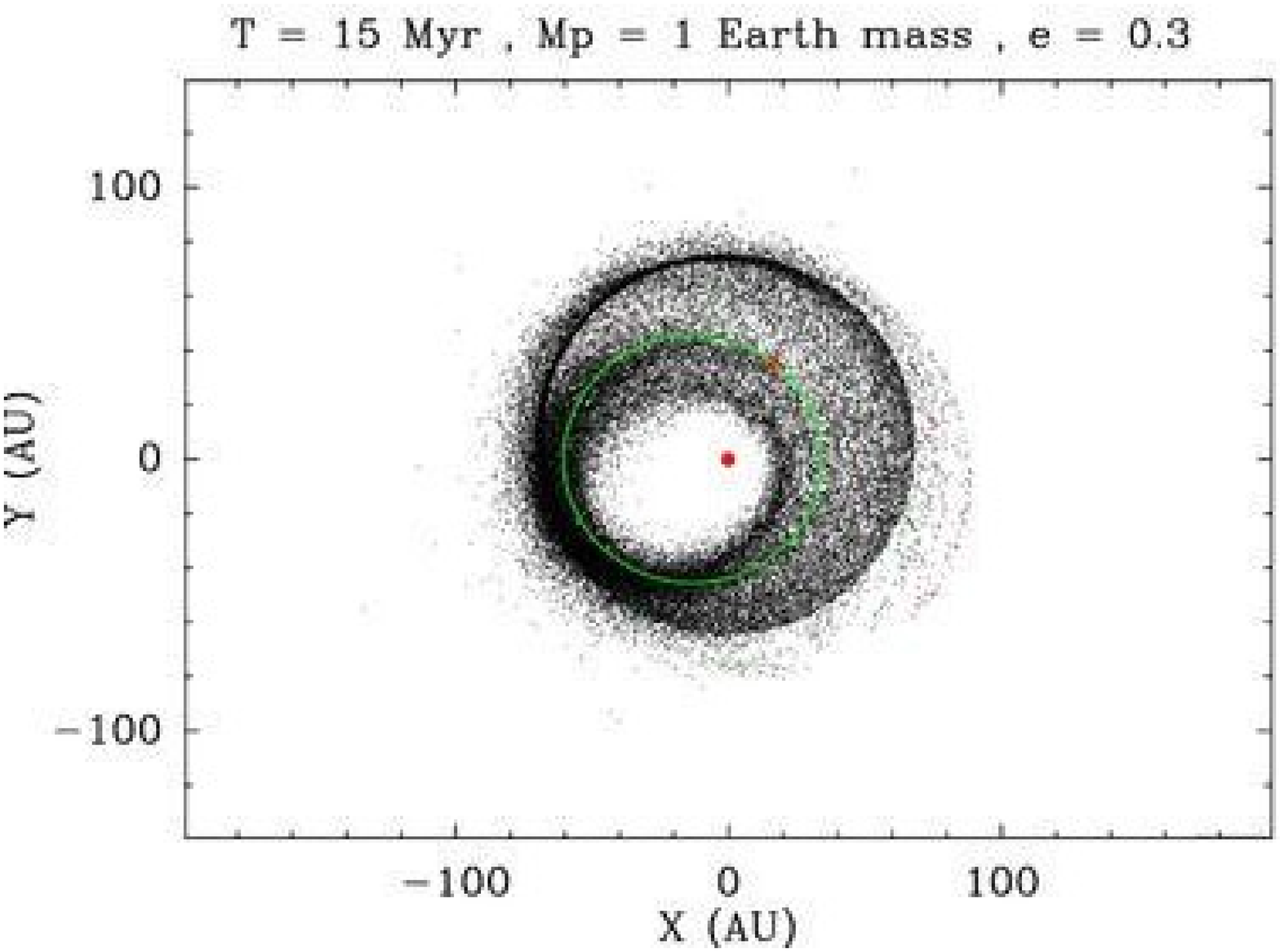} \hfil
\includegraphics[angle=-90,width=0.33\textwidth]{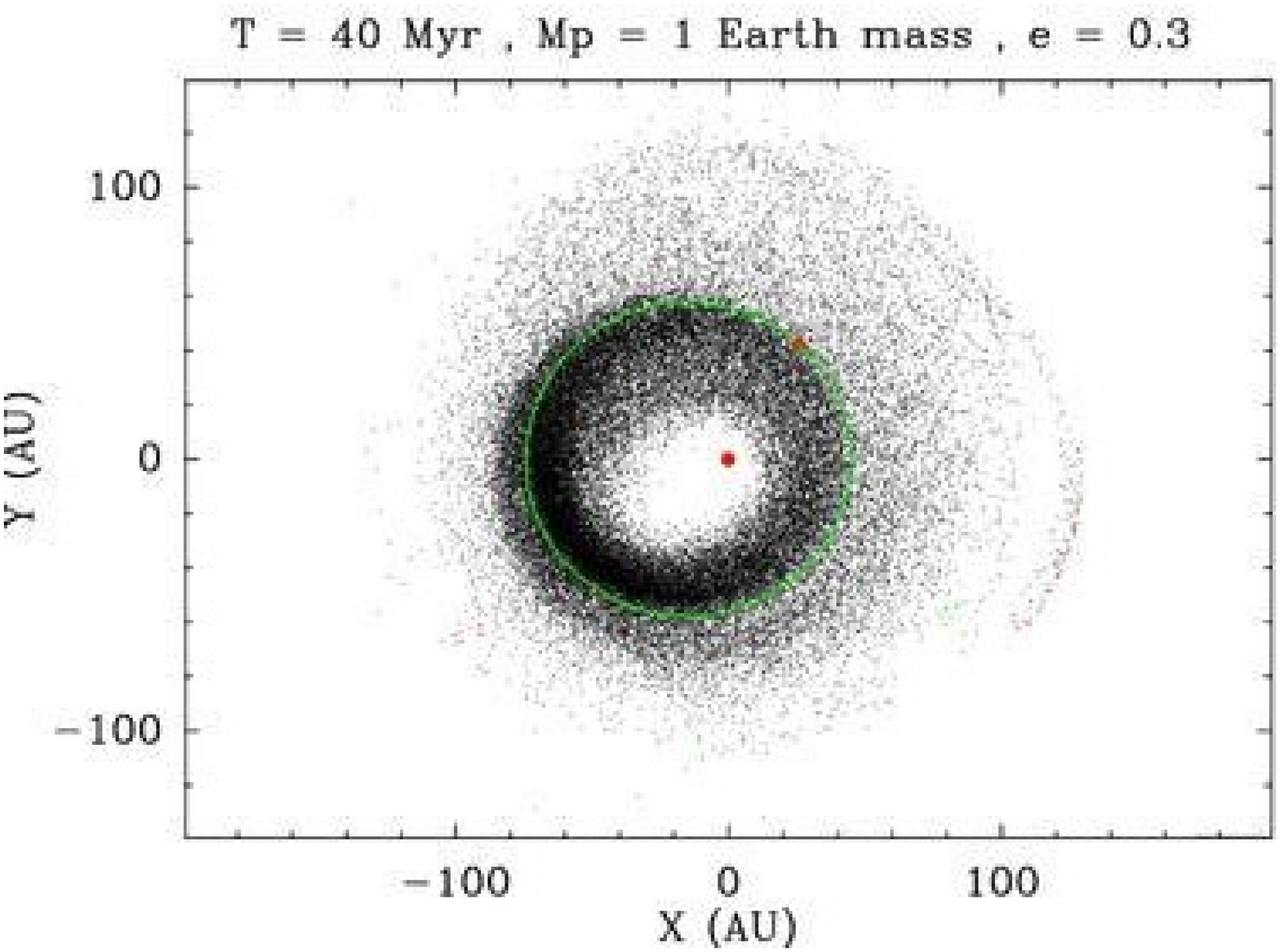}} \\
\makebox[\textwidth]{
\includegraphics[angle=-90,width=0.33\textwidth]{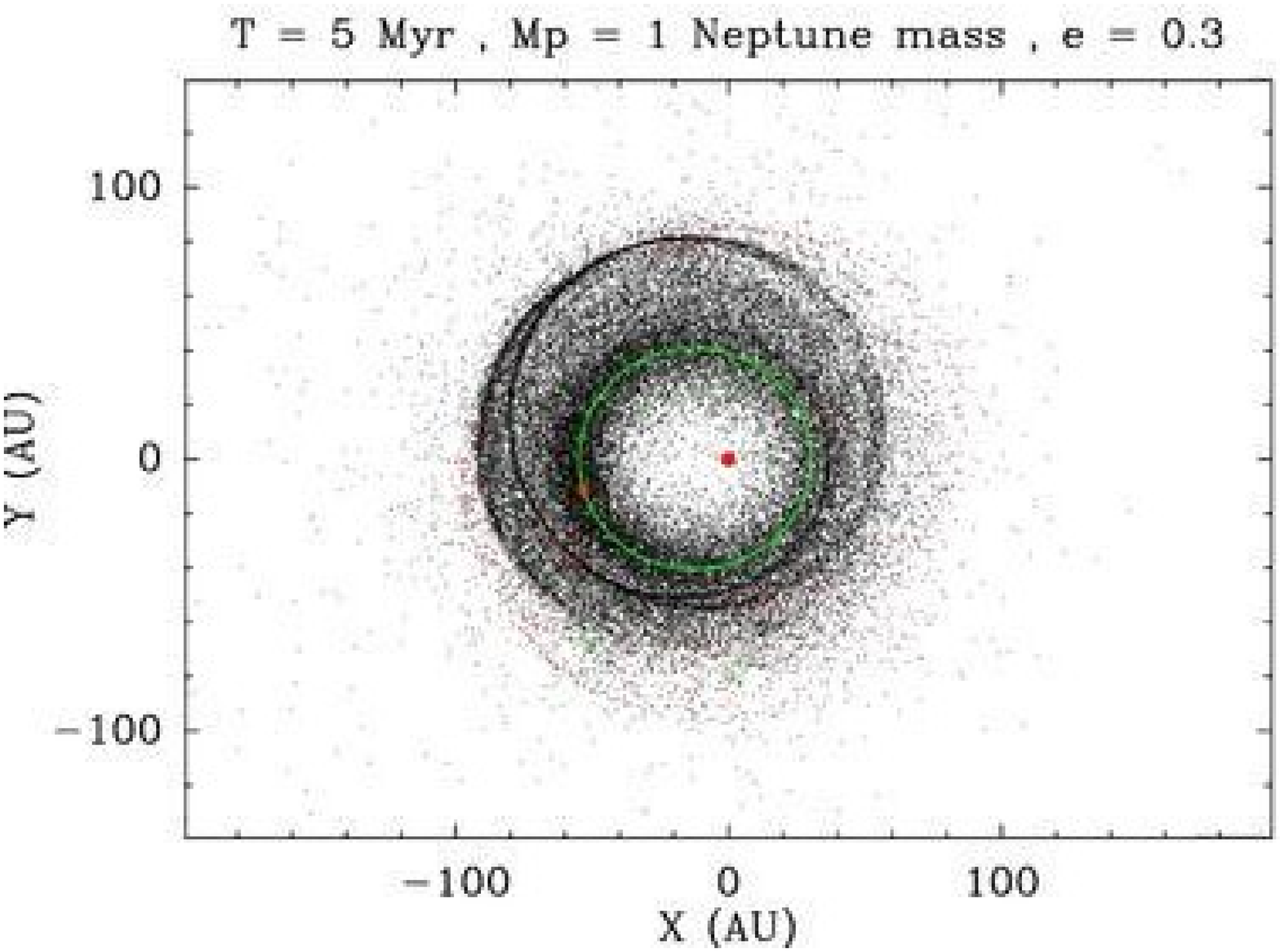} \hfil
\includegraphics[angle=-90,width=0.33\textwidth]{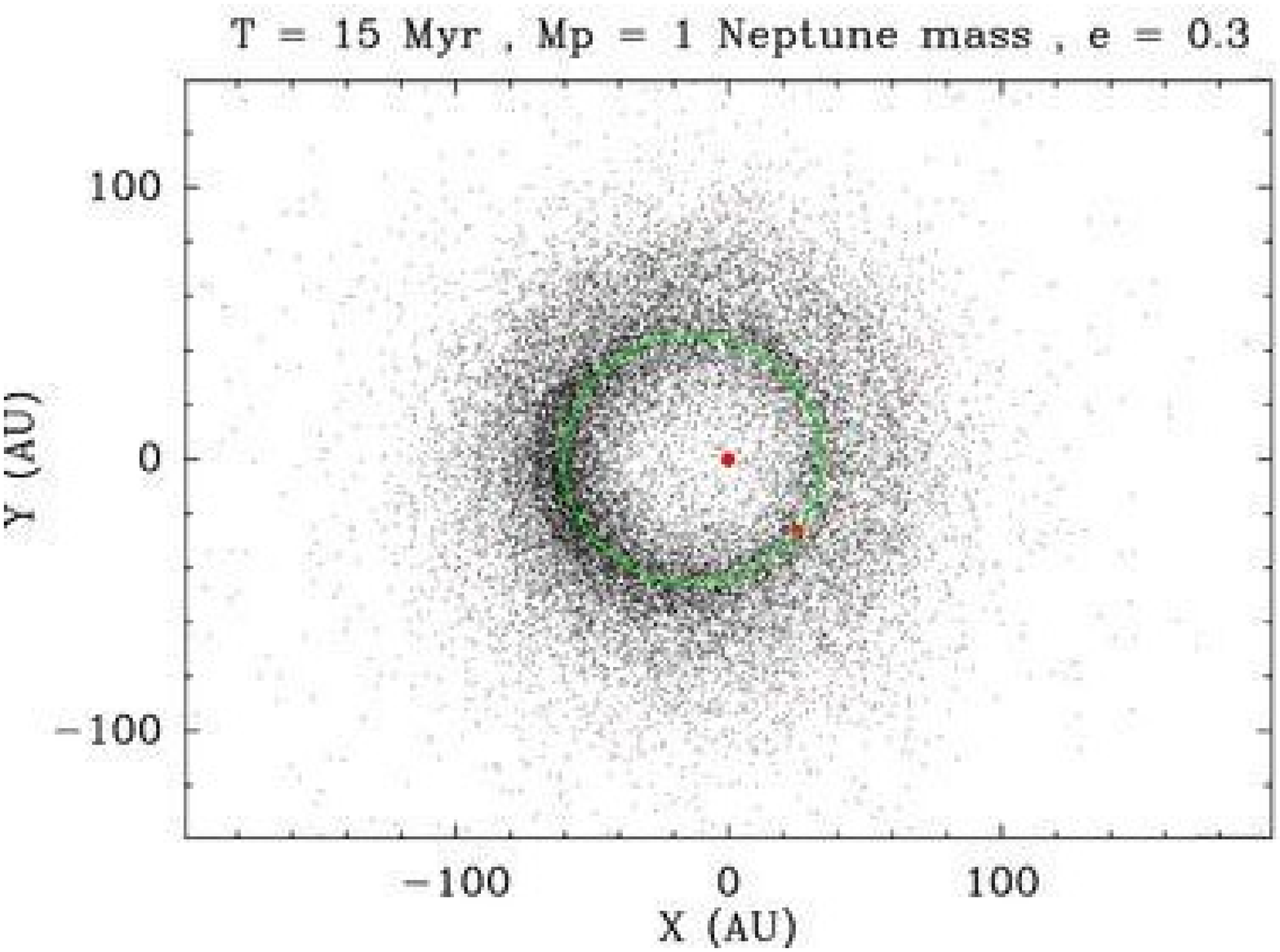} \hfil
\includegraphics[angle=-90,width=0.33\textwidth]{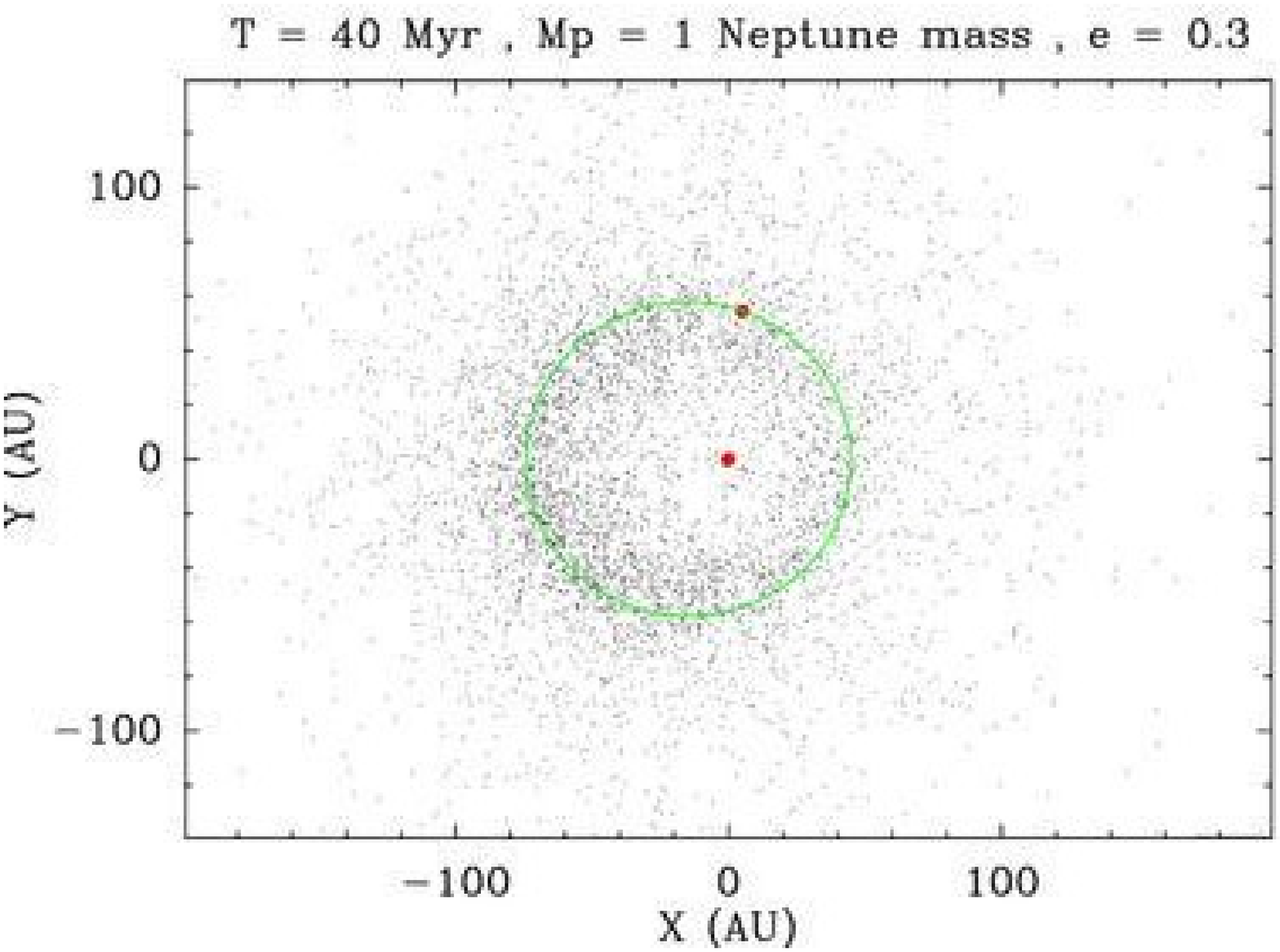}} \\
\makebox[\textwidth]{
\includegraphics[angle=-90,width=0.33\textwidth]{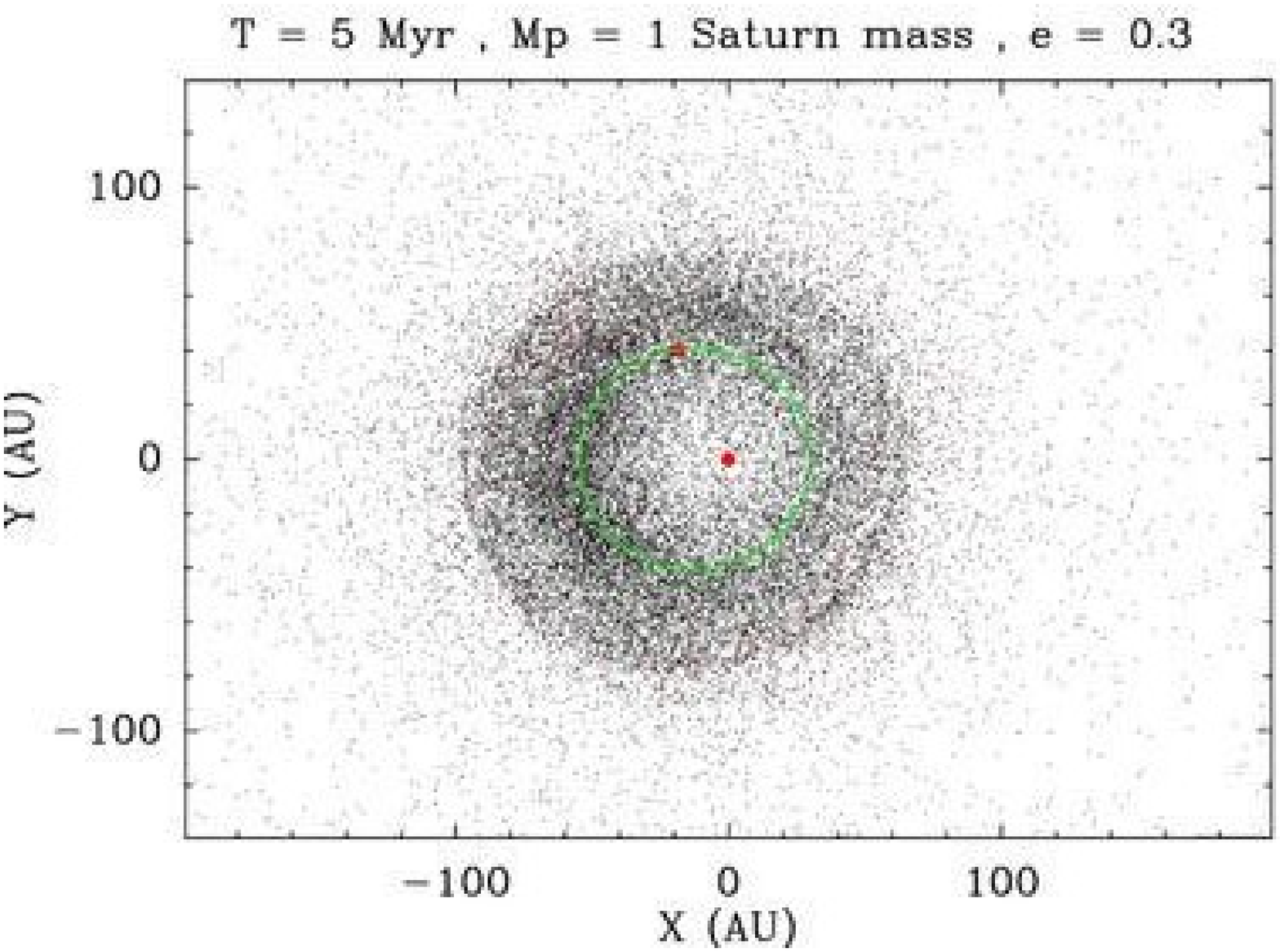} \hfil
\includegraphics[angle=-90,width=0.33\textwidth]{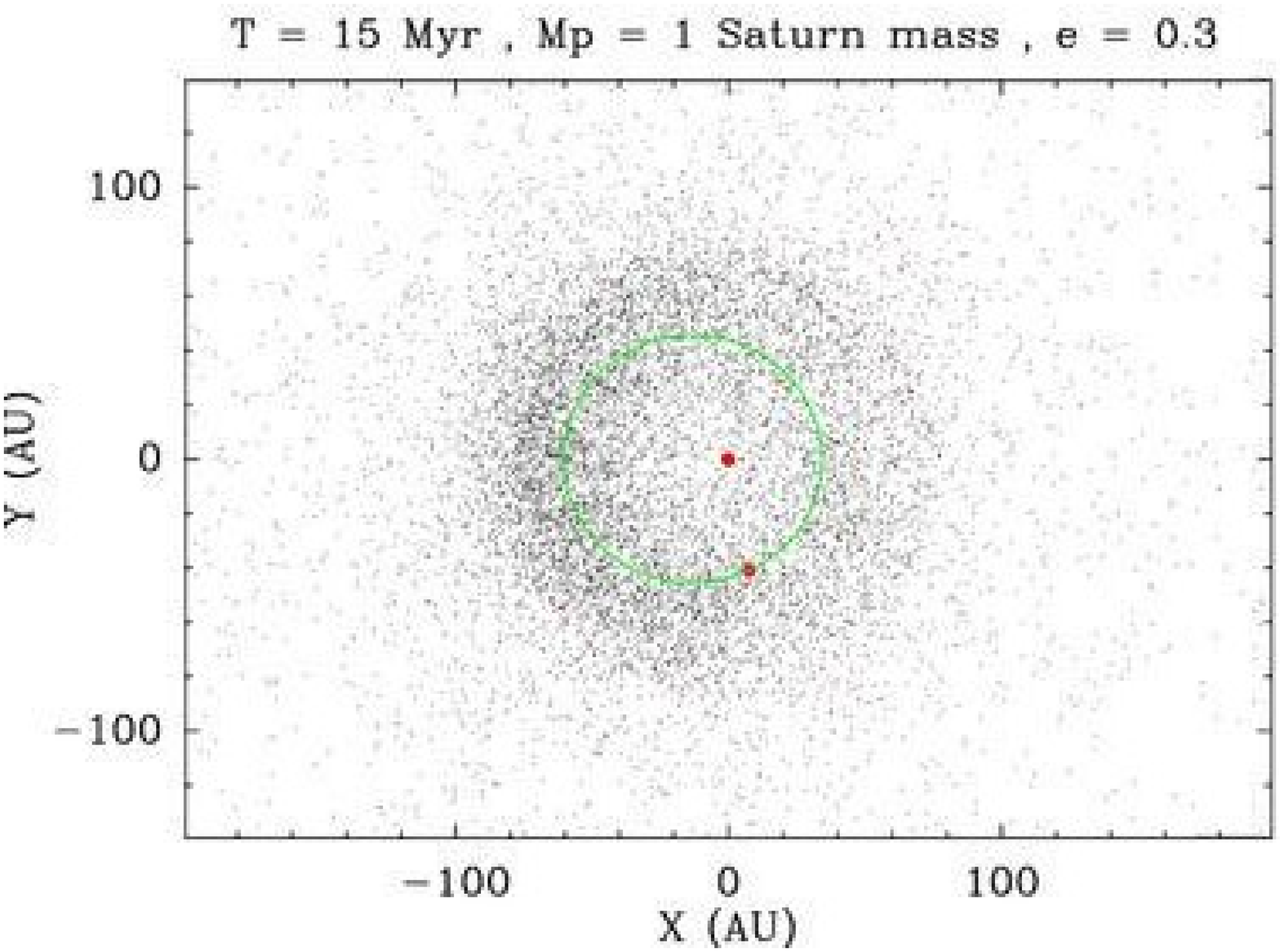} \hfil
\includegraphics[angle=-90,width=0.33\textwidth]{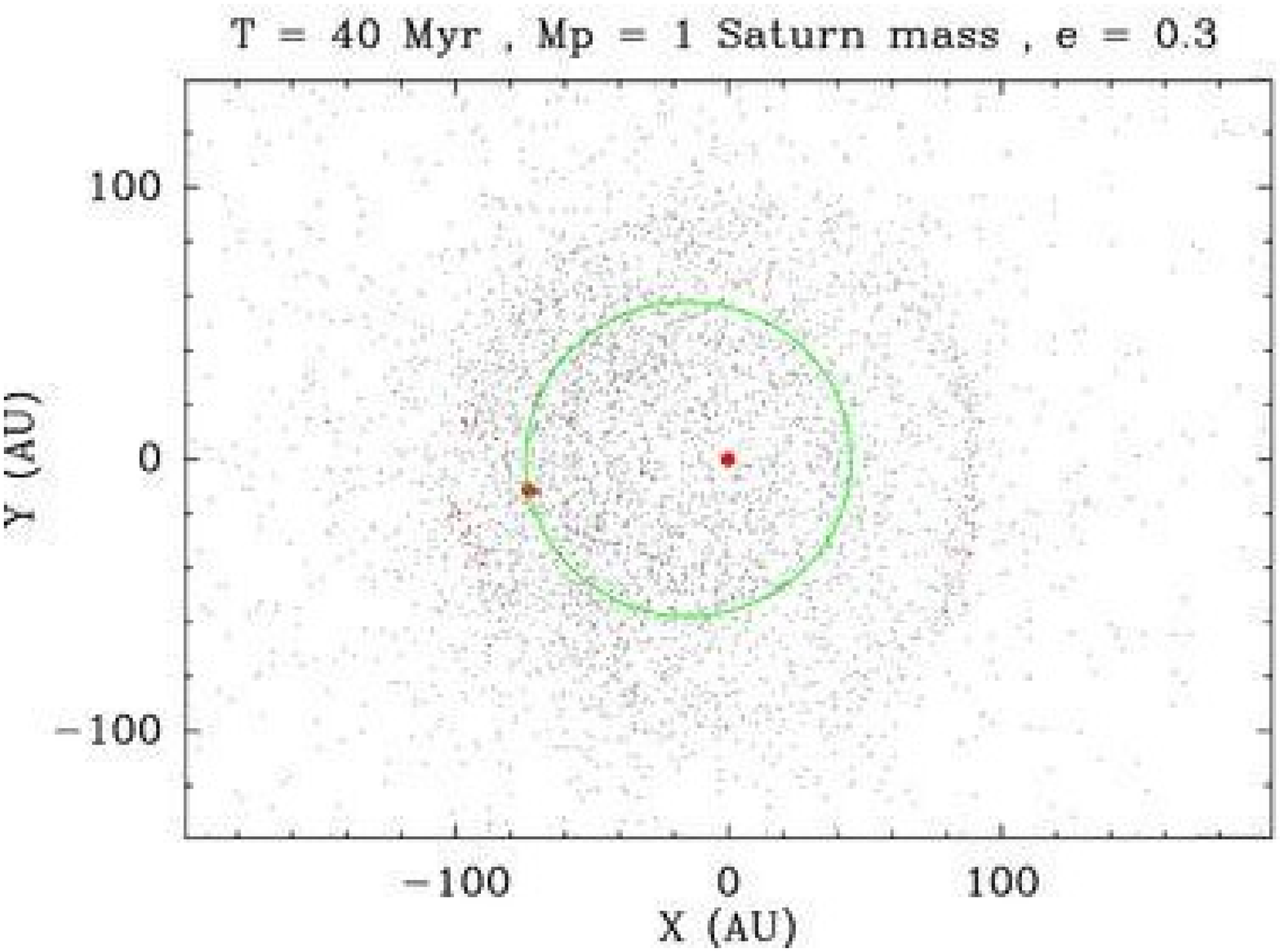}} \\
\makebox[\textwidth]{
\includegraphics[angle=-90,width=0.33\textwidth]{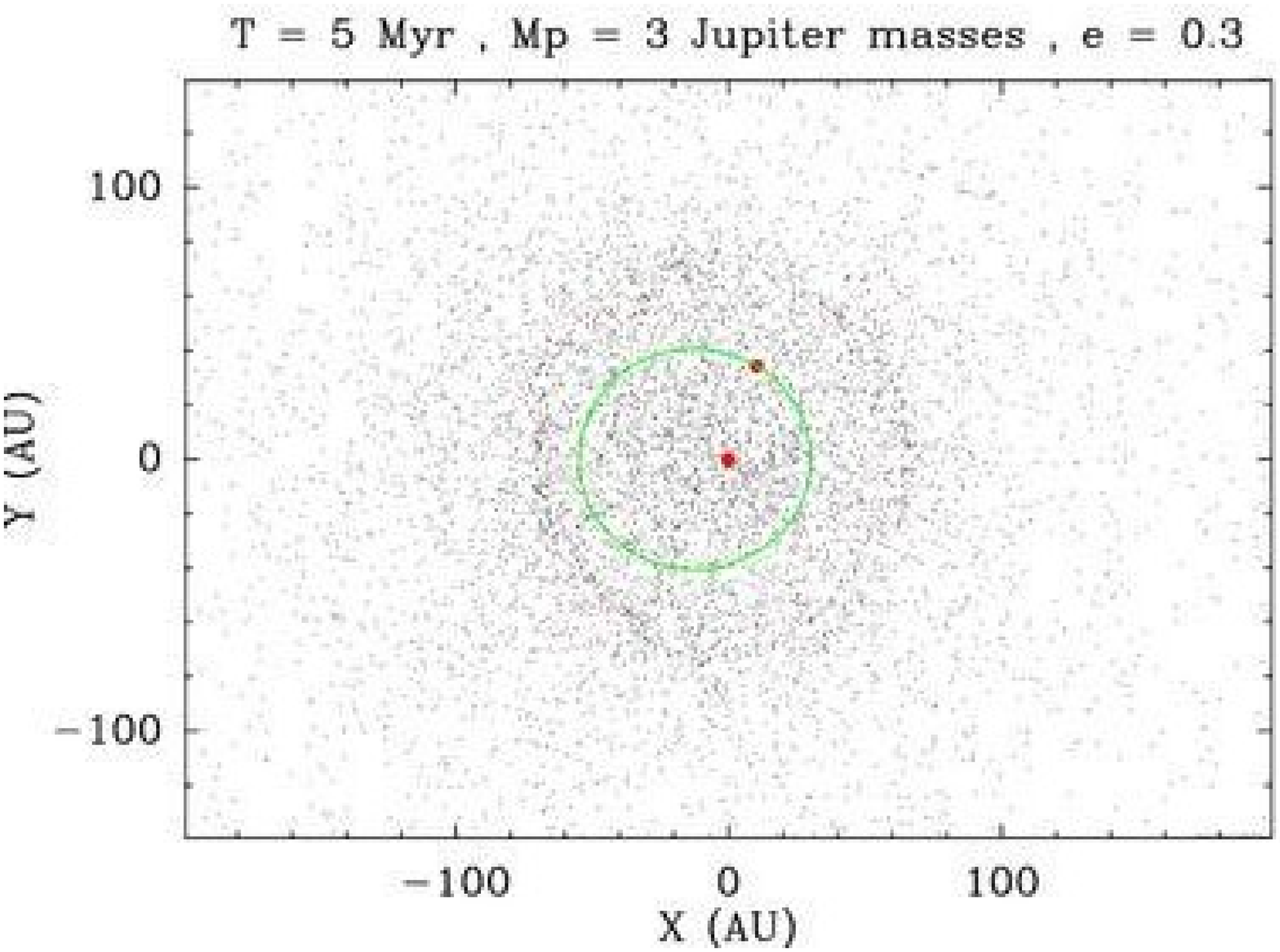} \hfil
\includegraphics[angle=-90,width=0.33\textwidth]{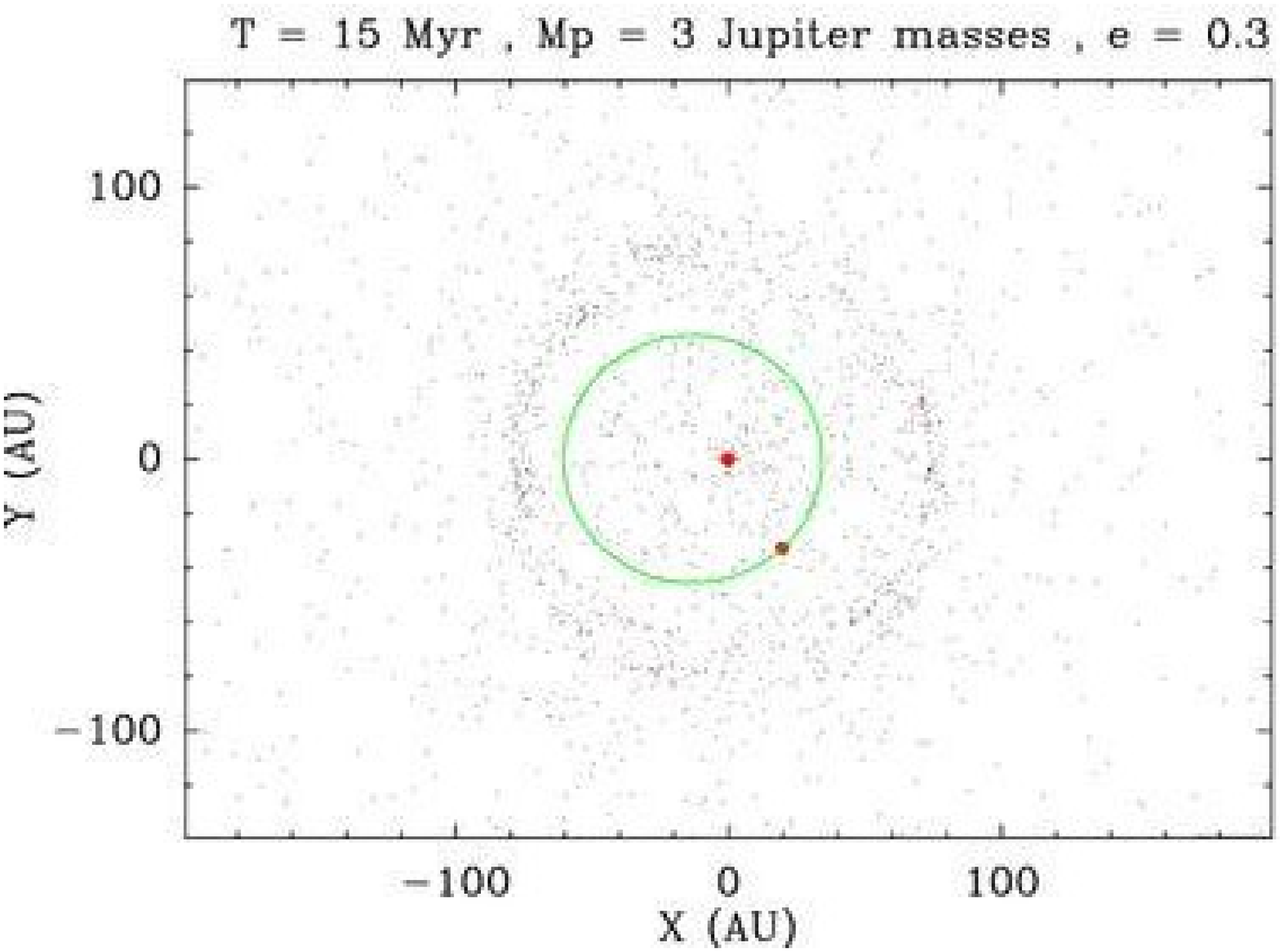} \hfil
\includegraphics[angle=-90,width=0.33\textwidth]{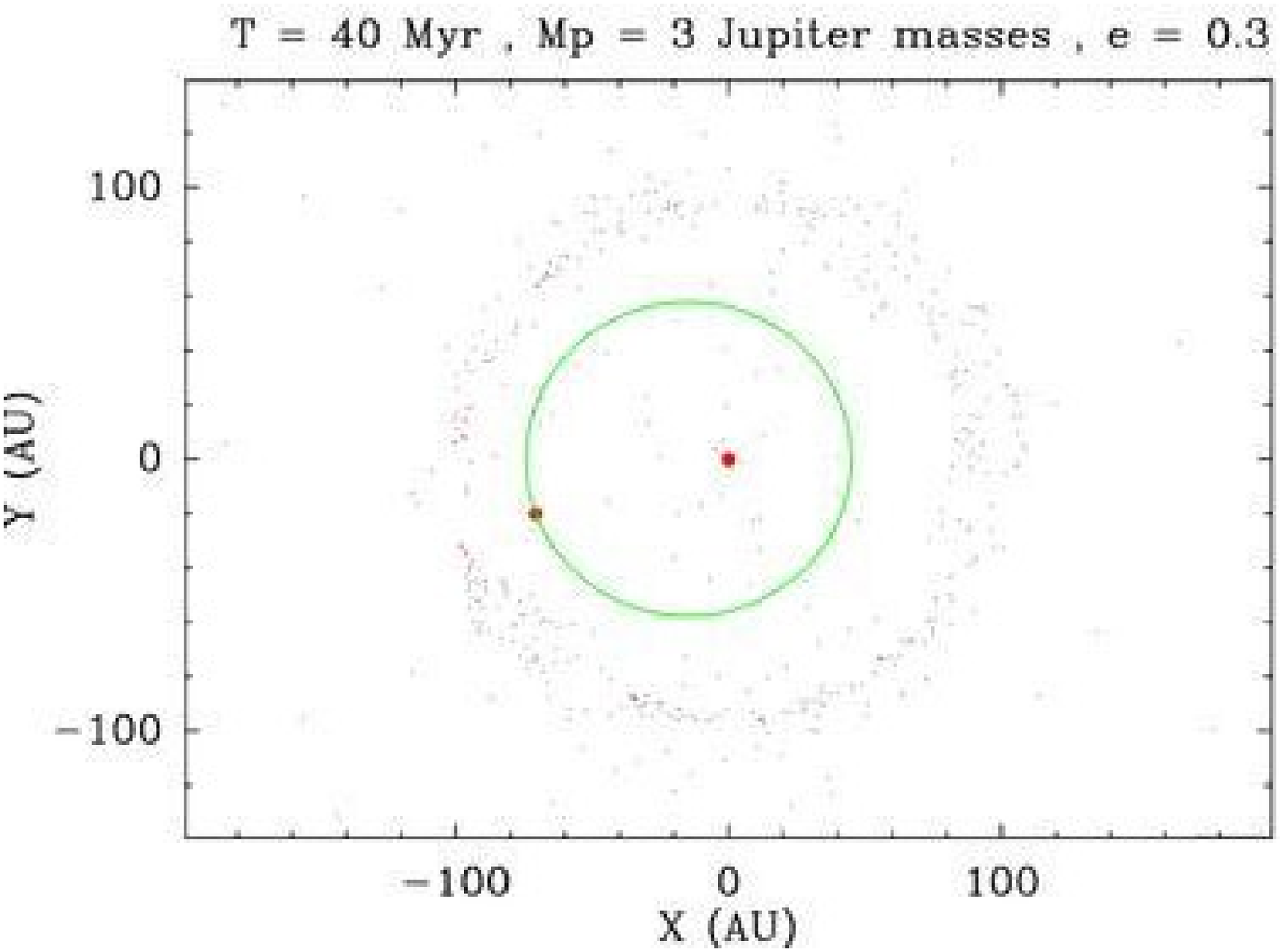}} \\
\caption{\label{figure_e3} Same as Fig. \ref{figure_e0}, for
  similar planets, but on a moderate eccentricity orbit ($e_p=0.3$). \thanks{See the electronic edition of the
    Journal for a color version of this figure.}}
\end{figure*}

\begin{figure*}
\makebox[\textwidth]{
\includegraphics[angle=-90,width=0.33\textwidth]{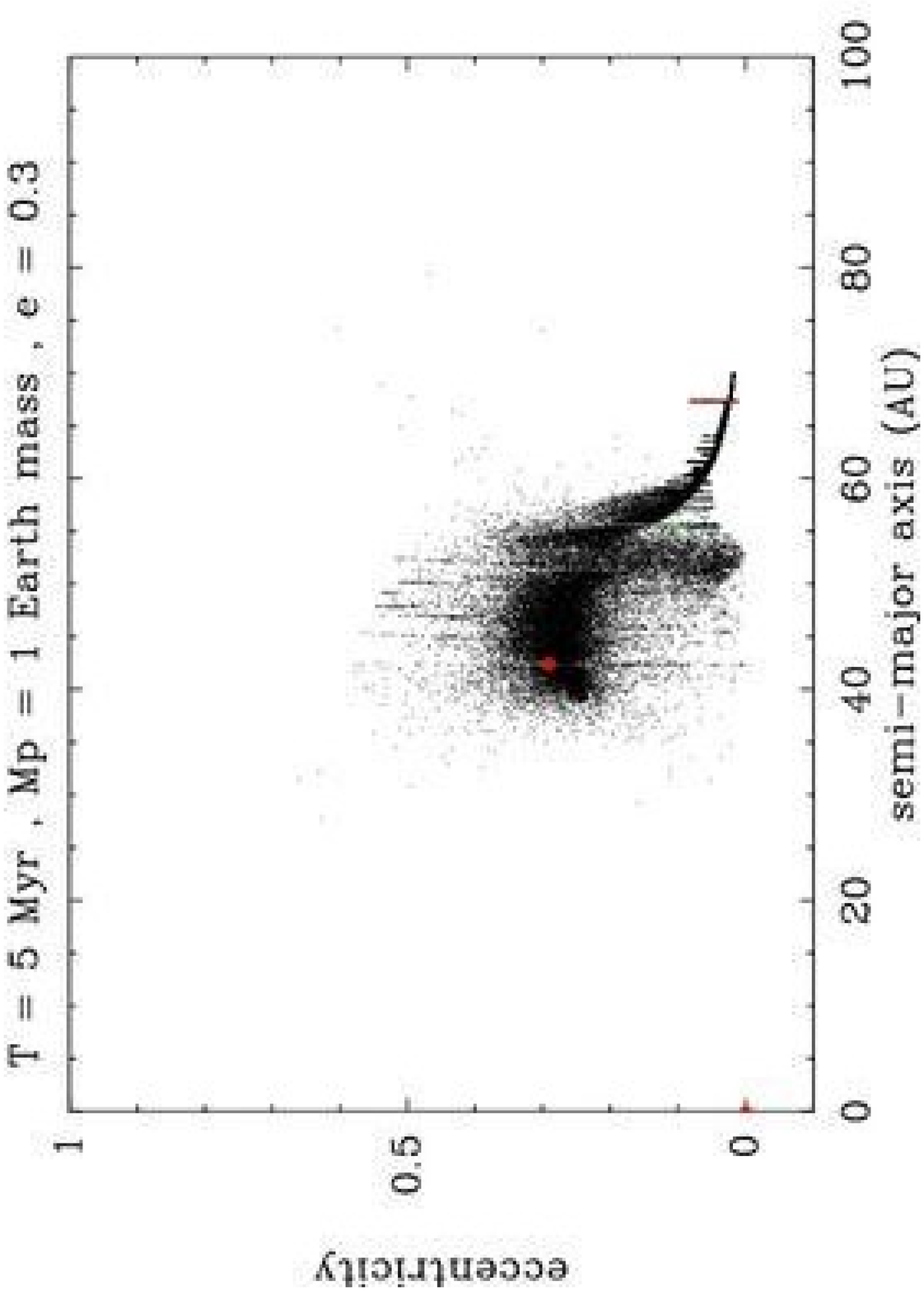} \hfil
\includegraphics[angle=-90,width=0.33\textwidth]{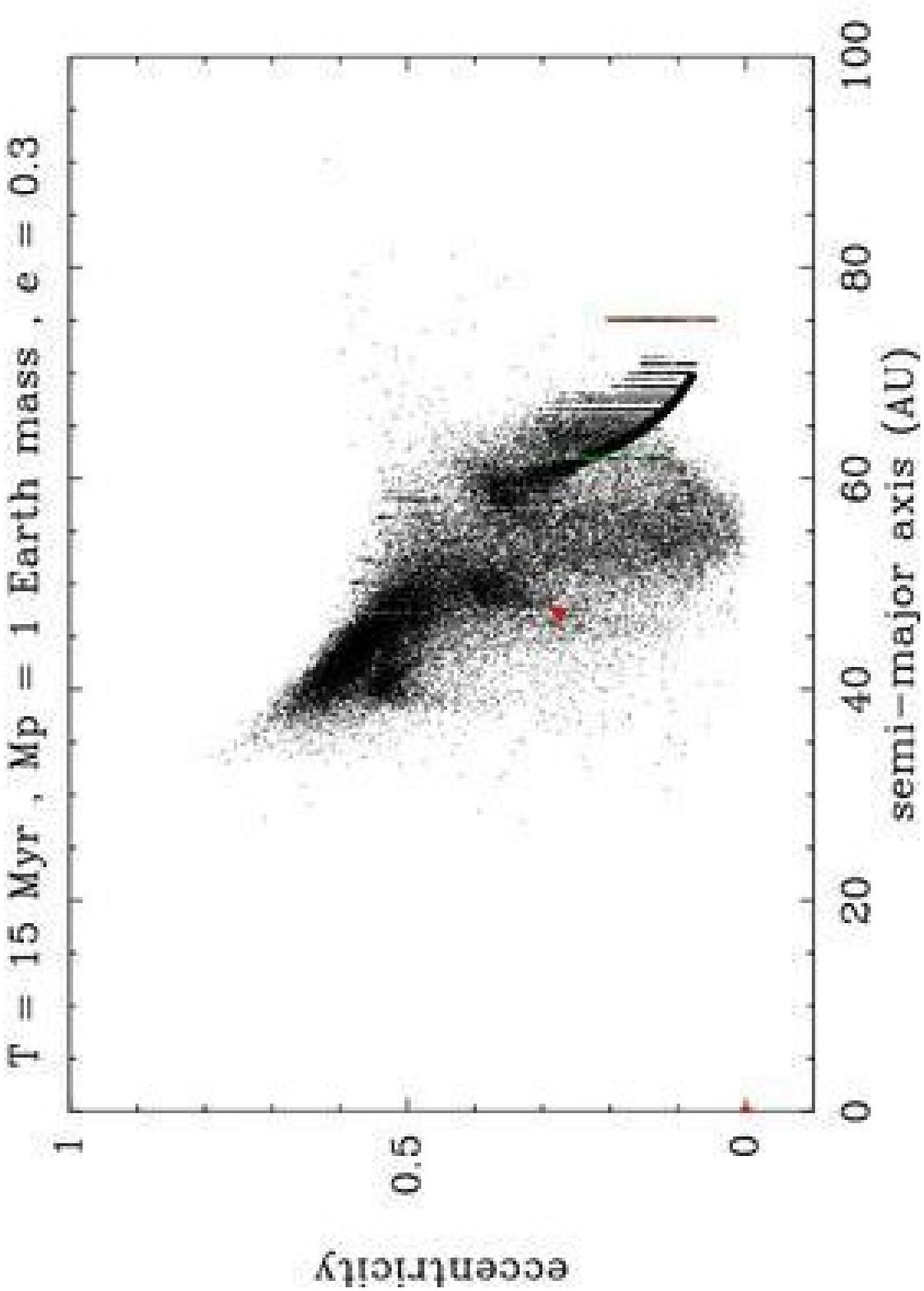} \hfil
\includegraphics[angle=-90,width=0.33\textwidth]{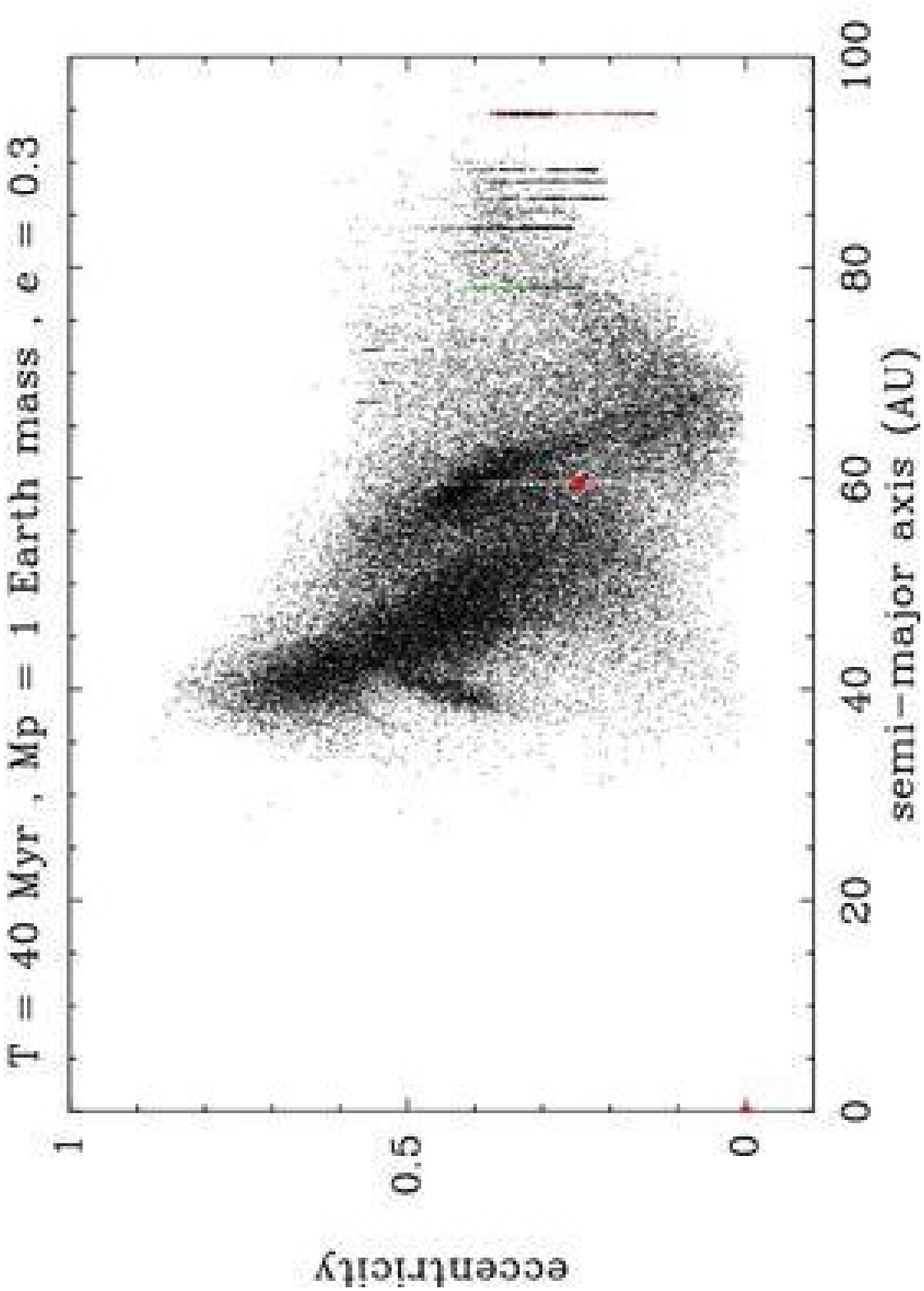}} \\
\makebox[\textwidth]{
\includegraphics[angle=-90,width=0.33\textwidth]{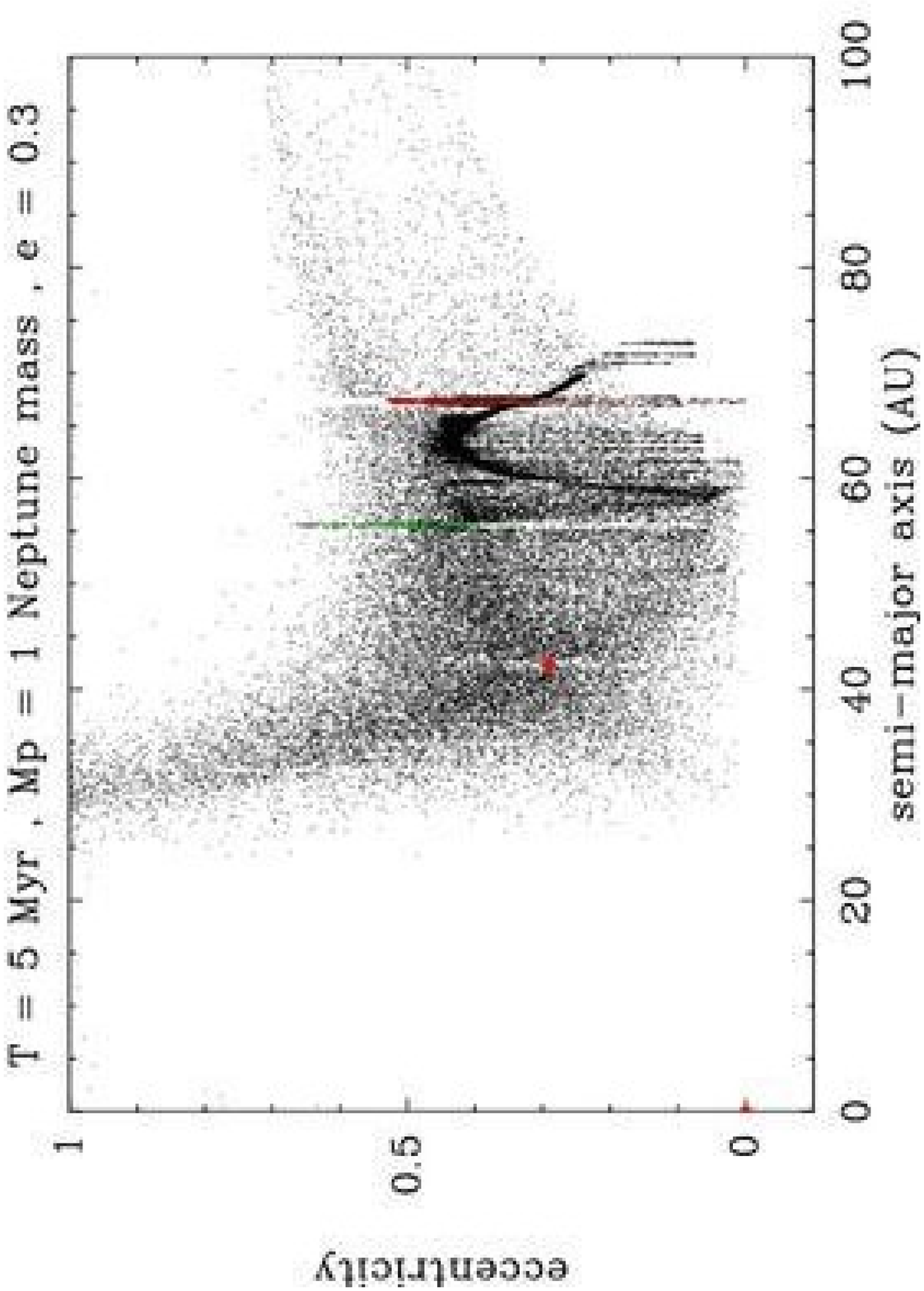} \hfil
\includegraphics[angle=-90,width=0.33\textwidth]{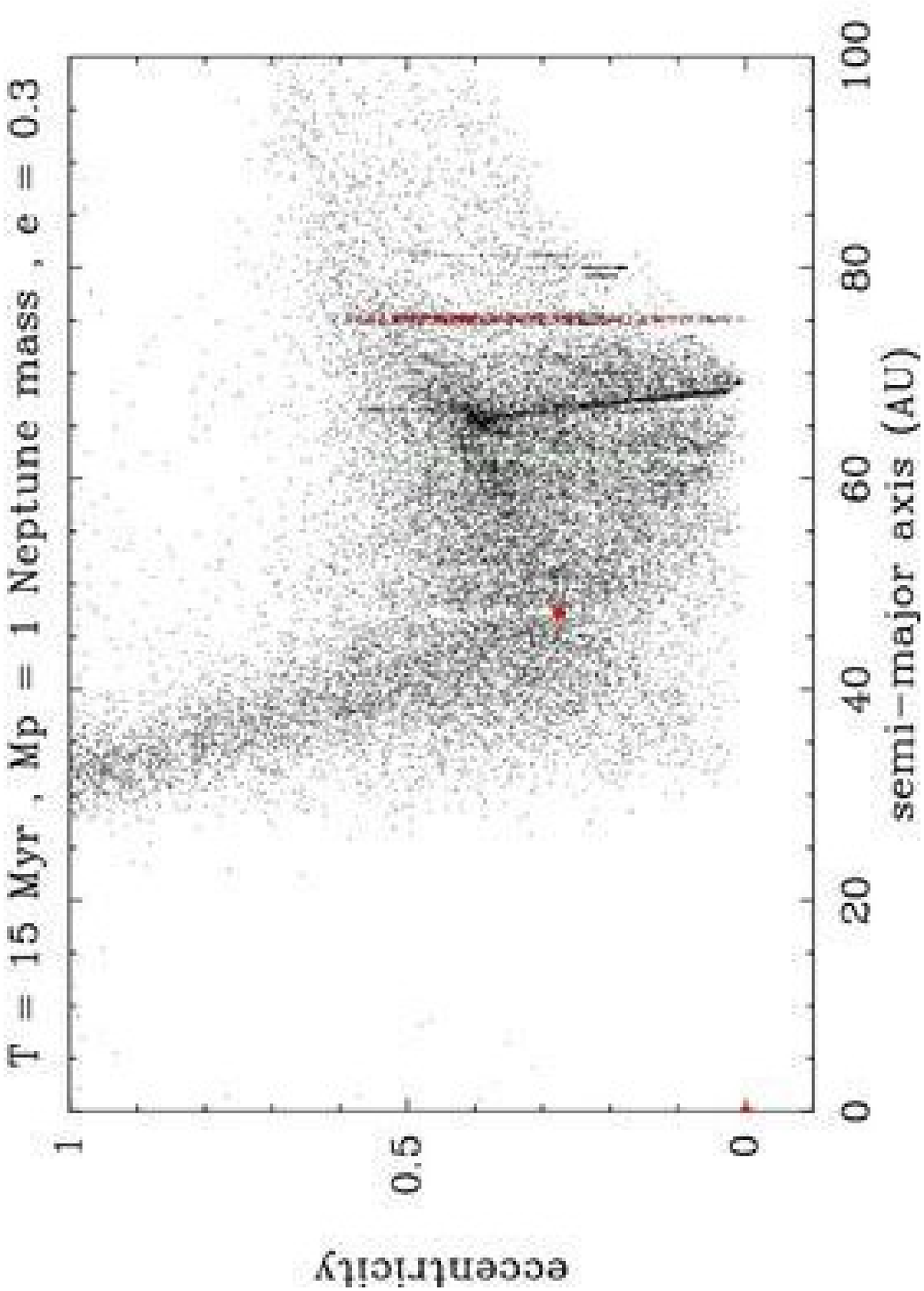} \hfil
\includegraphics[angle=-90,width=0.33\textwidth]{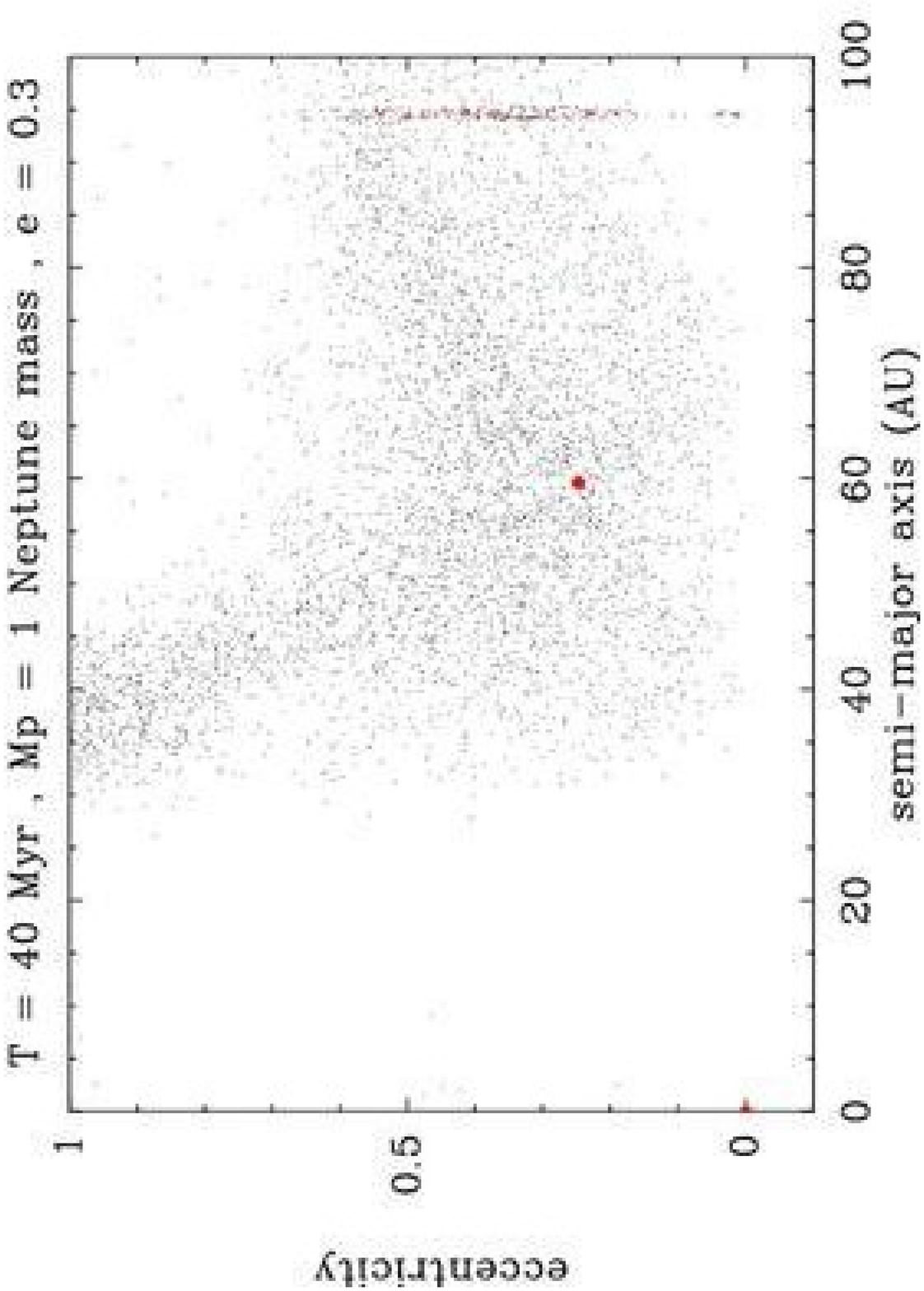}} \\
\makebox[\textwidth]{
\includegraphics[angle=-90,width=0.33\textwidth]{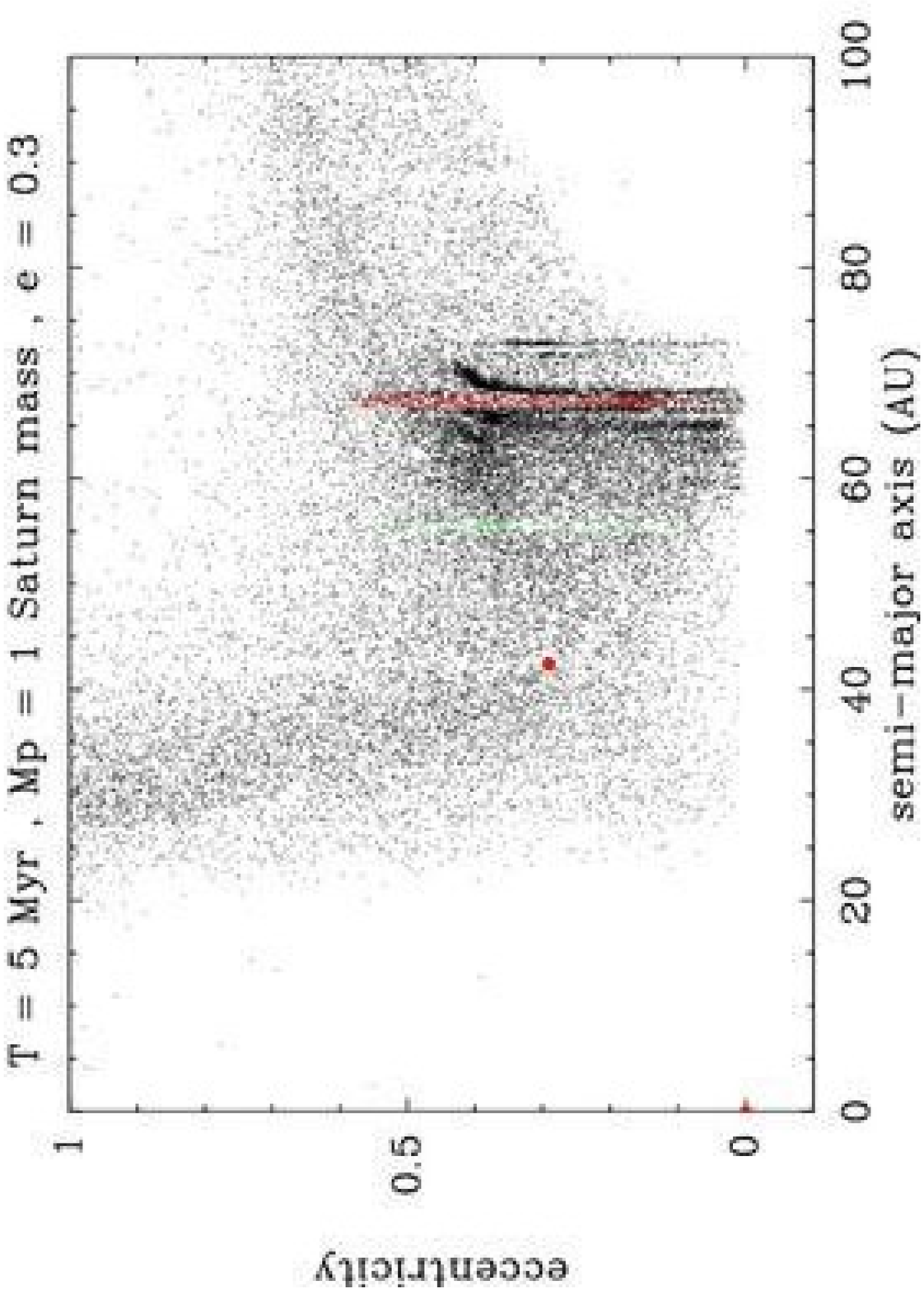} \hfil
\includegraphics[angle=-90,width=0.33\textwidth]{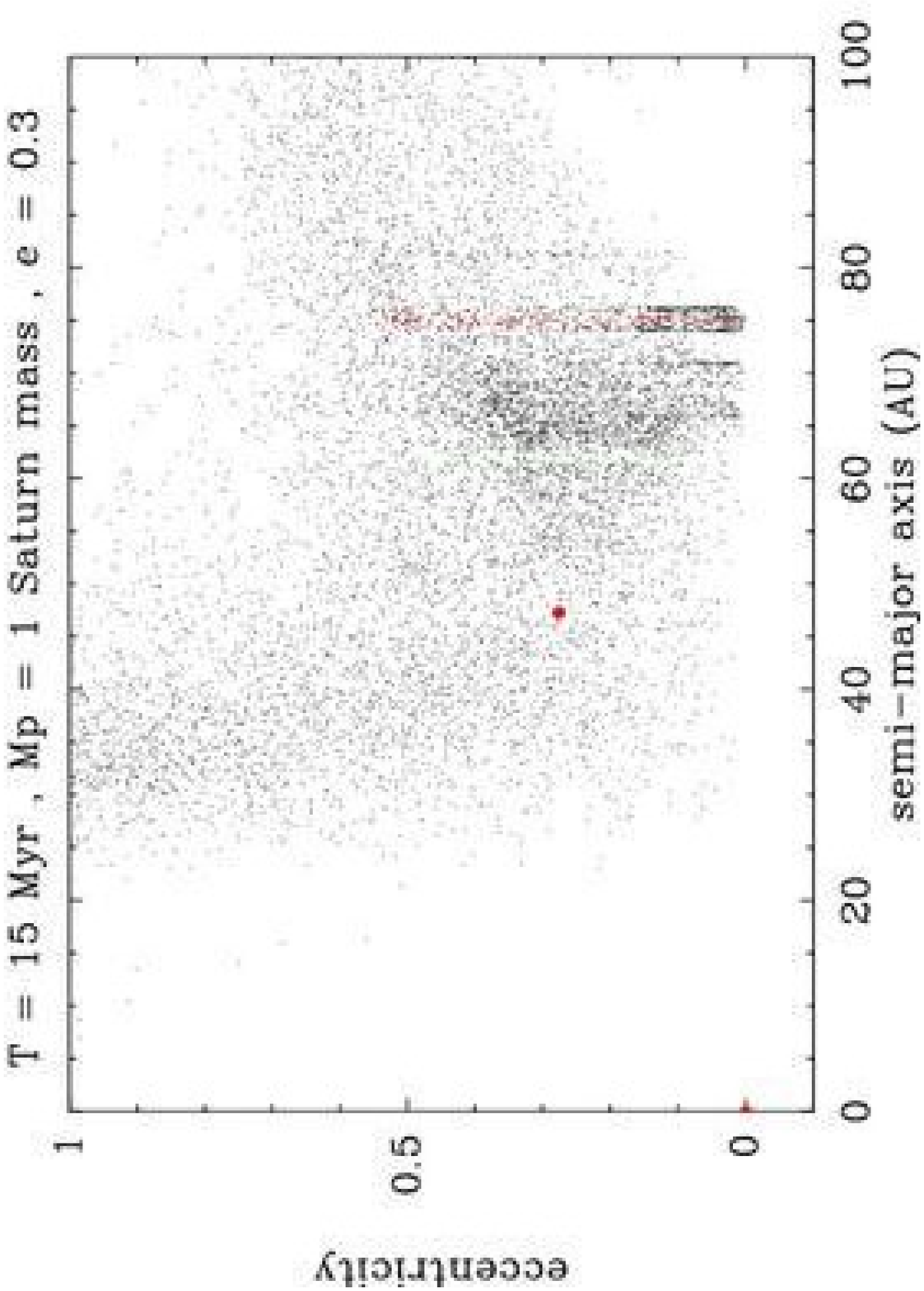} \hfil
\includegraphics[angle=-90,width=0.33\textwidth]{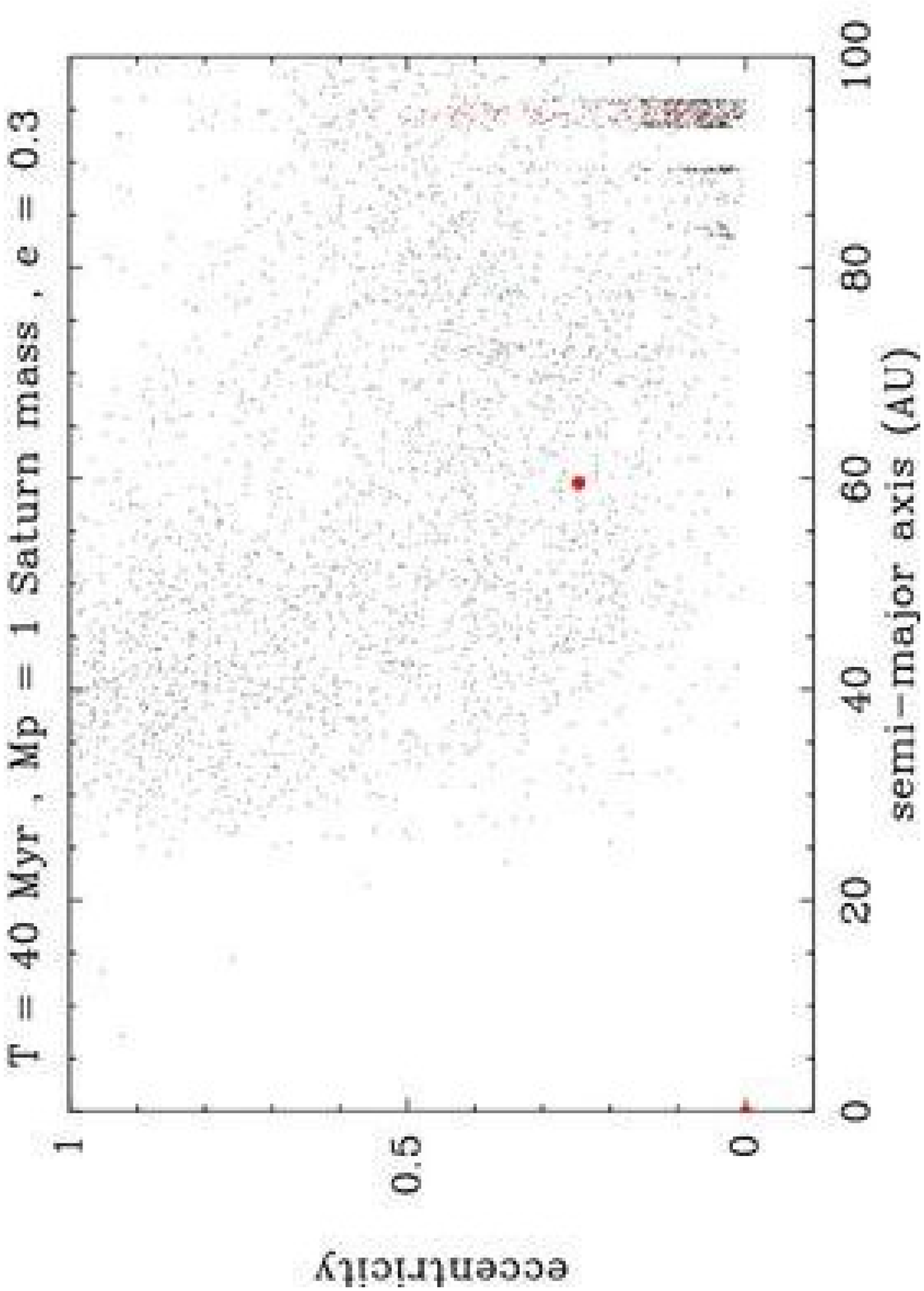}} \\
\makebox[\textwidth]{
\includegraphics[angle=-90,width=0.33\textwidth]{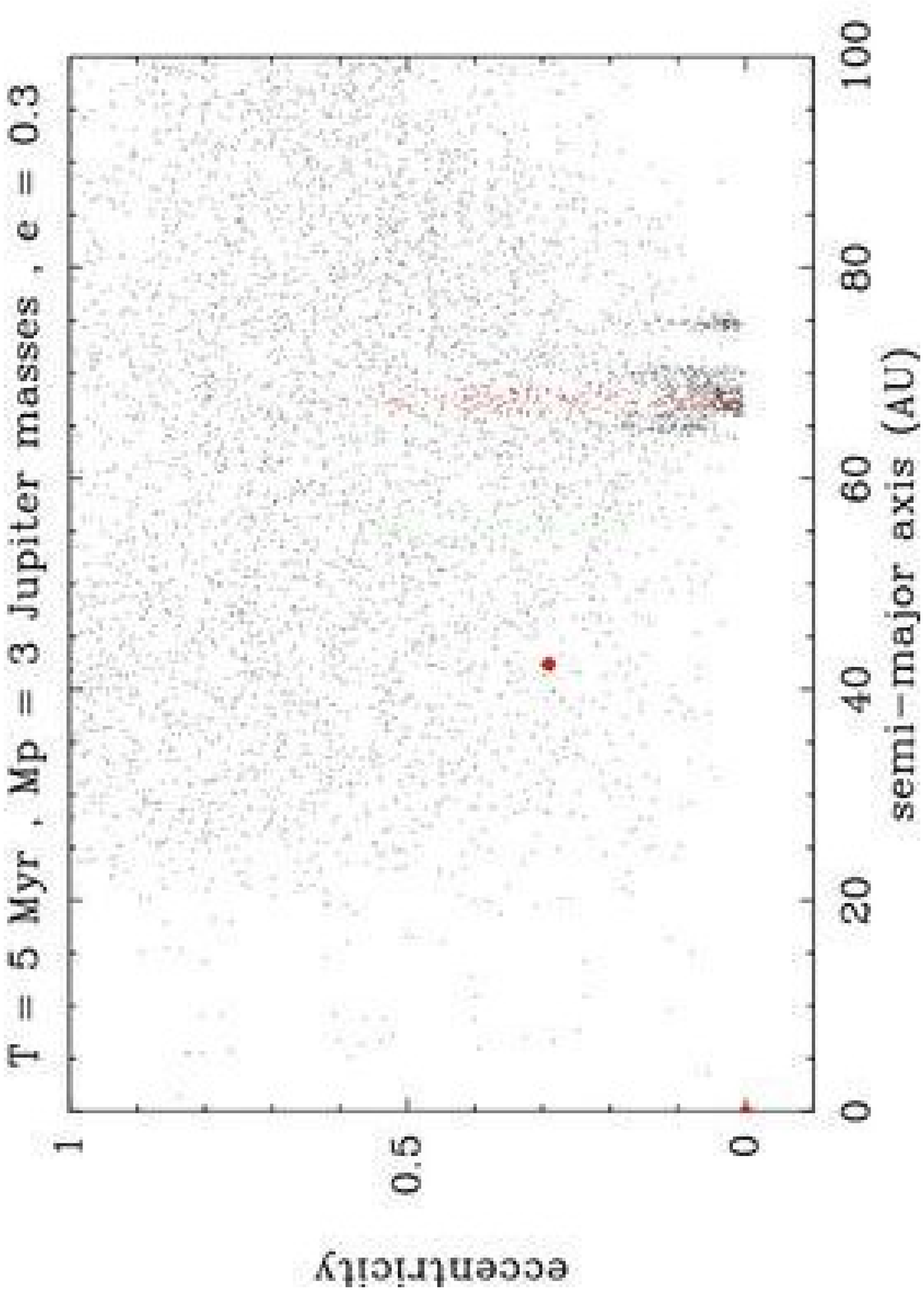} \hfil
\includegraphics[angle=-90,width=0.33\textwidth]{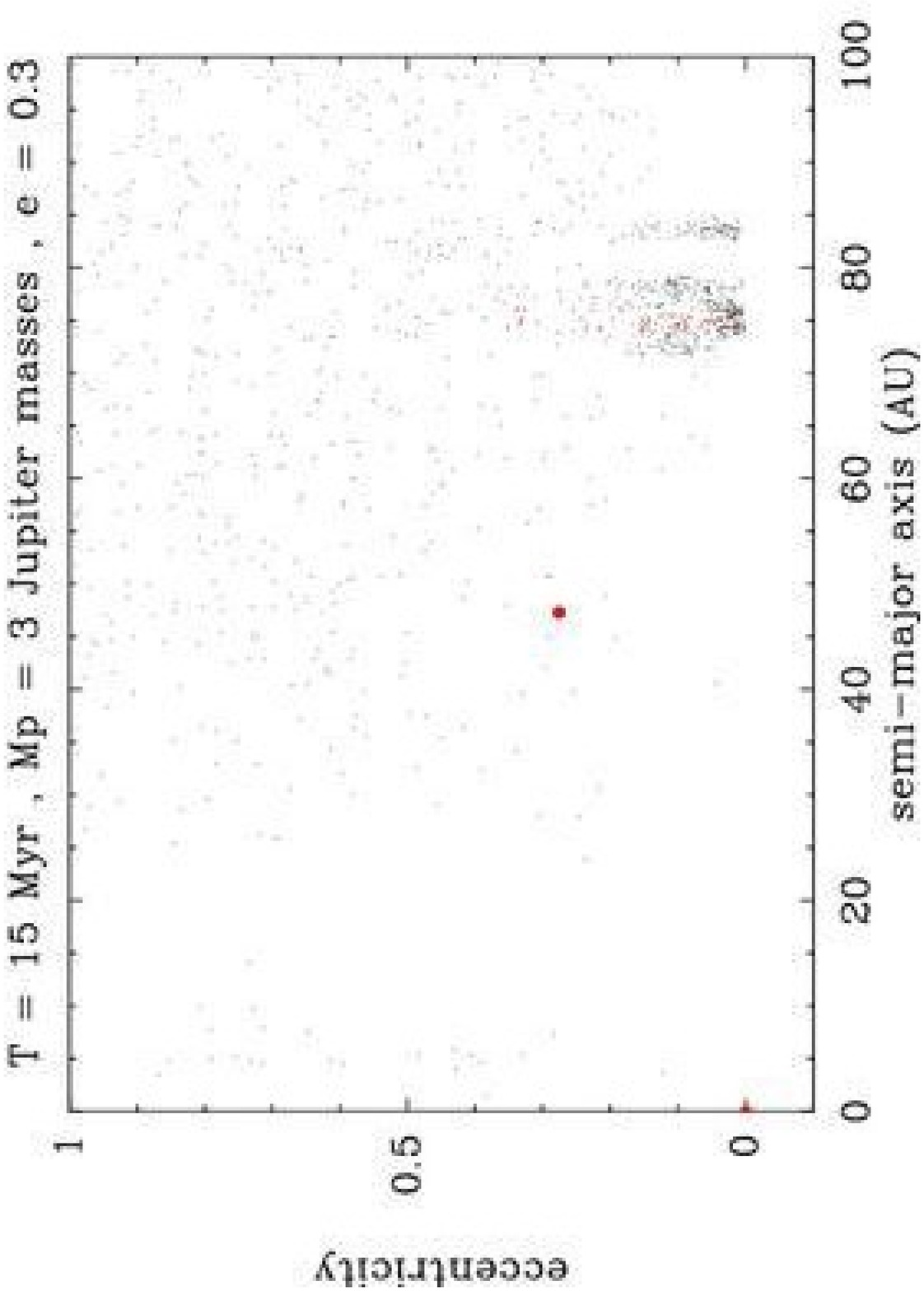} \hfil
\includegraphics[angle=-90,width=0.33\textwidth]{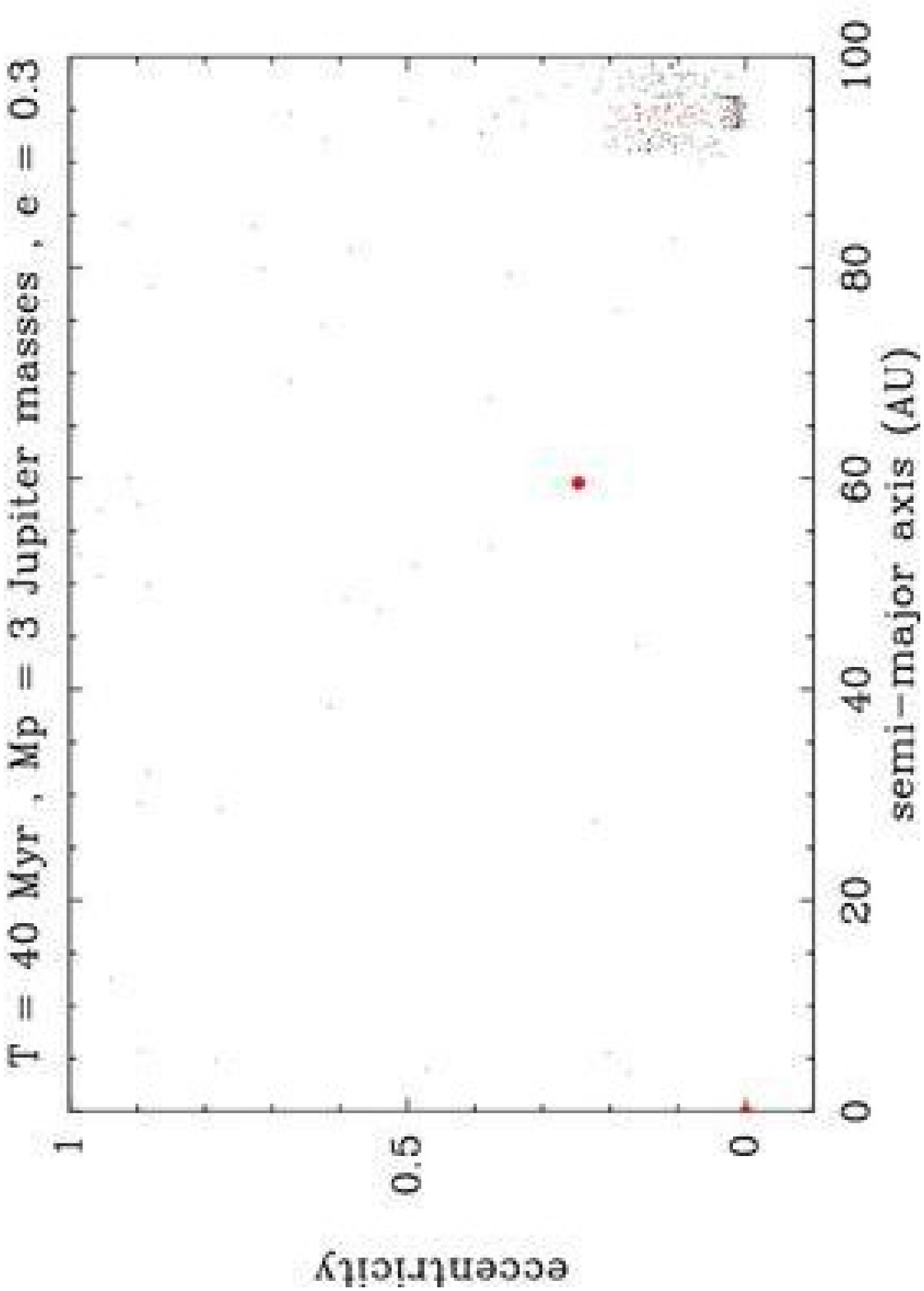}} \\
\caption{\label{figure_e3_ae} Same as Fig. \ref{figure_e0_ae}, for
  similar planets, but on a moderate eccentricity orbit ($e_p=0.3$).\thanks{See the electronic edition of the
    Journal for a color version of this figure.}}
\end{figure*}

At eccentricities larger than $0.1$, planets can significantly modify the
disk geometry and produce a dynamically warm disk where all
planetesimals may reach high eccentricities. Analytical developments are
more complex, but  are still feasible as shown by
\citet{2003ApJ...588.1110K} for resonant planetesimals and by
\citet{1999ApJ...527..918W} for non-resonant planetesimals. So far, only the P-R drag migration
scenario has been  tested for such planetary
orbits. We explore here the
planet migration scenario in the case of eccentric orbits (Fig. \ref{figure_e3} and \ref{figure_e3_ae}).

According to previous works, three phenomena should change in our simulations
with respect to the low-eccentricity orbit case. First,
close encounters between the planet and the planetesimals
are more frequent, increasing the depletion rate of the disk. As the
probability of ejection increases with the planetary mass, the
more massive the planet, the more depleted the disk. Second, even the non-resonant planetesimals see
their eccentricities rise significantly due to the gravitational perturbation of
the planet. Finally, the trapping probability is also modified
  \citep{2006MNRAS.365.1367Q}: increasing the planetary eccentricity
  decreases the
  trapping probability of the first order resonances but can increase
  it for higher order resonances. With a planet on an eccentric orbit, the resonant planetesimals
are also not well protected against close encounters, which limits even
more the number of planetesimals in MMRs. The migration thus cannot populate the MMRs enough to
generate detectable patterns: for structures
generated by a planet on a  moderate or high eccentricity orbit, the
MMRs, and thus the migration of the planet, is no longer an important
factor whatever the planet mass as the non-resonant dynamics dominates
the shape of the disk. 

The depletion rate is so efficient that  
  the disk is almost entirely depleted during our $40$ Myr simulations, except in the
 case of an Earth mass or Neptune mass planet on moderate eccentric orbits
(below $0.2$ or $0.3$). An Earth mass planet on a more eccentric orbit
(above $0.5$) can also produce transient collective
  non-resonant effects, spatially fixed, in the azimuthal distribution of the planetesimals
 (Fig. \ref{nonResonantStructure}) before the disk is depleted, as
 explained below. 

\subsection{Non-resonant structures}

The theoretical background for the dynamics of the planetesimals in our
simulations is the restricted three-body problem, i.e., a problem
where a mass-less test particle orbiting a star is perturbed by a
planet orbiting the star on an unperturbed Keplerian orbit.
We restrict ourselves to the planar case for simplicity.
In this framework,
the Hamiltonian of the problem is \citep[see, for instance][]{1993Icar..102..316M}
\begin{equation}
\mathcal{H}_0
=-\frac{\mathcal{G}M_{\ast}}{2a}-\mathcal{G}m_p\left(\frac{1}{\left|\vec{r}-\vec{r_p}\right|}
-\frac{\vec{r}\cdot\vec{r_p}}{r_p^3}\right)\qquad,
\end{equation}
where $a$ is the osculating semi-major axis of the orbit of the
planetesimal, $M_{\ast}$ is the mass of the star, $m_p$ that of the planet,
$\mathcal{G}$ is the constant of gravitation, and $\vec{r}$ 
and $\vec{r_p}$ are the heliocentric position vectors of the 
planetesimal and of the planet respectively.
If the test particle is not locked into a mean-motion
resonance with the planet, its secular motion is investigated
by performing a double temporal averaging of $\mathcal{H}_0$ over the orbital
motions of the planet and of the planetesimal \citep{2006A&A...446..137B}. In this context, the
semi-major axis $a$ is a secular constant, as it is canonically
coupled with the mean anomaly that has been removed from the
Hamiltonian by the averaging process. The secular Hamiltonian
of the planar problem turns out to have only one degree of freedom.
It depends for instance only on the eccentricity of the planetesimal $e$ and
of its longitude of periastron $\nu$ with respect to that of the
planet. For any given fixed values for $a$ and for the planet
eccentricity $e_p$, we can draw a phase portrait (i.e., level curves
of Hamiltonian) of the dynamics in an $(\nu, e)$ plane.

\begin{figure*}
\centering
\makebox[12truecm]{
\includegraphics[width=5.9truecm]{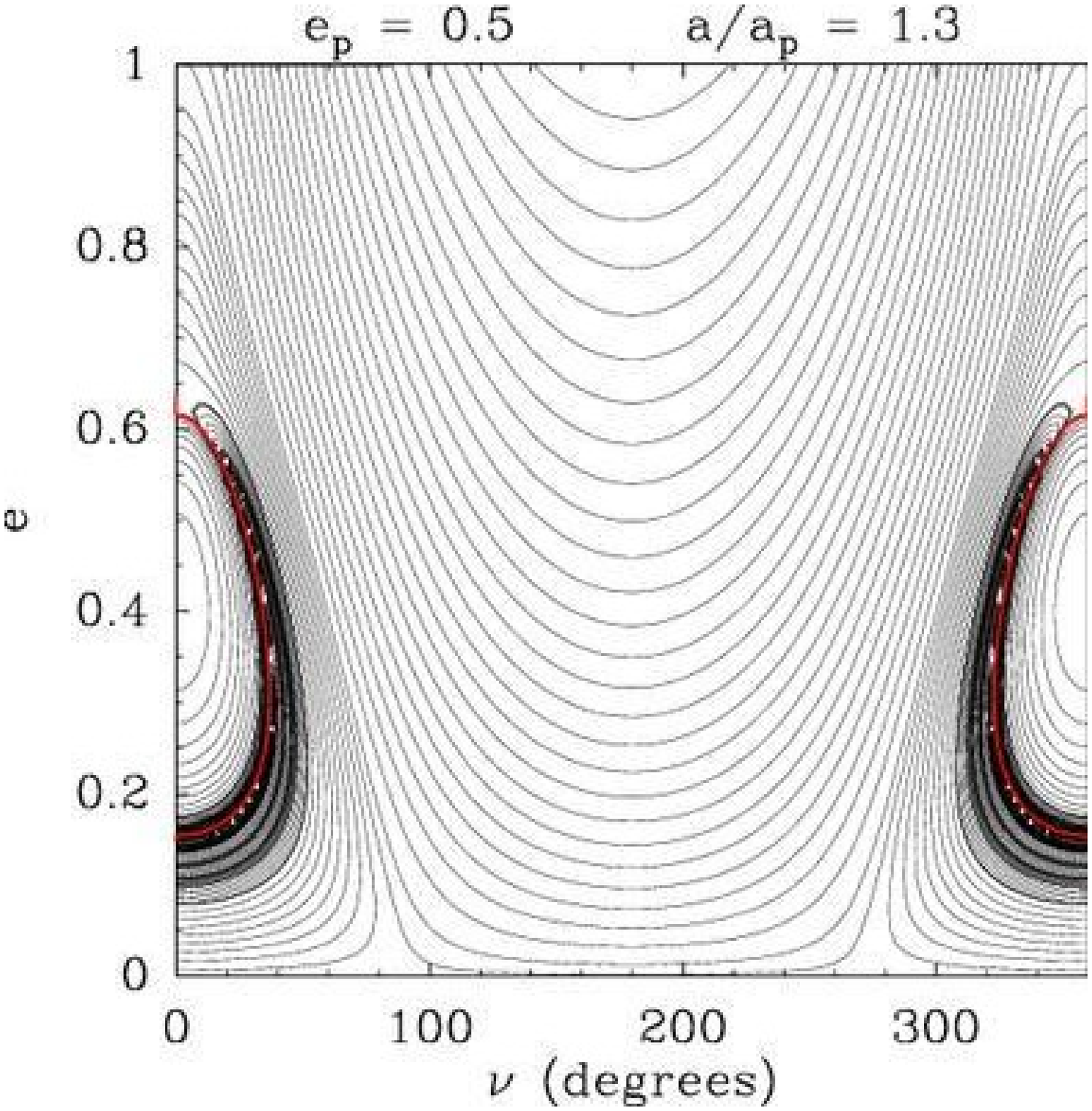}\hfil
\includegraphics[width=5.9truecm]{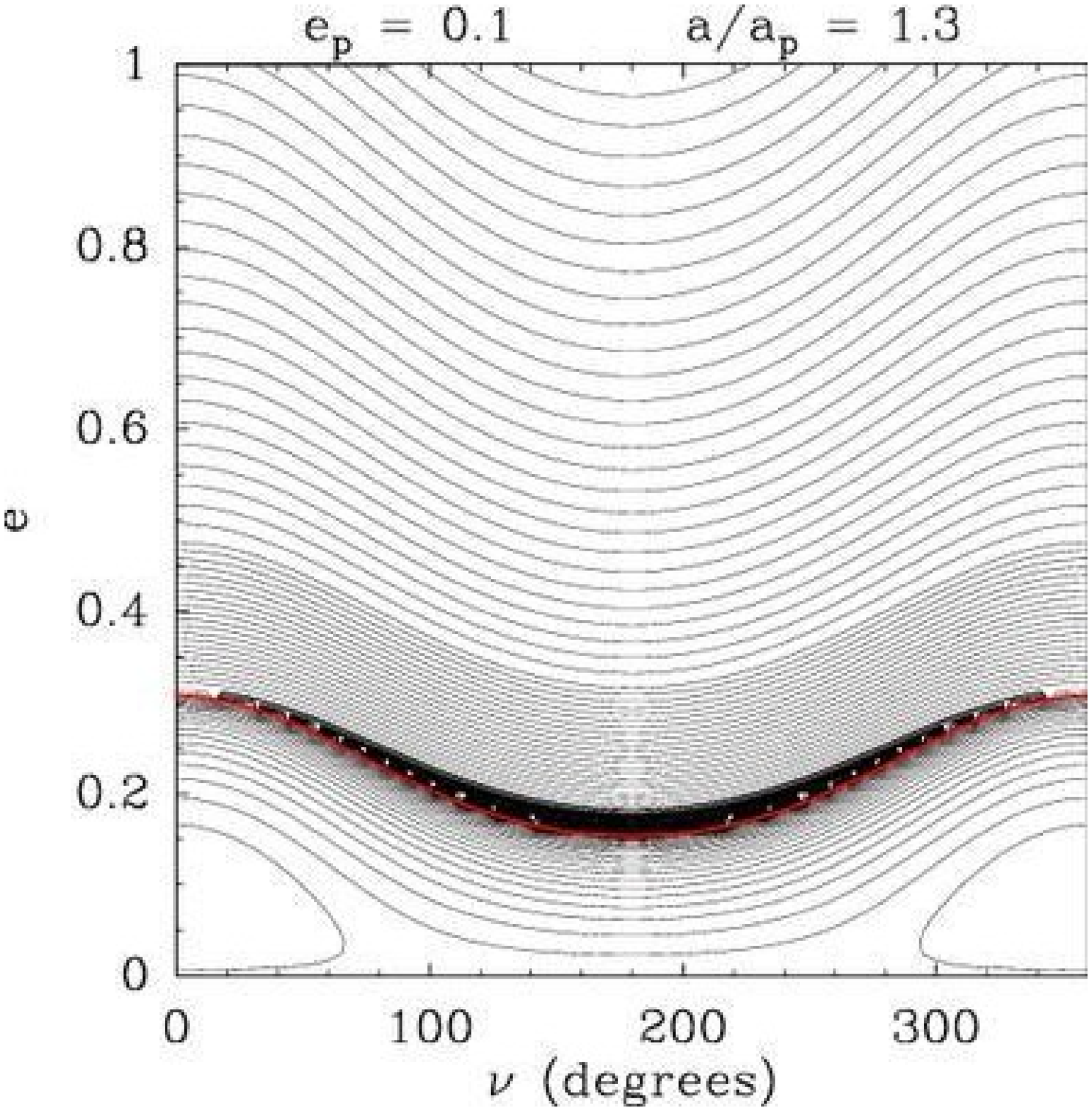}}
\caption{\label{hsec} Phase portraits (Hamiltonian level curves in an
$(\nu, e)$ plane)  of the secular non-resonant 
dynamics in the planar restricted three body problem, for a fixed
semi-major axis $a=1.3a_p$, and planet eccentricities $e_p=0.5$ (left
plot) and $e_p=0.1$ (right plot). The red line separates regions where
the planetesimal orbit does not cross that of the planet (low
eccentricities for $e_p=0.1$ and a small island around $\nu=0$ for
$e_p=0.5$) from regions where both orbits cross each other.\thanks{See the electronic edition of the
Journal for a color version of this figure.}}
\end{figure*}

Two examples for $e_p=0.1$ and $e_p=0.5$ are shown in Fig.~\ref{hsec}.
They both correspond to $a=1.3a_p$ ($a_p$ is the semi-major axis of
the planet). The whole dynamical problem obviously simply 
scales with $a_p$, so that $a$ only needs to be given in units of
$a_p$. The present case ($a=1.3a_p$) corresponds to a planetesimal orbiting outside the
planet's orbit, like those we are simulating. Note also that 
the topology of the Hamiltonian is independent of the mass $m_p$ of the
planet, because the non-constant part of $\mathcal{H}_0$ is proportional to
$m_p$. Hence the plots in Fig.~\ref{hsec} hold for any planetary mass. The planetary mass $m_p$ only affects the
speed at which the planetesimal
moves along the Hamiltonian level curves (the speed is $\propto m_p$).

\begin{figure}
\resizebox{\hsize}{!}{\includegraphics[angle=-90]{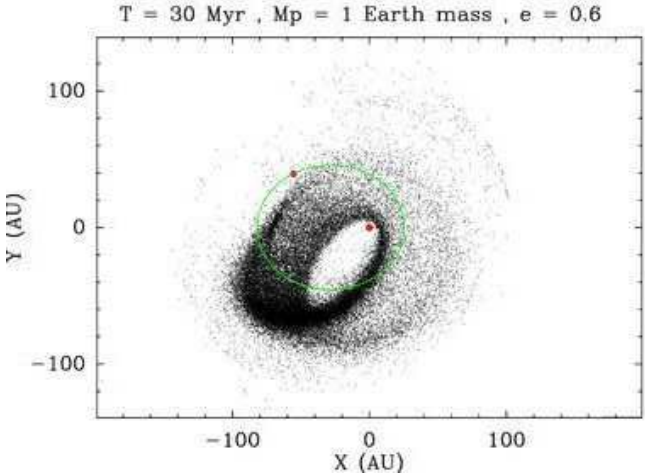}} 
\caption{\label{nonResonantStructure} An Earth mass planet on a very eccentric orbit
  ($e_p=0.6$). The plotting conventions are the same that
  Fig. \ref{withoutMigration}. The disk structure does not
    rotate with the planet and is spatially fixed.
\thanks{See the electronic edition of the
    Journal for a color version of this figure.}} 
\end{figure}

The plots in Fig.~\ref{hsec} hold for $a=1.3a_p$, but for other values
of $a$, we have similar plots. Conversely, the shape of the phase portrait
depends critically on $e_p$. We see that for a small $e_p$, a planetesimal
with a small initial $e$ will keep $e$ small for ever. For a large
$e_p$ however, any planetesimal with a small initial $e$ will be driven
to high $e$ values and $\nu\simeq 0$. Starting from a population of
planetesimals with negligible eccentricities, we end up after a certain
delay with many highly eccentric planetesimals with their lines of apsides
more or less aligned with that of the planet $\nu\simeq 0$.
This naturally generates a clump of planetesimals close to the apoastron
of their orbits, as due to Kepler's second law, the planetesimals
spend most of their time near apoastron. This is the origin of the
clumps we obtain in our simulations with low-mass planets
(Fig.~\ref{nonResonantStructure}).

So, why does this not hold for more massive planets? The secular dynamics described above
is valid as long as the planetesimal does not undergo any close encounter
with the planet. In the case of a close encounter, the orbit of the planetesimal
is suddenly changed, and it is often ejected. Many regions in
Fig.~\ref{hsec} correspond to a planet crossing orbit. The probability of
having a close encounter with the planet within a given timespan
is higher if the planet is more massive. It scales
as $m_p^{2/3}$, because the Hill radius $r_\mathrm{H}$ of the planet scales as
$m_p^{1/3}$, and the encounter cross-section is expected to scale as
$r_\mathrm{H}^2$. The mass ratio between a $3$ Jupiter mass planet and
an Earth-sized planet is $\sim 1\,000$. We thus expect a planetesimal to
undergo $100$ times more encounters with the first planet than
with the second. Finally, with massive planets, most of the planetesimals
are subject to a close encounter with the planet within the timespan
of the simulation described in Fig.\ref{figure_e3}.
This is why the disk appears so depleted at the end. 
Conversely, for low-mass planets, the close encounter probability
is so low that many planetesimals keep following the secular dynamics
until the end of the run. Therefore, they have enough time to generate
a strong asymmetric clump.

We stress here that this clump is not due to any mean-motion
resonance. There is thus no need for planet migration in this case and
this may appear as an alternative scenario to mean-motion
resonance for generating transient clumps. 
 Nevertheless, even with an Earth mass
   planet, these clumps do
 not last as long as the resonant clumps. Planetesimals are not
 protected against close encounters with the planet, which finally
 deplete most of the disk after $35$ Myr, in our simulations.

\section{Generalization}
\label{Generalization}

As explained in the previous sections, we have run many
simulations in order to sample correctly the parameter space of planetary
eccentricity versus planetary mass. We can thus address the question of
the visibility of asymmetric structures in a disk. As already mentioned, the shape of a disk is dominated either by resonant or
by non-resonant planetesimals. The region where MMRs
dominate the disk shape correspond to planets on low-eccentricity
orbits. In this region, two situations
can occur: planets can generate clear resonant patterns with several
visible clumps in the disk (generally while on circular or very low
eccentric orbits) or produce smooth patterns, with only a hole at the
planet location as the visible structure (generally while on an orbit with
an eccentricity between $0.05$ and $0.1$). Concerning the non-resonant
planetesimals, Earth mass planets on eccentric orbits can generate
observable structures by secular perturbations. Outside these
regions, the disks do not show any observable structures, when they
are not totally depleted. The results for all these simulations are
summarized in Table \ref{normal} and Fig. \ref{resume}. Three main
regions can be identified. In zones I and
  II, observable structures in the disk are generated by MMRs while in
  zone III, transient structures are generated by non-resonant mechanisms. In
  the remaining region, the disk does not show any structure. In
  zone I, MMRs create clumpy disks while in zone II they generate
  a smooth disk with a hole at the planet location.
This figure also shows the fraction of planetesimals still
bound to the system after $40$ Myr (background color). However, this quantity is sensitive to
several parameters (stellar mass, duration of the simulation, initial
distribution of the planetesimals ...) while the
limits of the three zones are quite independent.

However, we have assumed for these simulations a constant migration
rate of $0.5$ AU Myr$^{-1}$ and a disk of planetesimals initially on
circular orbits. With different assumptions the outcomes of the
simulations could be changed: we have thus investigated these
two parameters in order to discuss the robustness of our conclusions. For a given planetary
mass and eccentricity, we expect the
structures to change, as the trapping probability depends on the
migration rate and on the planetesimal eccentricity. But, again, our
main focus is to determinate if the resonant structures are visible or
not. For instance, in the low-eccentricity orbit case, Neptune mass and
Jupiter mass planets do not produce the same structures but they
have the same sensitivity to the planet eccentricity. Here, we investigate
if the migration rate or the initial planetesimal eccentricity
change these conclusions. The results of these additional simulations
are summarized in Tables \ref{fastslow} and \ref{complete_hot}, in the
same manner as in Table \ref{normal} for the nominal case. Table
\ref{fastslow} corresponds to simulations with different migration
rates but unexcited initial planetesimal disks. Table
\ref{complete_hot} describes  simulations with initially excited disks.

\subsection{Migration rate}

\begin{figure*}
\centering
\includegraphics[angle=-90,width=12cm,origin=br]{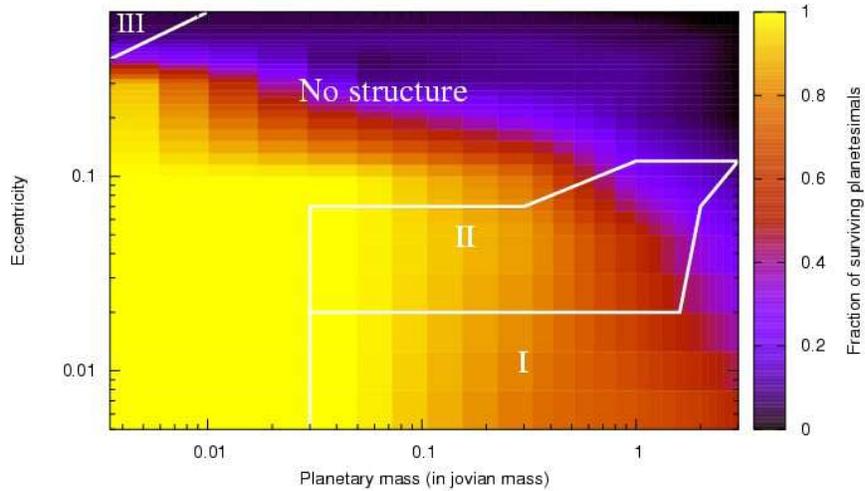}
\caption{\label{resume} Overview of the simulation outputs for
  planetesimal disks, in a (planet mass, planet eccentricity) plane. All
  simulations shown in this figure concern a planet with a constant
  migration rate of $0.5$ AU Myr$^{-1}$. Color scale indicates the
  fraction of planetesimals still bound after $40$ Myr. The
    fraction of surviving planetesimals is linearly interpolated
    between the simulations of the Table \ref{normal}. In zones I and
  II, observable structures in the disk are generated by MMRs while in
  zone III, transient structures are generated by non-resonant mechanisms. In
  the remaining region, the disk does not show any structure. In
  zone I, MMRs create clumpy disks while in zone II they generate
  a smooth disk with a hole at the planet location. \thanks{See the electronic edition of the
    Journal for a color version of this figure.}} 
\end{figure*}

\begin{figure}
\centering
\begin{tabular}{c}
\resizebox{\hsize}{!}{\includegraphics[angle=-90,width=0.48\textwidth]{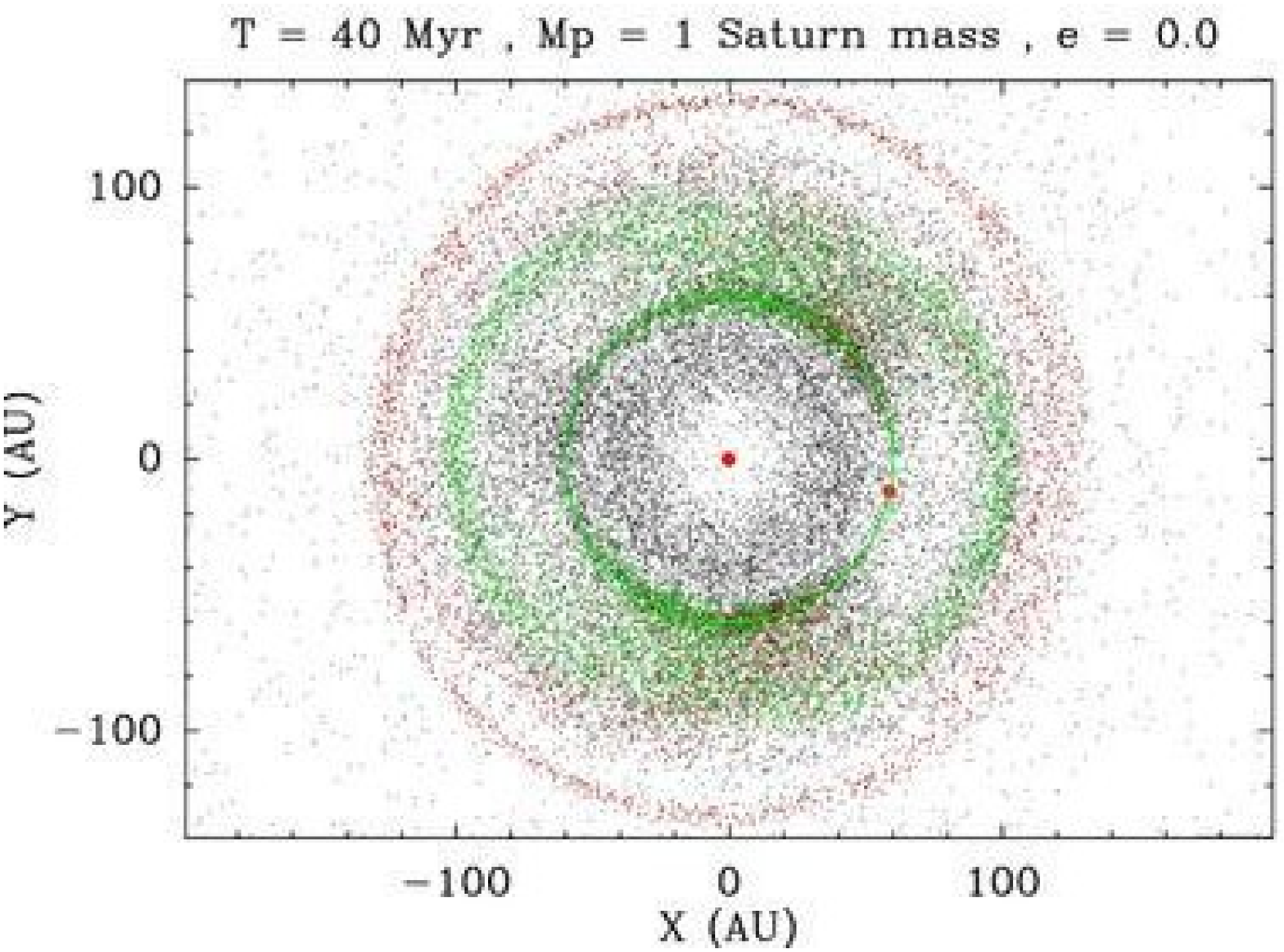}} \\
\resizebox{\hsize}{!}{\includegraphics[angle=-90,width=0.48\textwidth]{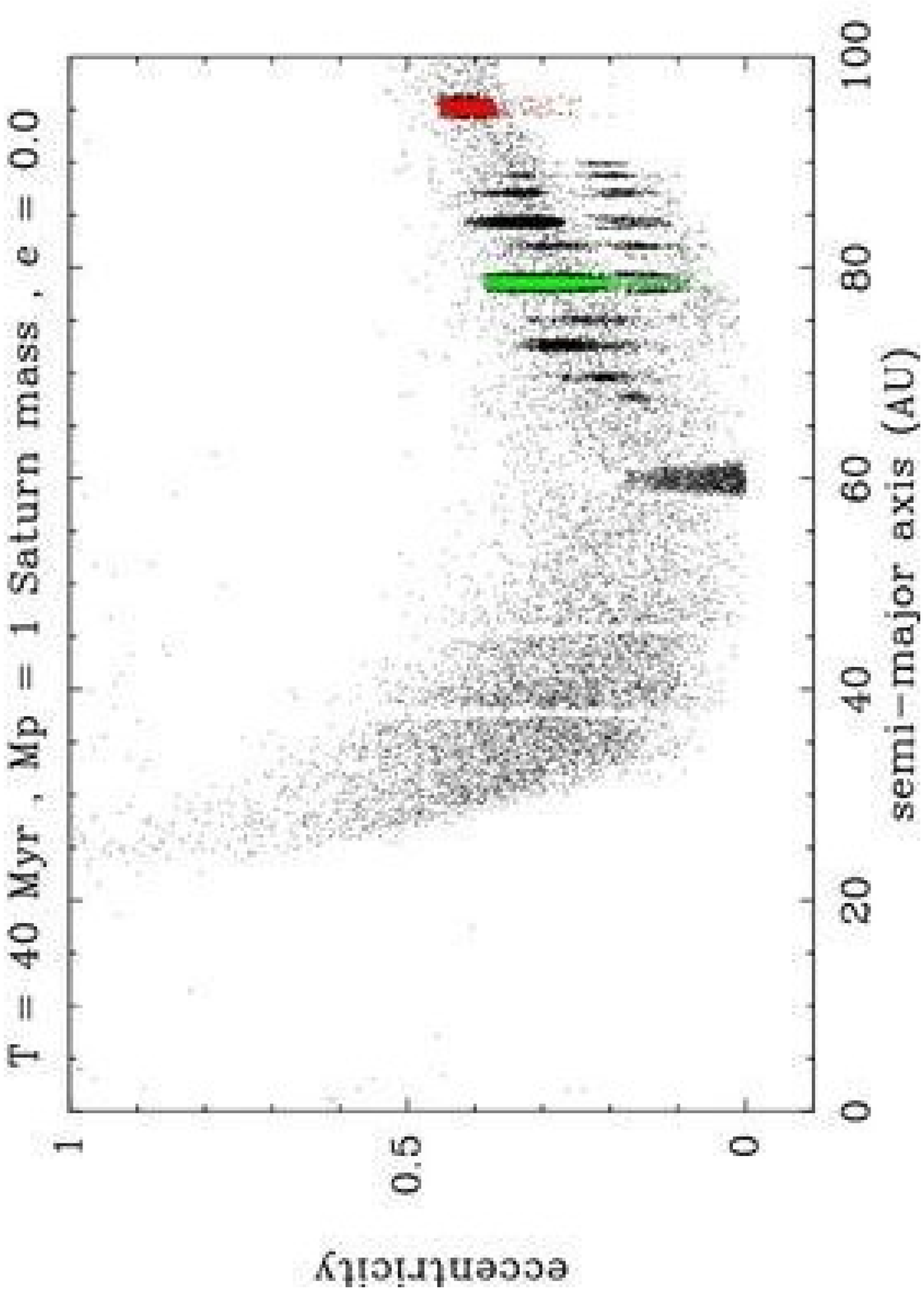}} 
\end{tabular}
\caption{\label{warmDisk} A Saturn mass planet on a circular
  orbit migrating outward a dynamically warm disk. Initial
  eccentricities of planetesimals are uniformly distributed between $0$
  and $0.2$. The plotting conventions are
  the same as in  Fig. \ref{figure_e0} and \ref{figure_e0_ae}. \thanks{See the electronic edition of the
    Journal for a color version of this figure.}}
\end{figure}

In our work, the migration rate parameter was
chosen to match the best fit obtained  by
\citet{2003ApJ...598.1321W} for the Vega debris disk to allow direct
comparison. In his paper, Wyatt discussed the impact of the
migration rate in the restrictive case of a planet on a circular orbit, and
has shown that the trapping probability in an MMR increases with decreasing
migration rates (or when the star is less massive).

We have thus performed simulations with a lower ($0.05$ AU Myr$^{-1}$)
and a higher ($5$ AU Myr$^{-1}$)  migration rate than previously
(Table \ref{fastslow}). Overall, our results are in good agreement with those of
\citet{2003ApJ...598.1321W}, even for eccentric orbits: the trapping probability increases when
the migration rate decreases. For example, with a low migration rate, a Neptune mass planet 
traps planetesimals in the $2$:$1$ MMR, while a Saturn mass planet populates the
second libration center of the $2$:$1$ resonance. 

Although the disk resonant shape is modified because the populated
MMRs change with the migration rate, the
dependence on the mass and eccentricity of the planet remains unchanged:
non-resonant planetesimals always dominate at moderate eccentricity and clear
resonant patterns are only visible with a planet on a quasi
circular orbit. We still note some subtle changes between
the different migration rates. With a higher migration rate,
low mass planets have less time to increase the eccentricities of the
non-resonant planetesimals which perturb less the resonant
structures. With a very low migration rate, planets have more time, during close encounters, to
eject non-resonant planetesimals, and
resonant planetesimals in the eccentric planet orbit case.
Nevertheless, as these changes are small, it is possible to summarize our simulations in the (planet
eccentricity; planet mass) parameter space and to discuss the visibility
of resonant clumps using only these two parameters, as in Fig. \ref{resume}.

\subsection{Warm disks}
We have assumed in our simulations that the planetesimal disk was initially
dynamically cold ($e_{disk}=0.0$), but \citet{2003ApJ...598.1321W}, while using
the same hypothesis, pointed out that dynamically warm disks are an
interesting alternative to be studied. We have thus
extended our work in this direction, assuming disks with planetesimal
eccentricities uniformly distributed between $0$ and $e_{limit}$
(Table \ref{complete_hot}).

It appears that an initially warm disk does not differ much from a cold
disk in the case of a planet on a moderate or high eccentricity
orbit. The secular perturbations, or the close encounters, due to such
a planet raise the planetesimal
eccentricities on a timescale shorter than the migration time. A
cold disk therefore becomes warm in the course of planetary
migration. However, with a
planet migrating on a circular orbit, the disk structures are
different depending on the initial eccentricities of the
planetesimals. A warm disk ($e_{limit}=0.1$) with a planet on a circular orbit is actually
roughly equivalent to a cold disk with a planet on a low-eccentricity
orbit ($0.05$ or $0.1$): the MMRs trap many planetesimals but the
clumps are smoothed by the large libration amplitude (Fig.
\ref{warmDisk}). From our simulations, it appears that, for a
  Saturn mass or Jupiter mass planet on a circular orbit, an $e_{limit}$ of $0.1$ is already
  too high to keep the resonant clumps visible. On the other hand,
  planets above $2$ Jupiter masses have a large enough depletion rate
  to keep the resonant clumps visible with an $e_{limit}$ up
  to $0.2$. These results show that the resonant structures are as
sensitive to the planet  eccentricity as to the planetesimals eccentricity.

\section{Comparison with previous works}
\label{Comparison}
\subsection{P-R drag scenario}
For both circular and low-eccentricity orbits, \citet{2003ApJ...588.1110K} obtain
much larger differences in the structures between low and high mass
planets that those observed in our simulations: the P-R drag
scenario seems to have a stronger dependence on the planetary
mass than the migration scenario. The particles
migrating inward due to P-R drag encounter
first the most external resonances, but the capture probability
for these resonances is low and the dust is more likely trapped in
resonances that are closer to the planet. With massive planets, however,
the probability of capture in distant resonances is high enough to trap a
large number of dust particles in these outer disk regions and stop them before
they can populate closer resonances. The MMRs populated by P-R drag,
and thus the structures generated, are therefore not the same,
depending on the planetary mass. Conversely, in the case of planetary
migration, all the resonances can be populated at the
same time, as planetesimals are initially present in the whole disk: as explained
in Section \ref{lowEccOrb}, the MMRs actually populated are thus roughly the
same, whatever the planet mass.

In practice, the differences between our results and those of  \citet{2003ApJ...588.1110K}
  are more due to the initial planetesimal
distribution in the disk than to the physical process. In their P-R drag scenario, the authors assume that the dust
starts migrating inwards far away from the planet, outside the $2$:$1$
resonance, while our
planetary migration scenario assumes the planetesimals to be initially
closer to the planet. If we
started with planetesimals at a larger distance, we
would have expected to see, as in the P-R drag scenario, more differences in the
resonant structures for planets of different masses.

\subsection{Planetary migration}

Our results show that, with the forced planetary migration scenario, it is easy
to distinguish a planet on a circular orbit from
another on a low-eccentricity orbit, except for
very low mass planets or very massive planets, because the resonant structures
are drastically different. Constraining the planetary mass is more
difficult than in the P-R drag scenario and only an order of magnitude
can be expected.

\citet{2003ApJ...598.1321W} used this scenario to reproduce Vega
disk observations at submillimetric wavelengths \citep{1998Natur.392..788H}. We must take into
account that, with the large SCUBA PSF, only the two major clumps can
be observed. However, this is enough to distinguish between our three
planet mass examples, at least for a migration rate of $0.5$
AU   Myr$^{-1}$:
\begin{itemize}
\item It is possible to distinguish between a Saturn mass
planet and a Neptune mass or Jupiter mass planet because it is the only
one out of the three that produces asymmetric clumps in density. 
\item We can
also distinguish between a Neptune mass planet and a Jupiter one
because, in the first case, the two clumps are in opposition with respect
to the star, while, in the second case, the two clumps are separated
in longitude by less than $180^\circ$.
\end{itemize}

It is thus possible to obtain an estimate of the planetary mass. The
situation is well summarized by Figure 11 of \citet{2003ApJ...598.1321W}. It defines
several regions in the (planetary mass, migration rate) parameter space
that can be observationally distinguished from each other, but
inside each region a wide range of planetary masses is possible.

Contradictions however appear between our simulation results and this
previous study. The asymmetry in
the emission of the two observed clumps was interpreted as the migration
of a Neptune mass planet by \citet{2003ApJ...598.1321W}. In his model, a Neptune can trap
planetesimals in the $3$:$2$ and $2$:$1$ resonances and generate two asymmetric
clumps, like a Saturn mass planet in our simulations. With our
numerical model, we have found that a Neptune mass planet cannot trap planetesimals
in the $2$:$1$ resonance, but only in the $3$:$2$ resonance: the two clumps are thus
symmetric and cannot reproduce the Vega
disk. A Neptune mass planet at a migration rate of about $0.5$ AU
Myr$^{-1}$ lies at the sharp transition between a $0$ and a
$100\% $ trapping probability \citep[][Fig. 4a]{2003ApJ...598.1321W}. A small change in the planetary mass or the migration
rate in this configuration produces a large modification in the
population of this resonance. As \citet{2003ApJ...598.1321W} uses a
scaling law to predict the trapping probability, differences between our results may be explained by the approximation of this scaling law.   

\begin{figure}
\centering
\begin{tabular}{c}
\resizebox{\hsize}{!}{\includegraphics[angle=-90,width=0.48\textwidth]{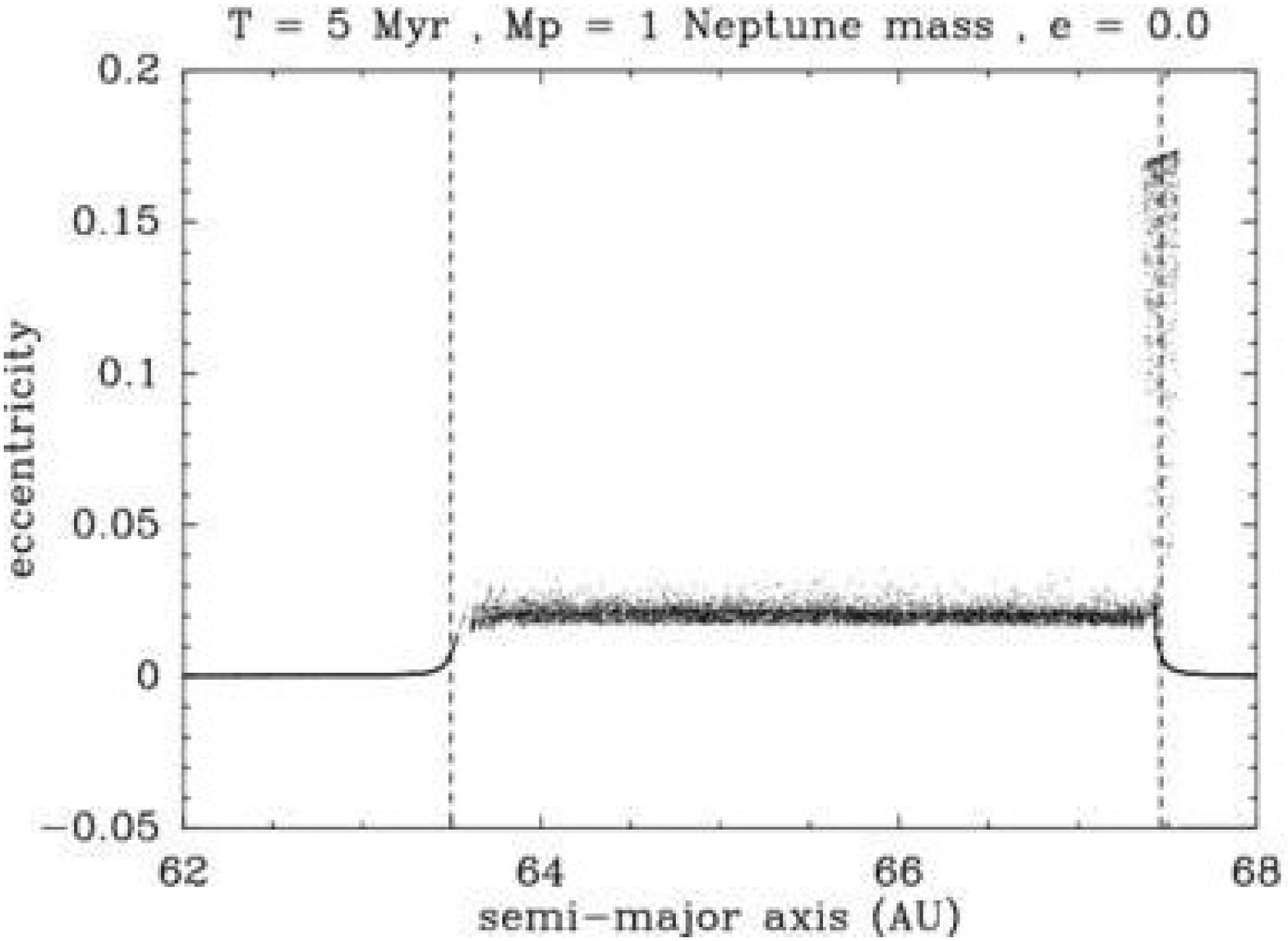}}\\
\resizebox{\hsize}{!}{\includegraphics[angle=-90,width=0.48\textwidth]{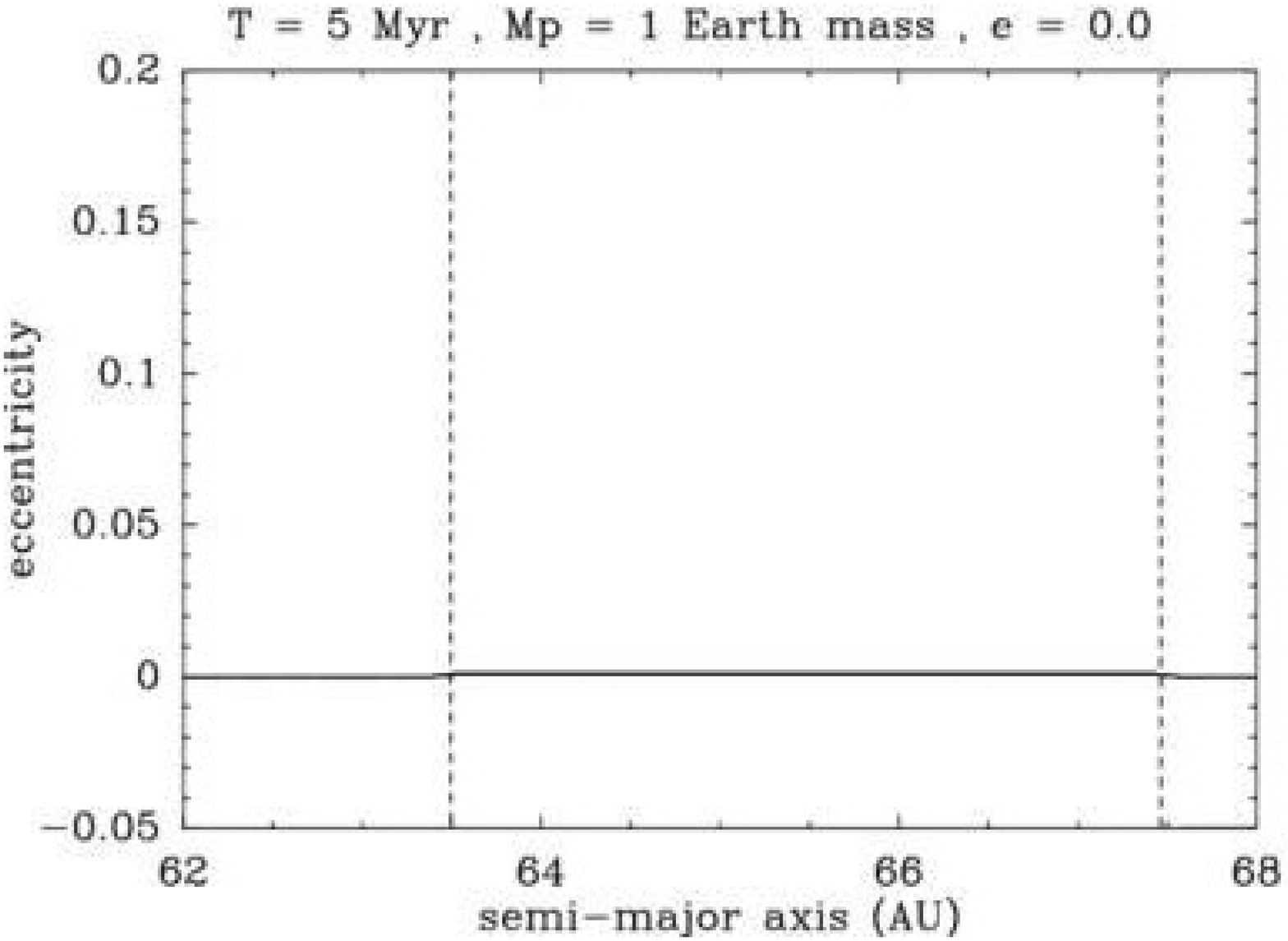}}\\
\resizebox{\hsize}{!}{\includegraphics[angle=-90,width=0.48\textwidth]{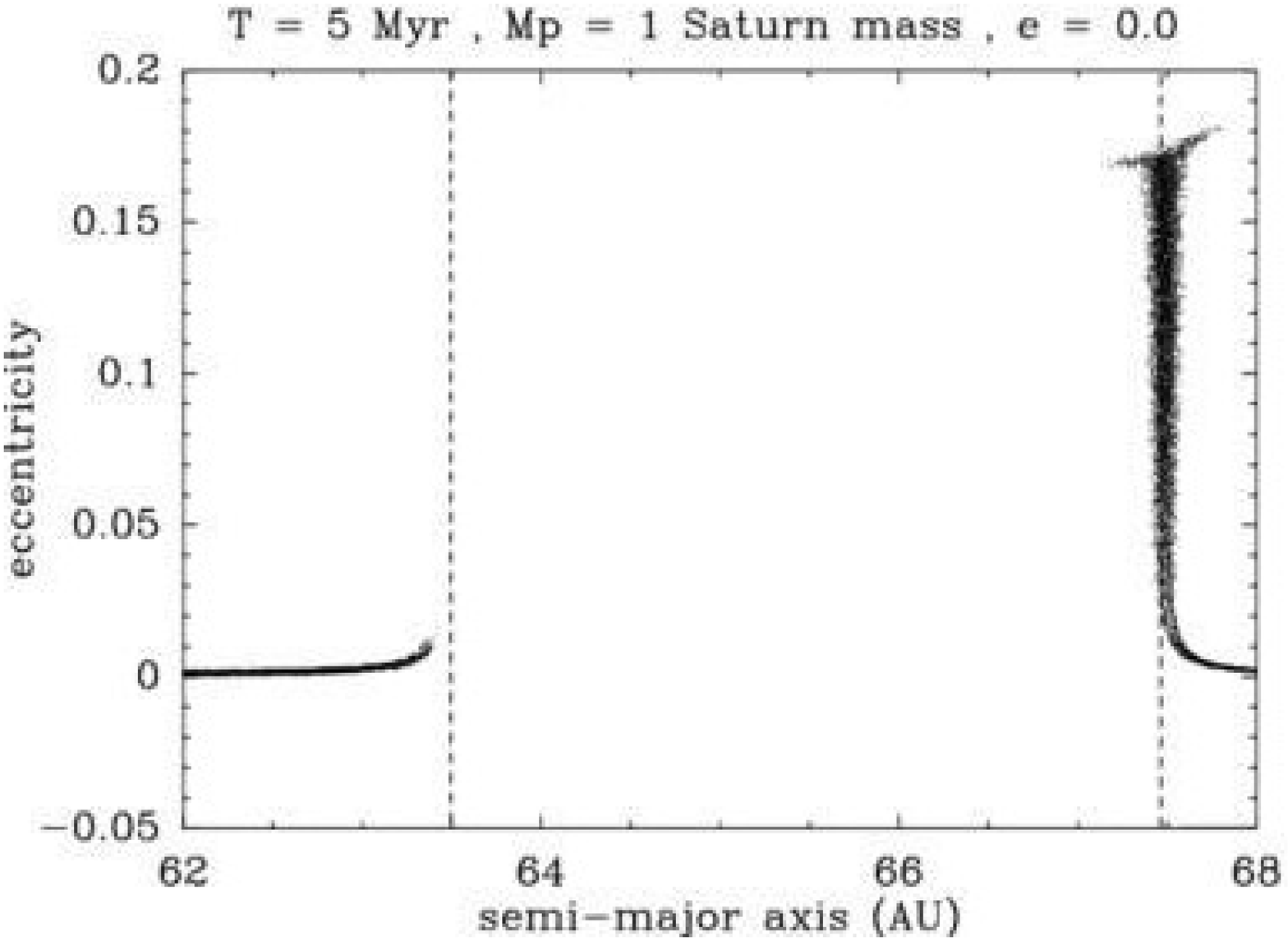}} 
\end{tabular}
\caption{\label{figure_discussion_wyatt} Capture in the $2$:$1$ resonance for a Neptune
  mass planet (top), en Earth mass planet (middle) and a Saturn mass planet (bottom), all on a circular
  orbit, after $5$ Myrs. Dashed lines show the position of the
  resonance at the beginning and at the end of the simulations. In
  those simulations, an
  Earth mass planet does not capture at all planetesimals while a
  Saturn mass planet traps all of them. A Neptune mass planet only
  traps a fraction of the planetesimals but gives a small kick in
  eccentricity for the others. 
  The plotting conventions are
  the same as in  Fig. \ref{figure_e0_ae}. \thanks{See the electronic edition of the
    Journal for a color version of this figure.}}
\end{figure}

Nevertheless, the $2$:$1$ resonance has an interesting behavior in the Neptune
mass planet case: it perturbs all  the planetesimals that cross it, but as
soon as they reach an eccentricity of about $0.02$ (in  $0.25$
Myr), most of them escape. While a Saturn mass planet (or a
more massive one)  cleans up the space between the initial and final
position of the $2$:$1$ resonance during the migration by trapping all
the planetesimals, a Neptune mass
planet only slightly rises the eccentricity of planetesimals
entering the $2$:$1$ resonance during the migration process (Fig.
\ref{figure_discussion_wyatt}).

\begin{figure}
\centering
\resizebox{\hsize}{!}{\includegraphics[angle=-90,width=0.48\textwidth]{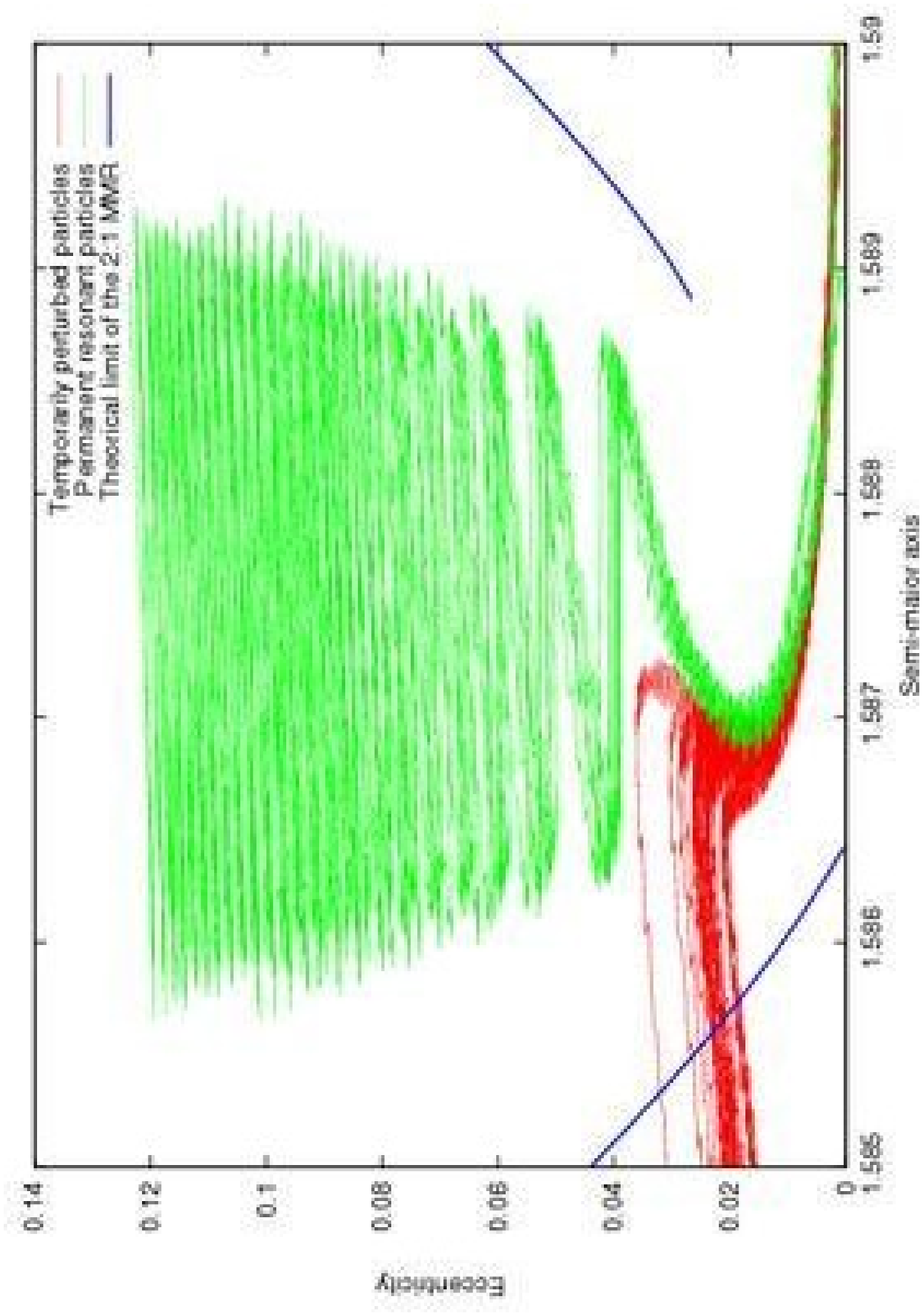}} \\
\caption{\label{figure_discussion_res} Projection on the (semi-major
  axis, eccentricity) plane of the trajectory of $40$ planetesimals
  near the $2$:$1$ resonance of a Neptune
  mass planet, migrating on a circular orbit. The semi-major axis is in
  units of the planet semi-major axis. Green lines are for permanently
  trapped planetesimals while the red lines are for the temporarily
  perturbed planetesimals. \thanks{See the electronic edition of the
    Journal for a color version of this figure.}}
\end{figure}

This phenomenon is better seen when the planetesimal trajectories are
drawn in a semi-major axis, eccentricity diagram where the semi-major
axis of the planetesimals are in units of that of the planet, in
order to hide the migration, as in Fig.~\ref{figure_discussion_res}.
All the planetesimals initially have
roughly the same trajectory (they move from right to left because they
do not migrate), but the small variations have a
strong impact when the planetesimals cross the resonance.
Some planetesimals remain trapped in the MMR, while others
escape after being temporarily perturbed. However, within each of these two subgroups, nearly all planetesimals 
have similar behavior: the permanently trapped planetesimals have the same
libration amplitude and the temporarily perturbed planetesimals escape the resonance
roughly at the same eccentricity.

 The width (in semi-major axis) of an
MMR is proportional to the square root of the planetary mass \citep{1996Icar..120..358B}. A Saturn
mass planet therefore has wider resonances than a Neptune mass planet:
its trapping probability is thus larger because it is less sensitive
to the orbital parameters of the planetesimals which cross the MMR.

\section{Discussion and limitations}
\label{Limitation}
\subsection{Dust content and collisional activity}

In this paper, we have only discussed the spatial distribution of the
planetesimals. However, it is the dust produced by collisions
between these planetesimals, and not the planetesimals themselves, that contributes to the
emitted flux and therefore to the observed disk shape. One has to investigate the influence of
the collisions on the dust spatial distribution in order to reproduce
the observations. Collisions can have several consequences for the dust
particles: 
\begin{itemize}
\item As collisional cascades produce dust particles of
all sizes, radiation pressure cannot be neglected any longer. This
has a strong incidence on the populations of the MMRs. As
explained by \citet{2006ApJ...639.1153W,2007A&A...462..199K}, depending on their sizes, the dust
particles can either stay in the resonance, leave the
resonance but still stay bound to the system, or be blown out of the
disk. Even in the case
where the MMRs dominate the dynamics of the planetesimals, the disk
can look rather smooth if the observations are sensitive to the smallest of the produced dust particles.
\item The collision rate is not the same in the whole disk: it is
enhanced in the MMRs, as shown by \citet{2007CeMDA.tmp...39Q}. These authors
have even calculated that the average collision rate in an MMR is the highest for
first-order resonances (by a factor about $2$ times the
non-resonant collision rate). This implies that at the wavelengths
where the observed
dust particles can still appear trapped in the MMRs (typically at $850\mu
$m), the resonant clumps are more visible than in our
simulations. This also means that the $3$:$5$ resonance, often
populated in the simulations, is actually not so prominent because its
clumps have a lower average collision rate than the first order
resonances.
\item The destructive collisions produce fragments with different
  velocities from the parents bodies. Since these velocities
  are usually
  small compared to the orbital velocities, most of the fragments
  cannot leave the resonances. However, analytic results from
  \citet{2007A&A...462..199K} show that small particles produced
  by the collisional cascade are fast enough to escape the MMRs. Dust particles are mainly lost by this ``velocity effect''
  before they become small enough to escape the MMR by radiation
  pressure.
\end{itemize}

In conclusion, predictions on the distribution of the  smallest dust particles, observable in
the near or mid-infrared wavebands, is difficult since it requires a
proper description of the collisional activity and of the dynamics of
particles influenced by radiation/wind forces. This is not the
case for larger dust grains which are observable at submillimeter
wavelengths: according to the previously quoted papers, these particles,
created by mutual collisions between the planetesimals, stay in the
MMRs. However, the most visible resonant clumps are not necessarily the
most populated, because the collision rate depends on the order of the
resonance. 

\subsection{Origin of planetary migration}
The described simulations obtained with our model raise numerous questions about the interaction
between a planetesimal disk and a planet. The structures
generated are very sensitive to the eccentricity of the planet, and to
a lesser extent to the migration speed. Moreover, some of the
generated structures (especially those obtained for high eccentricity, low-mass
planets) appear only after a fairly long time. A key issue to address
in this question is the temporal evolution of the eccentricity (and
the migration rate) and the combined evolution of the disk. A low
mass, high-eccentricity planet can only generate the features
described above (Fig. \ref{nonResonantStructure}) if its eccentricity remains high for a sufficiently long
time. This may appear unrealistic, as the planet eccentricity should
decrease due to the interactions with the planetesimals. Another
important question is the transition between circular and
low-eccentricity orbits: we have shown in the previous section that
resonant patterns change significantly when the eccentricity deviates
from $0$. What will be the resulting disk structures if the planet
eccentricity undergoes periodic secular modulation between these two
regions (like the giant planets of the Solar System) ? And how do the particles stay
at an eccentricity lower than $0.1$ all the time ?

 These questions can be resolved by better modeling the origin
and the evolution of planetary migration.
In any realistic simulation, the orbit  of the planet will be subject to secular
evolution. There is even no need for other planets for this. The
interaction with the disk particles themselves can be sufficient to
significantly affect the planetary orbit. Thus taking into
account the influence of the disk on the planets is necessary to
derive realistic simulations. Several studies have already discussed
the origin of planetary migration either by ``planet-planetesimals'' \citep{2000ApJ...534..428I,2004Icar..170..492G}
interactions or by ``planet-planet'' interactions
\citep{2006DPS....38.5403M}. Depending on the scenario and on the
initial conditions, one can observe migration on low-eccentricity
orbit or more chaotic migration after a short time on an eccentric orbit.

\section{Conclusion}

We have studied the problem of the presence of observable structures
in planetesimal disks due to mean motion resonance with an unseen
planet migrating outward in the disk. Using numerical simulations, we
have explored a large range of parameters for the planet (mass and
orbital eccentricity) and the disk (initial distribution of
planetesimal eccentricities). In the case of a planet on a circular orbit
migrating inside a dynamically cold disk, our results are in agreement
with previous analytical studies.

In the cases not already addressed, namely planets on eccentric orbits
or dynamically warm disks, we have found that the observability of
resonant structures demands very specific
orbital configurations. The clumps produced by MMRs with a planet on a
circular orbit are smoothed in the case of a planet on an even
moderately eccentric orbit. An eccentricity as low as $0.05$ is enough
to smooth all the resonant structures, except for the most massive
planets. These results indicate that although trapping planetesimals
in MMRs is an efficient mechanism to generate clumpy disks, stringent
conditions must be fulfilled for this scenario to occur. Theoretical modeling
  of the origin of the planetary migration therefore will have
  to explain how planetary systems can remain under these
  conditions. Moreover, we only consider a planet migrating at a constant rate. A more realistic
model with a variable, stochastic migration rate can reduce the
population of resonances and thus their observability. A better model
of planet migration  thus should be developed in
future studies.

\begin{table*}
\centering
\caption{\label{normal} Summary of results for all
  simulations done in the present study with an initially unexcited
  disk (initial planetesimal eccentricities are equal to zero) and a
  standard migration rate of $0.5$ AU Myr$^{-1}$. For each simulation,
we list the migration rate, the planet mass and eccentricity, the
fraction of surviving planetesimals at the end of the simulation ($40$
Myr) and the resulting disk shape, following the convention of Fig. \ref{resume}.}
\begin{tabular}{ccccc||ccccc}
Mig. rate$^a$ & Mass$^b$ & Ecc. &
Surv. planetesimals$^c$ & Disk shape$^d$& Mig. rate$^a$ & Mass$^b$ & Ecc. &
Surv. planetesimals$^c$ & Disk shape$^d$\\
\hline
0.5 & 0.0035 &0.0&$100\%$&None&0.5 & 0.05 &0.0&$100\%$&I\\
 & &0.01&$100\%$&None& & &0.01&$100\%$&I\\
 & &0.05&$100\%$&None& & &0.05&$100\%$&II\\
 & &0.1&$100\%$&None&  & &0.1&$100\%$&None\\
 & &0.2&$100\%$&None& & &0.2&$50\%$&None\\
 & &0.3&$100\%$&None& & &0.3&$10\%$&None\\
 & &0.4&$35\%$&III& & &0.4&$5\%$&None\\
 & &0.5&$10\%$&III& & &0.5&$10\%$&None\\
 & &0.6&$5\%$&III& & &0.6&$5\%$&None\\
 & &0.7&$5\%$&III& & &0.7&$5\%$&None\\
0.5 & 0.33 &0.0&$75\%$&I&0.5 & 1 &0.0&$70\%$&I\\
 & &0.01&$70\%$&I& & &0.01&$70\%$&I\\
 & &0.05&$85\%$&II& & &0.05&$55\%$&II\\
 & &0.1&$75\%$&None& & &0.1&$25\%$&II\\
 & &0.2&$25\%$&None& & &0.2&$10\%$&None\\
 & &0.3&$10\%$&None& & &0.3&$10\%$&None\\
 & &0.4&$10\%$&None& & &0.4&$10\%$&None\\
 & &0.5&$10\%$&None& & &0.5&$5\%$&None\\
 & &0.6&$5\%$&None& & &0.6&$5\%$&None\\
 & &0.7&$5\%$&None& & &0.7&$0\%$&None\\
0.5 & 2 &0.0&$65\%$&I&0.5 & 3 &0.0&$55\%$&I\\
 & &0.01&$60\%$&I& & &0.01&$50\%$&I\\
 & &0.05&$30\%$&I& & &0.05&$20\%$&I\\
 & &0.1&$15\%$&II& & &0.1&$10\%$&I\\
 & &0.2&$5\%$&None& & &0.2&$0\%$&None\\
 & &0.3&$5\%$&None& & &0.3&$0\%$&None\\
 & &0.4&$5\%$&None& & &0.4&$0\%$&None\\
 & &0.5&$5\%$&None& & &0.5&$0\%$&None\\
 & &0.6&$0\%$&None& & &0.6&$0\%$&None\\
 & &0.7&$0\%$&None& & &0.7&$0\%$&None\\
\hline
\multicolumn{10}{l}{$^a$In AU Myr$^{-1}$.}\\
\multicolumn{10}{l}{$^b$In Jovian mass.}\\
\multicolumn{10}{l}{$^c$Fraction of surviving planetesimals at the end of the simulation, i.e. $40$ Myr.}\\
\multicolumn{10}{l}{$^d$As in Fig. \ref{resume}.}\\
\end{tabular}
\end{table*}

\begin{table*}
\centering
\caption{\label{fastslow} Same as Table \ref{normal}, but for
  different migration rates.}
\begin{tabular}{ccccc||ccccc}
Mig. rate$^a$ & Mass$^b$ & Ecc. &
Surv. planetesimals$^c$ & Disk shape$^d$& Mig. rate$^a$ & Mass$^b$ & Ecc. &
Surv. planetesimals$^c$ & Disk shape$^d$\\
\hline
5 & 0.035 &0.0&$100\%$&None&5 & 0.05 &0.0&$100\%$&I\\
 & &0.01&$100\%$&None& & &0.01&$100\%$&II\\
 & &0.05&$100\%$&None& & &0.05&$100\%$&II\\
 & &0.1&$100\%$&None& & &0.1&$100\%$&None\\
 & &0.2&$100\%$&None& & &0.2&$100\%$&None\\
5 & 0.33 &0.0&$100\%$&I&5 & 1 &0.0&$100\%$&I\\
 & &0.01&$100\%$&I& & &0.01&$100\%$&I\\
 & &0.05&$100\%$&II& & &0.05&$100\%$&II\\
 & &0.1&$100\%$&II& & &0.1&$95\%$&II\\
 & &0.2&$95\%$&None& & &0.2&$75\%$&None\\
5 & 2 &0.0&$85\%$&I&5 & 3 &0.0&$75\%$&I\\
 & &0.01&$85\%$&I& & &0.01&$75\%$&I\\
 & &0.05&$80\%$&I& & &0.05&$60\%$&I\\
 & &0.1&$65\%$&II& & &0.1&$40\%$&I\\
 & &0.2&$50\%$&None& & &0.2&$25\%$&None\\
0.05 & 0.035 &0.0&$100\%$&II&0.05 & 0.05 &0.0&$90\%$&I\\
 & &0.01&$100\%$&II& & &0.01&$85\%$&I\\
 & &0.05&$100\%$&None& & &0.05&$85\%$&II\\
 & &0.1&$100\%$&None& & &0.1&$75\%$&II\\
 & &0.2&$80\%$&None& & &0.2&$10\%$&None\\
0.05 & 0.33 &0.0&$75\%$&I&0.05 & 1 &0.0&$70\%$&I\\
 & &0.01&$75\%$&I& & &0.01&$70\%$&I\\
 & &0.05&$65\%$&II& & &0.05&$40\%$&II\\
 & &0.1&$40\%$&II& & &0.1&$20\%$&II\\
 & &0.2&$5\%$&None& & &0.2&$5\%$&None\\
0.05 & 2 &0.0&$60\%$&I&0.05 & 3 &0.0&$55\%$&I\\
 & &0.01&$60\%$&I& & &0.01&$55\%$&I\\
 & &0.05&$25\%$&I& & &0.05&$15\%$&I\\
 & &0.1&$15\%$&II& & &0.1&$10\%$&None\\
 & &0.2&$0\%$&None& & &0.2&$0\%$&None\\
\hline
\multicolumn{10}{l}{$^a$In AU Myr$^{-1}$.}\\
\multicolumn{10}{l}{$^b$In Jovian mass.}\\
\multicolumn{10}{l}{$^c$Fraction of surviving planetesimals at the end
of the simulation, i.e. $4$ Myr for $5$ AU Myr$^{-1}$ migration rate and $200$ Myr
for $0.05$ AU Myr$^{-1}$.}\\
\multicolumn{10}{l}{$^d$As in Fig. \ref{resume}.}\\
\end{tabular}
\end{table*}

\begin{table*}
\centering
\caption{\label{complete_hot} Same as Table \ref{normal} but for
  simulations with initially excited disks. The maximum initial
  eccentricity of the planetesimals is mentioned for all
  simulations. The migration rate is $0.5$ AU Myr$^{-1}$ for all simulations.}
\begin{tabular}{ccccc||ccccc}
Max. ecc.$^a$ & Mass$^b$ & Ecc. &
Surv. planetesimals$^c$ & Disk shape$^d$ &Max. ecc.$^a$ & Mass$^b$ & Ecc. &
Surv. planetesimals$^c$ & Disk shape$^d$ \\
\hline
0.1 & 0.035 &0.0&$100\%$&None&0.1 & 0.05 &0.0&$100\%$&II\\
 & &0.01&$100\%$&None& & &0.01&$100\%$&None\\
 & &0.05&$100\%$&None& & &0.05&$100\%$&None\\
 & &0.1&$100\%$&None& & &0.1&$100\%$&None\\
 & &0.2&$100\%$&None& & &0.2&$55\%$&None\\
0.1 & 0.33 &0.0&$70\%$&II&0.1 & 1 &0.0&$60\%$&I\\
 & &0.01&$70\%$&II& & &0.01&$55\%$&II\\
 & &0.05&$85\%$&II& & &0.05&$50\%$&II\\
 & &0.1&$70\%$&None& & &0.1&$25\%$&II\\
 & &0.2&$20\%$&None& & &0.2&$5\%$&None\\
0.1 & 2 &0.0&$50\%$&I&0.1 & 3 &0.0&$40\%$&I\\
 & &0.01&$45\%$&I& & &0.01&$40\%$&I\\
 & &0.05&$25\%$&I& & &0.05&$15\%$&I\\
 & &0.1&$15\%$&II& & &0.1&$10\%$&I\\
 & &0.2&$5\%$&None& & &0.2&$0\%$&None\\
0.2 & 2 &0.0&$35\%$&I&0.2 & 3 &0.0&$30\%$&I\\
 & &0.01&$35\%$&I& & &0.01&$25\%$&I\\
 & &0.05&$20\%$&I& & &0.05&$10\%$&I\\
 & &0.1&$10\%$&I& & &0.1&$5\%$&I\\
 & &0.2&$5\%$&None& & &0.2&$0\%$&None\\
\hline
\multicolumn{10}{l}{$^a$For planetesimals.}\\
\multicolumn{10}{l}{$^b$In Jovian mass.}\\
\multicolumn{10}{l}{$^c$Fraction of surviving planetesimals at the end
of the simulation, i.e. $40$ Myrs.}\\
\multicolumn{10}{l}{$^d$As in Fig. \ref{resume}.}\\
\end{tabular}
\end{table*}

\begin{acknowledgements}
We are grateful to Philippe Thebault, Alexander V. Krivov and Martina Queck for 
enlightening discussions about dust collisions in debris disk. We also
thank Alessandro Morbidelli, Jens
Rodmann and the anonymous referee for helpful comments on this paper. Most of the
computations presented  in this paper were performed at the Service
Commun de Calcul  Intensif de l'Observatoire de Grenoble (SCCI).
\end{acknowledgements}

\bibliographystyle{aa}
\bibliography{biblio}

\end{document}